\documentclass{ws-ijmpd}
\usepackage[super,compress]{cite}
\newcommand{\be}{\begin{equation}}
\newcommand{\ee}{\end{equation}}
\newcommand{\bea}{\begin{eqnarray}}
\newcommand{\eea}{\end{eqnarray}}
\newcommand{\bse}{\begin{subequations}}
\newcommand{\ese}{\end{subequations}}
\newcommand{\bei}{\begin{itemize}}
\newcommand{\ei}{\end{itemize}}

\newcommand{\beq}{\begin{eqnarray}}
\newcommand{\eeq}{\end{eqnarray}}

\newcommand{\nn}{\nonumber}

\usepackage[all]{xy}
\usepackage{hyperref}
\usepackage{epsfig}
\usepackage{color}
\usepackage{verbatim}
\usepackage{graphicx}
\usepackage[active]{srcltx}
\usepackage{color}
\makeatletter \@addtoreset{equation}{section}
\renewcommand{\theequation}{\thesection.\arabic{equation}}

\begin{document}

\markboth{E. Ebrahimi, H. Golchin, A. Mehrabi, S. M. S. Movahed}
{Consistency of Nonlinearly Interacting Ghost Dark Energy with
Recent Observations}

%
\catchline{}{}{}{}{}
%

\title{Consistency of nonlinear interacting ghost dark energy with recent observations}

\author{E. Ebrahimi}

\address{Faculty of Physics, Shahid Bahonar University,\\
Kerman, PO Box 76175, Iran;\\
Research Institute for Astronomy and Astrophysics of
Maragha (RIAAM), Maragha, Iran\\
eebrahimi@uk.ac.ir}

\author{H. Golchin}

\address{Faculty of Physics, Shahid Bahonar University,\\
Kerman, PO Box 76175, Iran\\
h.golchin@.uk.ac.ir}

\author{A. Mehrabi}

\address{Department of Physics, Bu-Ali Sina University, Hamedan
    65178, 016016, Iran\\
Mehrabi@basu.ac.ir}

\author{S. M. S. Movahed}

\address{Department of Physics, Shahid Beheshti University, G.C., Evin, Tehran 19839, Iran\\
 School of Physics, Institute for Researches in Fundamental
Sciences (IPM), P.O.Box 19395-5531,Tehran, Iran\\
m.s.movahed@ipm.ir}

\maketitle

\begin{history}
\received{Day Month Year}
\revised{Day Month Year}
\end{history}

\begin{abstract}
In this paper we investigate ghost dark energy model in the
presence of non-linear interaction between dark energy and dark
matter. We also extend the analysis to the so called generalized
ghost dark energy (GGDE) which $\rho_D=\alpha H+\beta H^2$. The
model contains three free parameters as $\Omega_D,
\zeta(=\frac{8\pi G \beta}{3})$ and $b^2$ (the coupling
coefficient of interactions). We propose three kinds of non-linear
interaction terms and discuss the behavior of equation of state,
deceleration and dark energy density parameters of the model. We
also find the squared sound speed and search for signs of
stability of the model. To compare the interacting GGDE model with
observational data sets, we use more recent observational
outcomes, namely SNIa from JLA catalog, Hubble parameter, baryonic
acoustic oscillation and the most relevant CMB parameters
including, the position of acoustic peaks, shift parameters and
redshift to recombination. For GGDE with the first non-linear
interaction, the  joint analysis indicates that
$\Omega_D=0.7192\pm0.0062$, $b^2=0.146^{+0.030}_{-0.026}$ and
$\zeta=0.104\pm0.047$ at 1 optimal variance error. For the second
interaction, the best fit values at $1\sigma$ confidence are
$\Omega_D=0.72091\pm0.0065$, $b^2=0.0395\pm0.0080$ and
$\zeta\le0.0173$. According to combination of all observational
data sets considered in this paper the best fit values for third
non-linearly interacting model are $\Omega_D=0.7287\pm0.0062$,
$b^2=0.0109\pm0.0023$ and $\zeta\le0.00764$ at $1\sigma$
confidence interval.  Finally we found that the presence of
interaction is compatible  in mentioned models via current
observational datasets.\end{abstract}


\ccode{PACS numbers:}


\section{Introduction}

Accelerating expansion of the universe is a mysterious event in
the modern cosmology. The most relevant observational results
based on SNIa as robust standard candle around 1998 demonstrated
that our universe is experiencing an accelerative phase of
expansion at the late time
\cite{Riess:1998cb,Perlmutter:1998np,bernardis,perl2}. After that,
other observational data sets provided by cosmic microwave
background (CMB) \cite{hanany,netter,spergel,Ade:2015rim} and
large scale structure
 as well as baryon acoustic oscillation
\cite{coll,teg,cole,springel,Percival:2009xn,Blake:2011rj,Reid:2012sw}
favored mentioned acceleration phase. On the other hand, there are
many models to achieve the late time acceleration epoch for
evolution of our universe. In a more general and famous approach,
the responsible for mentioned phase is generically relegated to
the influence of a strange form of energy which is the so-called
dark energy (DE). The simplest candidate for this approach is
cosmological constant
\cite{Sahni:1999gb,Carroll:2000fy,Copeland:2006wr,luon}.  Such proposal
is noticeable in the case that the right underlying theory of
gravity is supposed to be general relativity (GR). Moreover there
are some approaches to modify gravity in cosmological scales, in
order to include the late time accelerating epoch
\cite{DeFelice:2010aj,Tsujikawa:2010zza,Clifton:2011jh,Amendola12a}.
Dynamical models of dark energy (models with time varying equation
of state parameter) are also allowed from  theoretical point of
view, such models are supported by observational evidences. An
incomplete list of examples includes time varying cosmological
constant\cite{sola},
quintessence\cite{wett,Ratra,duta,saridakis,Rahvar:2006tm},
phantom fields \cite{caldwell,nojiri1,singh,hu,setare,saridakis2},
holographic dark energy which originate from quantum gravity
approach\cite{hsu,li,li1,nojiri2,saridakis3,saridakis4}, age
graphic dark energy \cite{cai,weicai,weicai2} and so on. An
interesting problem which potentially can be resolved in dynamical
approach is the so-called ``coincidence problem", which simply
asks ``why dark energy component becomes dominant at present
time?".

Ghost dark energy (GDE), is another dynamical DE model introduced
in \cite{urban,ohta}. GDE is based on the Veneziano ghost field
which has already been introduced to solve the $U(1)$ problem in
$QCD$ theory \cite{kawar,witten,Veneziano,rosen,nath}. In a
Minkowskian spacetime there is not any observable consequence from
the ghost field but turning into a dynamical spacetime the ghost
field affects the vacuum energy density.

In the lowest level, ghost dark energy contribution to the vacuum
energy density can be taken as $\Lambda^3_ {QCD} H$, where $H$ is
the Hubble parameter and $\Lambda^3_ {QCD}$ is QCD mass scale
\cite{ohta}. In \cite{CaiGhost,urban}, authors showed that this
contribution is enough to drive an acceleration in the cosmic
background. Considering the values $\Lambda_{\rm QCD}\sim 100 MeV$
and $H \sim 10^{-33}eV$ one obtains $\Lambda^3_{\rm QCD}H\sim
(3\times10^{-3}eV)$ which solve the fine tuning
problem\cite{ohta}. Another interesting feature of this model is
that there is no need to introduce new degrees of freedom since
GDE is totally embedded in the known physics. Different features
of GDE is extensively studied
\cite{sheykhi1,sheykhi2,sheykhi3,sheykhi4,chao1,chao2,chao3}.

The influence of the ghost field to the vacuum energy density was
reconsidered in reference \cite{caighost2}.  They found that,
there exists a second order term proportional to $H^2$ which can
contribute to the vacuum energy density. Accordingly, the ghost
dark energy density can be redefined as \cite{caighost2} $\rho_D=\alpha H+\beta
H^2$  where $\alpha$ and $\beta$ are constants.
The dark energy model which is based on this relation is called
generalized ghost dark energy (GGDE). According to previous
results represented in \cite{caighost2,ebrsheyggde} considering GGDE model, the second order term ($\beta H^2$) has opposite
dynamical influence with respect to the linear term ($\alpha H$).
Subsequently, one can expect that some problems with GDE will be
removed in the generalization approach.

As mentioned before, cosmological constant or in brief $\Lambda$
is one of the first candidates can explain dynamical evolution of
the universe with accelerating expansion rate. In the standard
model of cosmology entitled $\Lambda$CDM, dark sectors namely, DE
$(\Lambda)$ and dark matter (DM) evolve separately and do not
interact. However, because of the uncleared nature of both DE and
DM, it is possible to consider interaction between the dark
sectors in the universe and there is not any theoretical reason
against such an interaction \cite{wett}. Interacting models are
also capable to solve the coincidence problem. There also exists
observational evidences which support the interacting models
\cite{interact1,oli}.

 Considering a typical interaction between DE and DM, there will
arise a simple question that what the form of the interaction term
is? Phenomenologically one can choose the interaction terms as a
linear combination of $\rho_D$ and $\rho_m$ at the simplest level.
However one can ask why just linear choices? when we work on GDE
and GGDE models with a linear regime they suffer a negative sound
speed. Seeking a solution to this problem and also looking for
better consistency with observations, one can consider non-linear
interacting models. Non-linear interacting models of DE have been
introduced and studied in the literature \cite{mangano,baldi}.
Also authors of \cite{jian} , showed that a product coupling,
i.e., an interaction term which is proportional to the product of
$\rho_D$, $\rho_m$ and $\rho_D+\rho_m$ can be consistent with
observations. Next in \cite{Arevalo:2011hh}, Arevalo et al.
proposed a general form of non-linear interaction term and studied
the cosmic dynamics of the universe in presence of such an
interaction term. For special cases of the non-linear interaction
term they found analytic solutions which is consistent with the
supernova type Ia (SNIa) data from the Union2 set. We will try
almost a same choices of non-linear interaction terms in the GDE
frame work.

Our aim in this paper is to consider a new form of non-linear
interaction term seeking better agreement with observations and
improving model defects such as stability. Observational
consistency check has vital role not only for either to accept or
to rule out underlying model but also for doing reliable
comparison between different proposed theoretical models for
describing the universe. Therefore, in this paper we put
observational constraints on the free parameters of our
interacting dynamical dark energy model. The most recent data for
SNIa Joint Light-curve Analysis (JLA) sample will be considered.
Also to make more completeness, we use Hubble parameter for
different redshift from OHD dataset. Various measurements for
determining the Baryon Acoustic Oscillation (BAO) including Sloan
Digital Sky Survey (SDSS), Baryon Oscillation Spectroscopic Survey
(BOSS), WiggleZ survey and 6dFGS survey are included. The major
contribution of DE is at late epoch so here we consider the
position of sound horizon, shift parameter and redshift of
recombination from CMB observation, nevertheless it is possible to
use CMB power spectrum to do much more completed evaluation.
Finally, the joint analysis of all mentioned observational data to
determine the best fit values for model free parameters has been
done.  As an other interesting investigation, one can explore the
influence of GDE on the large scale structure in the universe,
which will be done in separate work. One can find some examples
for comparison of  dark energy models with different observational
data sets in
\cite{afsh,qiang,cuzin,avil,rani,mores,dela,chen,semiz}.

This paper is outlined as follows. In the next section, we review
linear interacting and non-interacting GGDE models in a flat
universe. Section \ref{nli} is devoted to extension from linear to
non-linear interaction. In section \ref{observational} we will
rely on the most recent observational data to check the
consistency of our model. To this end, following indicators will
be taken into account: SNIa from JLA observations, Hubble parameter, Baryonic Acoustic oscillation and
CMB observational quantities including shift parameter, location
of the first peak and redshift to recombinations. We summarize our
results in section \ref{sum}.

\section{GGDE model with a general interaction term in flat universe}
In this section we review GGDE model in general case. We will
focus on the absence and presence of an interaction term between
dark sectors of our universe. In this paper we limit ourself to
influence of GGDE on the background evolution of universe and for
the sake of clarity, as a beginning task we start with the
Friedmann equation in a flat universe:
\begin{eqnarray}\label{Fried}
H^2=\frac{8\pi G}{3} \left( \rho_r+\rho_m+\rho_D \right),
\end{eqnarray}
where $\rho_r$, $\rho_m$ and $\rho_D$ are the energy densities of
radiation, pressure-less matter (including dark and baryonic) and
dark energy, respectively. According to the GGDE model
\cite{caighost2}, the energy density of the dark energy defined
by: \be \label{eden} \rho_D=\alpha H+\beta H^2. \ee The
dimensionless energy density of various components are:
\begin{eqnarray}\label{Omega}
 \Omega_m&\equiv&\frac{\rho_m}{\rho_{cr}}= \frac{8\pi G
\rho_m}{3 H^2}\,, \nn \\
\Omega_D&\equiv&\frac{\rho_D}{\rho_{cr}}=\frac{8\pi
G(\alpha+\beta H)}{3H}\,, \\
\Omega_r&\equiv&\frac{\rho_r}{\rho_{cr}}= \frac{8\pi G \rho_r}{3
H^2}\,. \nn
\end{eqnarray}
 where $\rho_{cr}=3H^2/(8\pi G)$, so the first Friedmann equation can be written
 as $\Omega_r+\Omega_m+\Omega_D=1$.
Here we assume that a coupling term ($Q$) exists between dark
energy and dark matter components. So the energy conservation
equations read as: \bea
\dot\rho_m+3H\rho_m&=&Q, \label{consm}\\
\dot\rho_D+3H\rho_D(1+w_D)&=&-Q,\label{consd}\\
\dot\rho_r+3H\rho_r(1+w_r)&=&0,\nonumber
 \eea
where $w_D=\frac{P_D}{\rho_D}$ and $w_r=1/3$. Summing above
equations one finds the total energy density conservation
equation: $\label{totrho} \dot\rho+3H(\rho+P)=0 $, where
$\rho=\rho_r+\rho_m+\rho_D$ and the total pressure equals
$P=P_r+P_D$. However we consider interaction between dark sectors
of our cosmos, since pressure is determined by kinetic
considerations, therefore it is possible to have particle
production (annihilation) without changing to functional form of
pressure. Also according to {\it a prior} information from
observations we consider pressureless dark matter in our analysis.
The effective equation of state for dark matter determined by
continuity equation, Eq. (\ref{consm}), equates to $w_m^{{\rm
eff}}\equiv-Q/3H\rho_m$ and for dark energy according to Eq.
(\ref{consd}),  we have $w_D^{\rm eff}\equiv w_D+Q/3H\rho_D$
\cite{gavela09,Majerotto09}. Taking the time derivative of
(\ref{Fried}) we find:
 \be\label{hdot}
\dot{H}=-4\pi G\rho_D[y(1+w_r)+1+u+w_D], \ee where
$y=\rho_r/\rho_D$ and $u=\rho_m/\rho_D$. Differentiating
(\ref{eden}) with respect to time and also replacing $\dot H$ from
(\ref{hdot}), we reach the following relation \be \label{rhodot}
\dot \rho_D=-4\pi
G\rho_D\left[y(1+w_r)+1+u+w_D\right](\alpha+2\,\beta H)\,, \ee
 so the conservation equation for DE (\ref{consd}), takes to the form
 \be
\label{rhod2} 4\pi G\,u(\alpha+2\beta
H)-\frac{Q}{\rho_D}=(1+w_D)\big[3H-4\pi G\,(\alpha+2\beta
H)\big]-4\pi G y(1+w_r)(\alpha+2\beta H), \ee now introducing
$\zeta\equiv\frac{8\pi G \beta}{3}$ and using the following
relation \be \label{e1} \frac{4\pi G}{3H}\,(\alpha+2\beta
H)=\frac{\Omega_D+\zeta}{2}\,, \ee
 which is obtained from $\Omega_D$ in (\ref{Omega}), it is possible to
rewrite (\ref{rhod2}) as \be
u(\Omega_D+\zeta)-\frac{2Q}{3H\rho_D}=(1+w_D)(2-\Omega_D-\zeta)-
y(1+w_r)(\Omega_D+\zeta), \ee finally, noticing that
$u=\frac{\Omega_m}{\Omega_D}=\frac{1}{\Omega_D}-1$, the equation
of state parameter $w_D$ can be obtained as \bea \label{wdg}
w_D&=&\frac{\zeta-\Omega_D-Q'\,\Omega_D-\Omega_r\Omega_D-\zeta\Omega_r+y(1+w_r)
(\Omega_D+\zeta)\Omega_D}{\Omega_D(2-\Omega_D-\zeta)}\,,\\
Q'&=&\frac{2Q}{3H \rho_D}.\nonumber \eea
Now let us turn to the deceleration parameter which is defined in
terms of scale factor $a$, as: \be \label{dec2}
q=-\frac{a\ddot{a}}{\dot{a}^2}=-1-\frac{\dot{H}}{H^2}, \ee
substituting $\dot H$ from (\ref{hdot}) and using the definition
of $\Omega_D$ in (\ref{Omega}), one can finds

\be \label{decel} q=-1+\frac32\, \Omega_D
\left[y(1+w_r)+1+u+w_D\right]. \ee

Using the relation ${\dot\Omega_D}= H \frac{d\Omega_D}{d\ln a}$
one obtains the evolution of DE density parameter as
$\frac{d\Omega_D}{d\ln a}= -\frac{8\pi G \alpha}{3H}\,\frac{\dot
H}{H^2}$,  noting (\ref{e1}), the evolution of $\Omega_D$ can be
written as: \bea \label{doda} \frac{d\Omega_D}{d\ln
a}&=&\frac{4\pi G\alpha}{H^2(\alpha+\beta
H)}\left[y(1+w_r)+1+u+w_D\right]\,.
 \eea
Since we are interested in late time universe, the contribution of
radiation is negligible. In this case, the above equation can be
written as: \bea
 \frac{d\Omega_D}{d\ln
a}&=&\frac32\,(\Omega_D-\zeta)(1+w_D \Omega_D). \eea

Hereafter we consider $\Omega_r\to 0$ in all relevant equations,
however for observational consideration, where the radiation has
important contribution we will keep radiation density in relevant
equations. Till now, we have shown the results in terms of  a
general form $Q$ for interaction between DE and DM. To test the
validity of the above results, we find the cosmological parameters
$w_D$ and $q$ for some definite interactions. In the case of
non-interacting model, setting $Q=Q'=0$ in (\ref{wdg}),
(\ref{decel}); we find \be \label{niwq}
w_D=\frac{\zeta-\Omega_D}{\Omega_D(2-\Omega_D-\zeta)}\,,\qquad
q=\frac12+
\frac32\,\frac{\zeta-\Omega_D}{\,\,2\!-\zeta\!-\Omega_D}\,, \ee
which are in agreement with the respective relations obtained in
\cite{ebrsheyggde}. Another model for interaction is $Q=3b^2 H
(\rho_m+\rho_D)=3b^2 H \rho_D(1+u)$\,. This interaction is linear
in terms of $\rho$, with a coupling constant $b^2$. In this case
one finds that $Q'=\frac{2b^2}{\Omega_D}$. Substituting this value
to (\ref{wdg}) and (\ref{decel}) leads to: \bea \label{wqli}
w_D&=&\frac{\zeta-\Omega_D-2b^2}{\Omega_D(2-\Omega_D-\zeta)}\,,\\
q&=&\frac12+\frac32\,\frac{\zeta-\Omega_D-2b^2}{\,\,2\!-\zeta\!-\Omega_D}\,,
\eea and the evolution equation for energy density of DE is given
by: \be \label{dedli} \frac{d\,\Omega_D}{d\ln
a}=3\,\Omega_D\left[\frac{1-\Omega_D}{2-
\Omega_D-\zeta}\left(1+\frac{2b^2}{\Omega_D}-\frac{\zeta}{\Omega_D}\right)-\frac{b^2}{\Omega_D}\right],
\ee which are the same as the results in
\cite{ebrsheyggde,ebrinsggde} . In the next section we will
consider some new non-linear functional forms for interaction
between dark matter and dark energy.

\begin{figure}
\center
\includegraphics[height=55mm,width=55mm,angle=0]{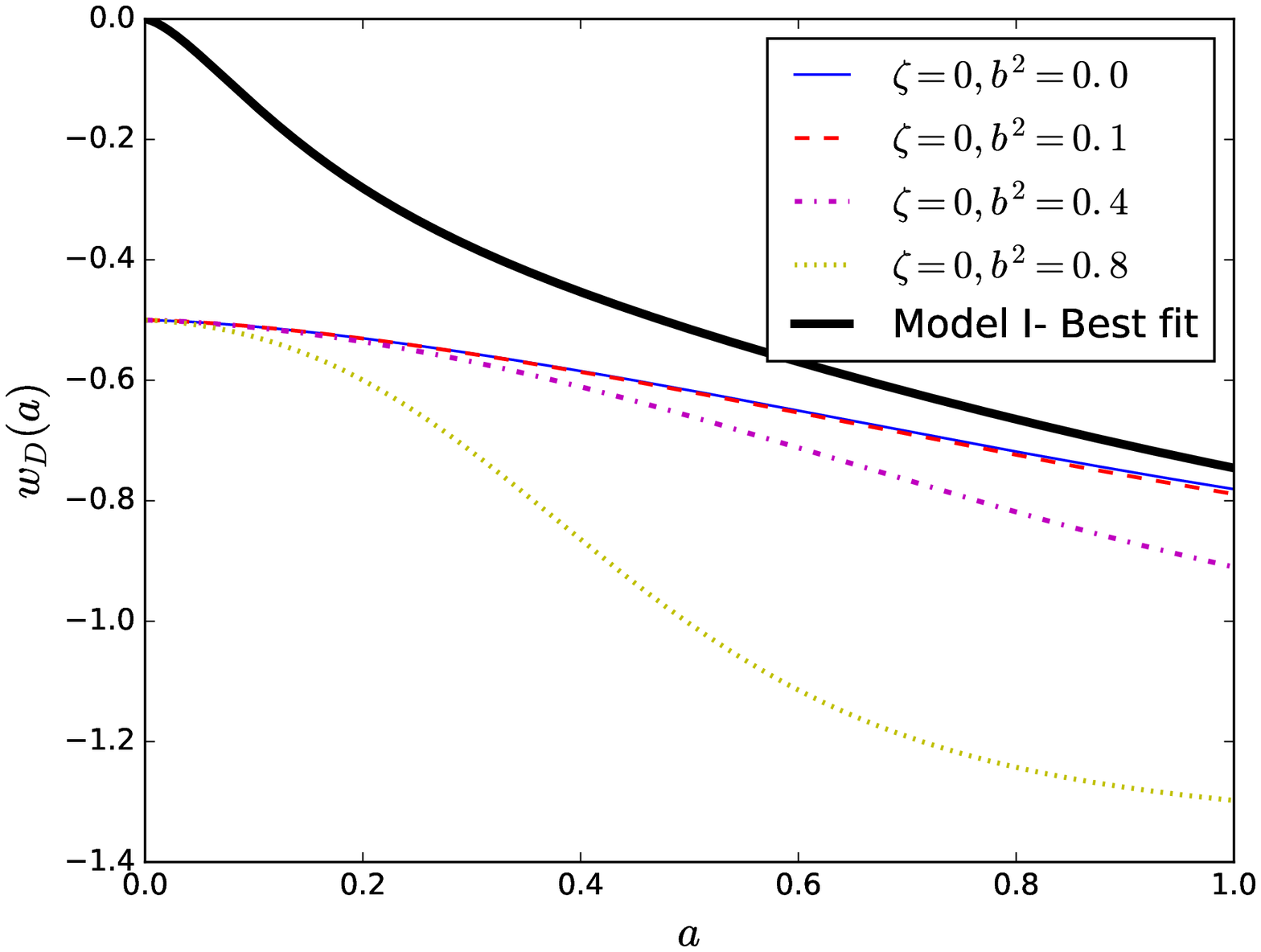}
\includegraphics[height=55mm,width=55mm,angle=0]{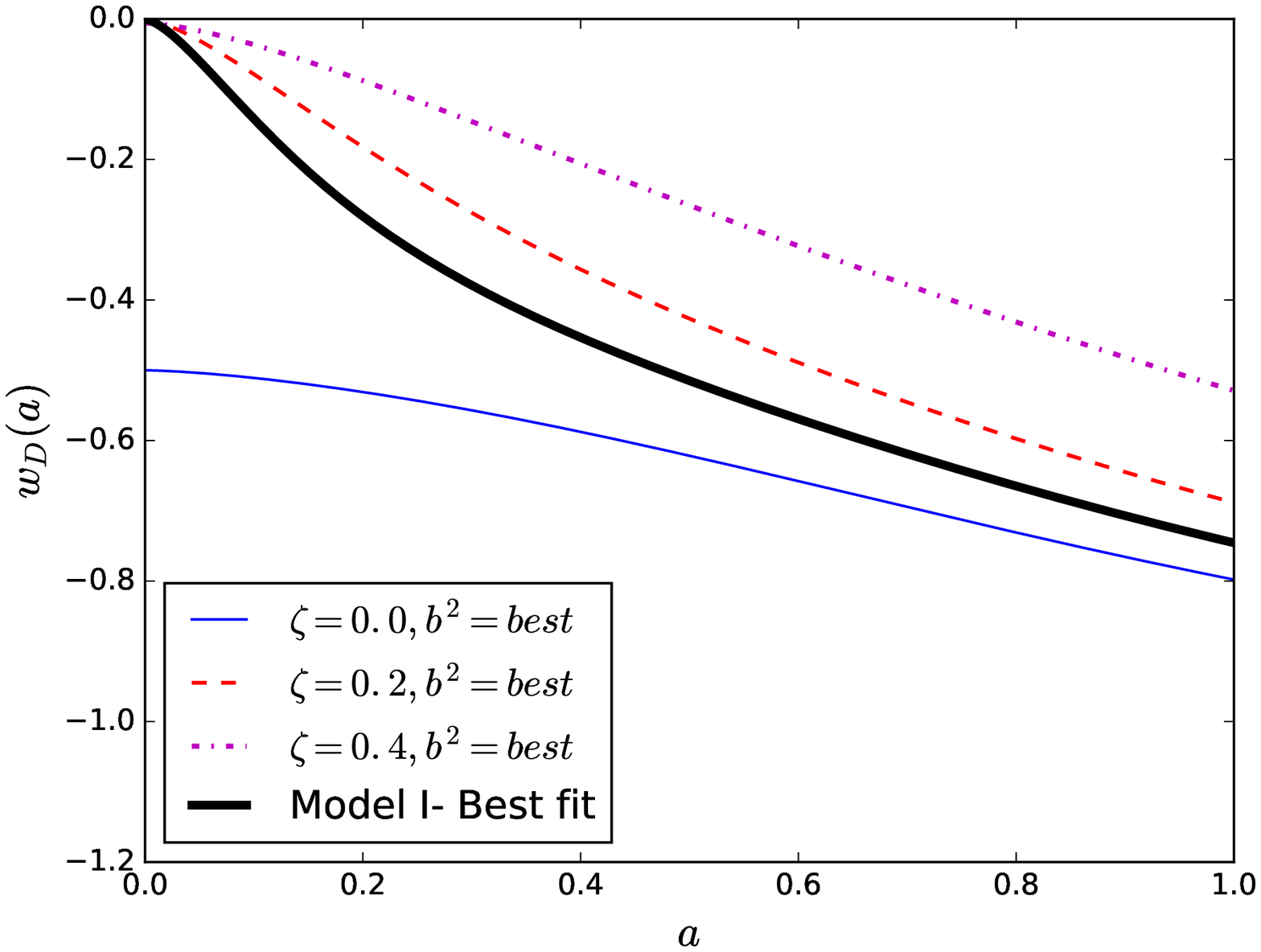}
\includegraphics[height=55mm,width=55mm,angle=0]{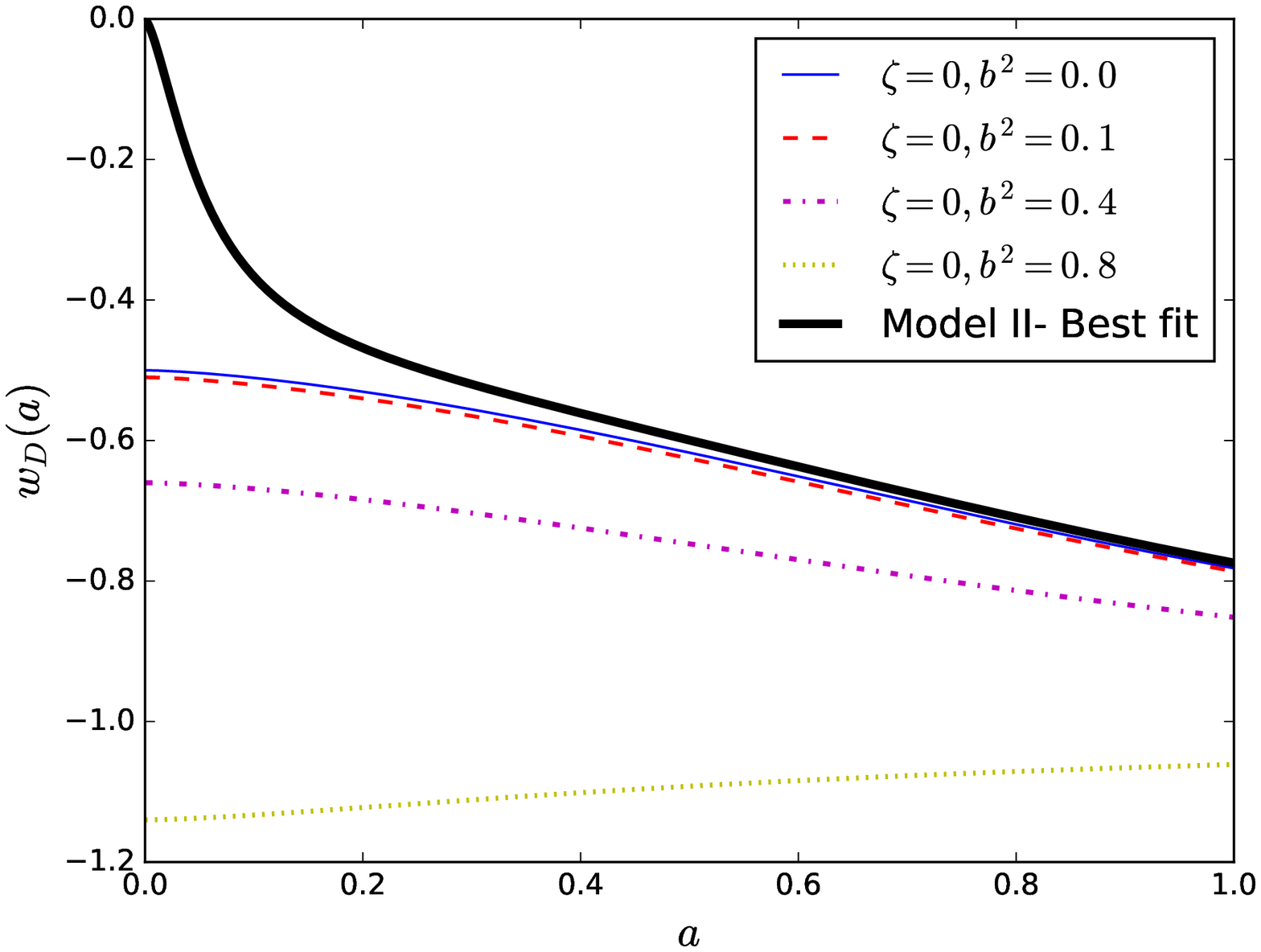}
\includegraphics[height=55mm,width=55mm,angle=0]{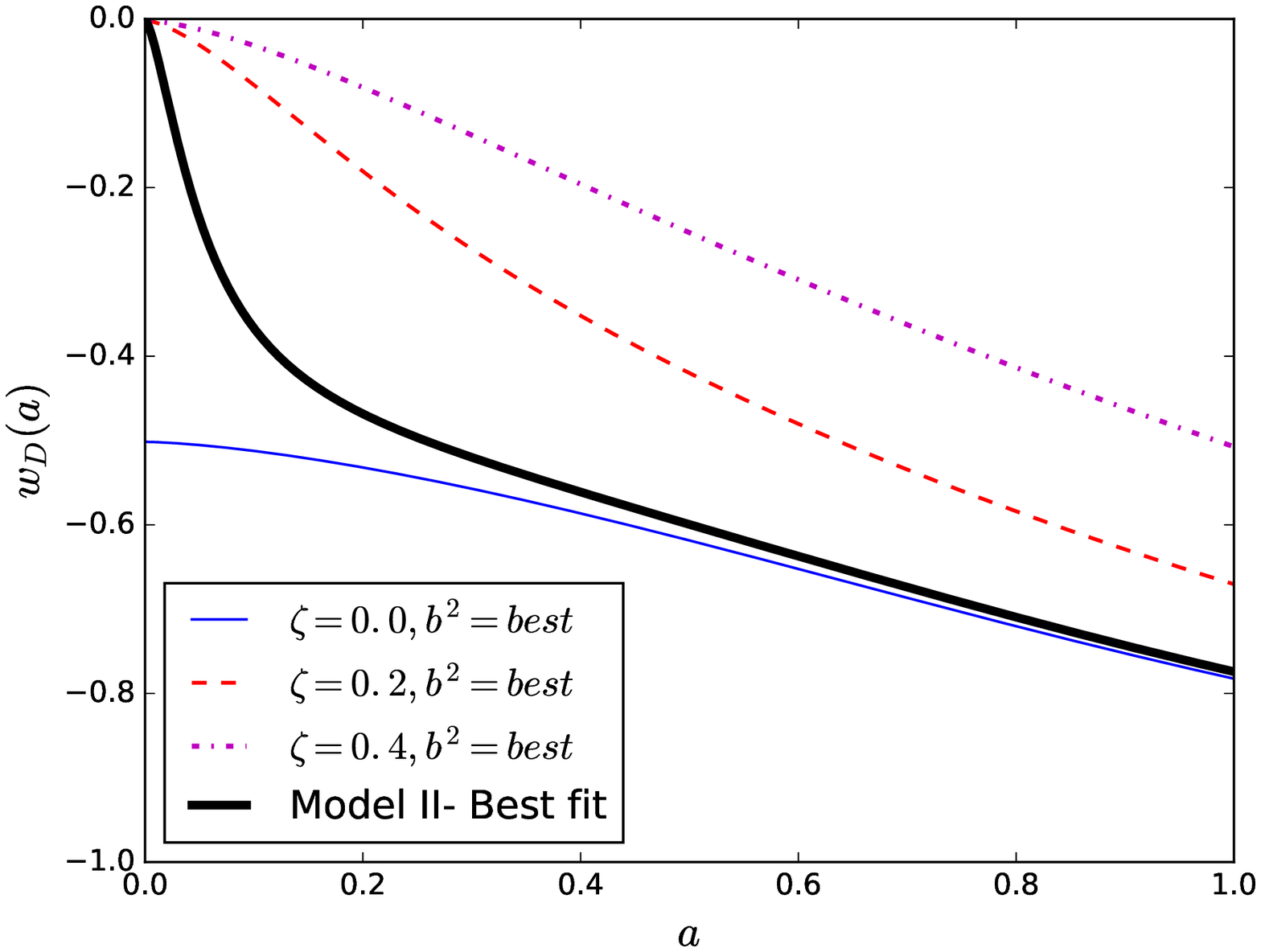}
\includegraphics[height=55mm,width=55mm,angle=0]{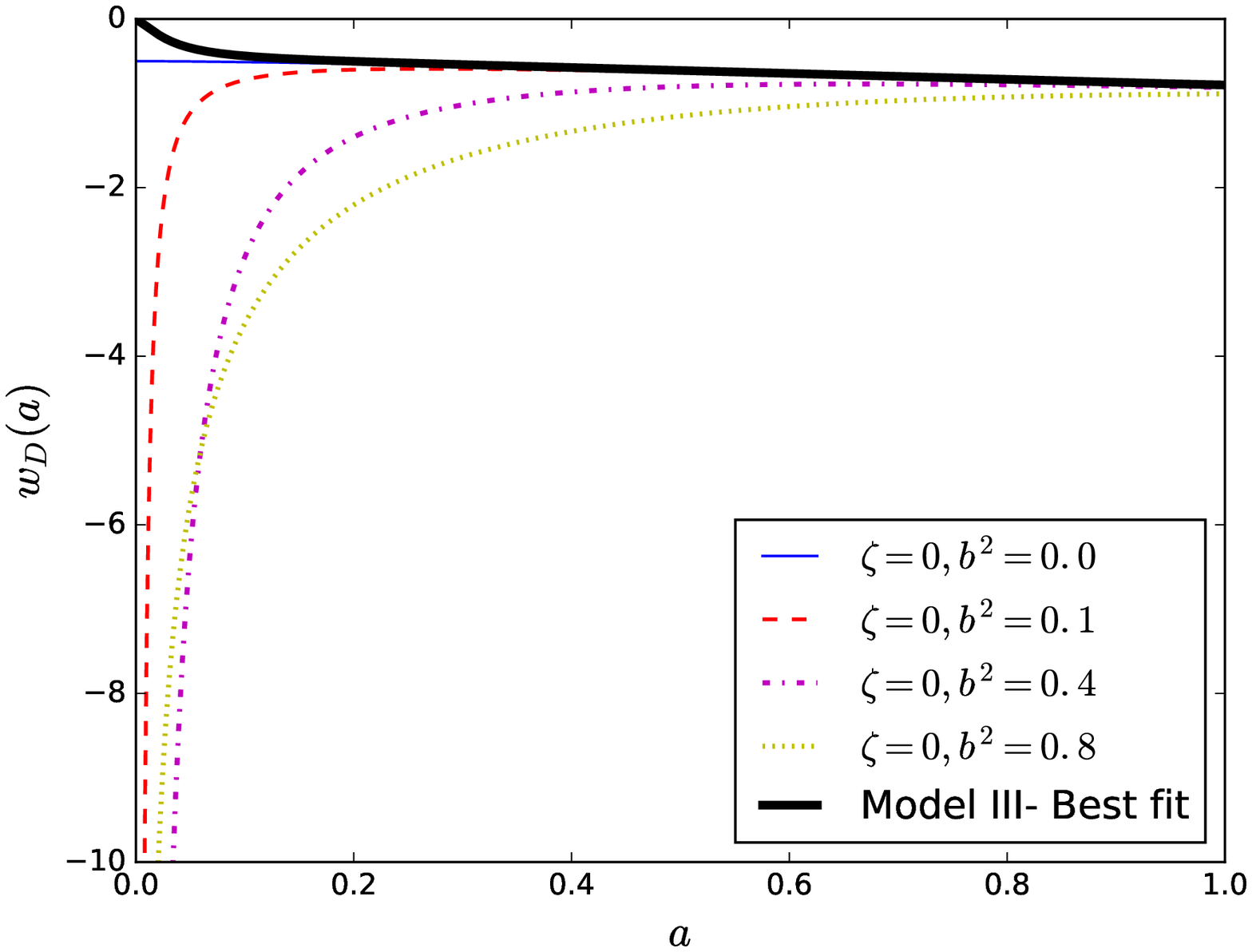}
\includegraphics[height=55mm,width=55mm,angle=0]{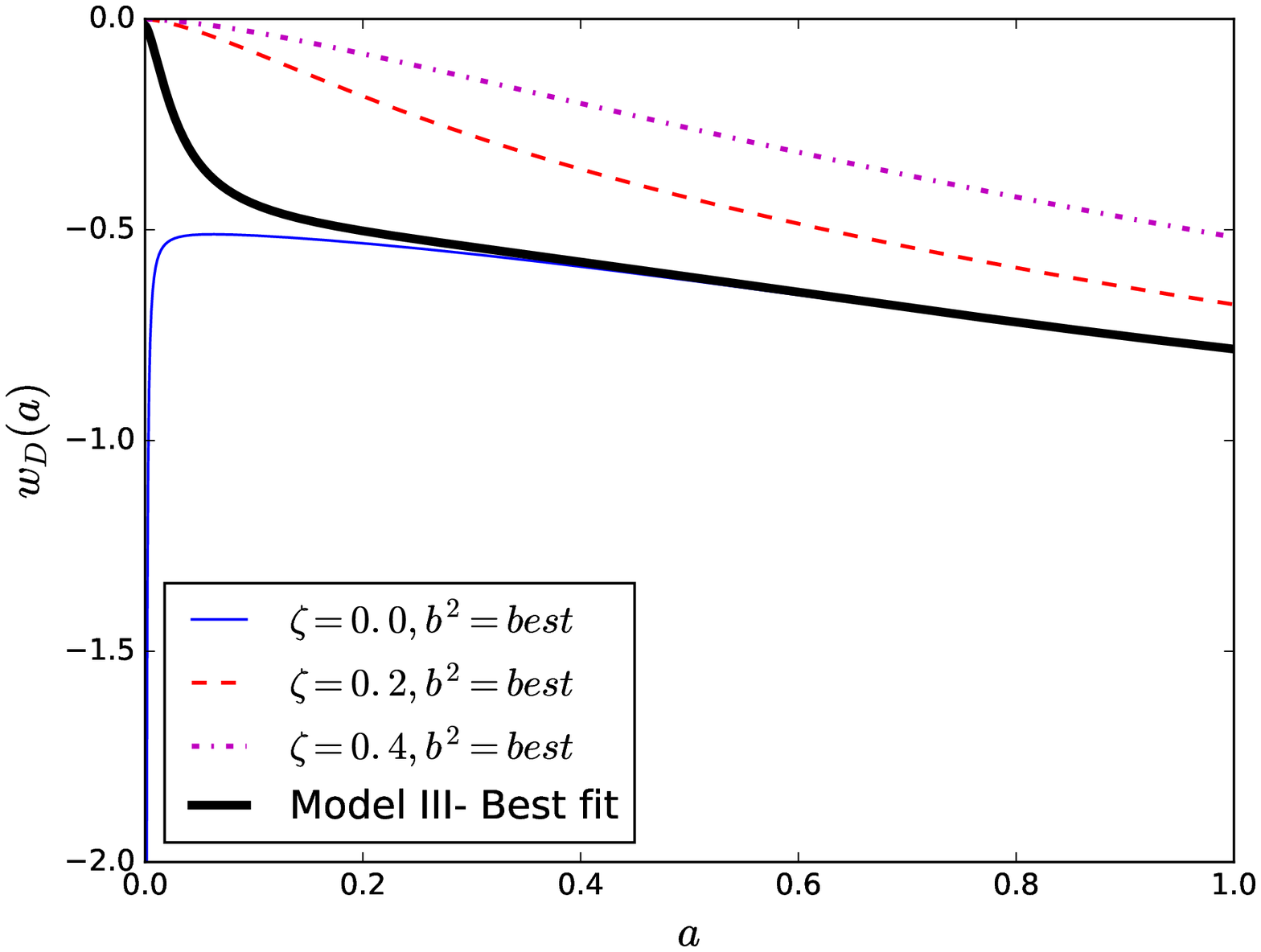}
\vspace{2mm} \caption{{\footnotesize The evolution of $w_D$ as a
function of scale factor for three models explained in the text.
Left panels are devoted to $\zeta=0$, while right panels
correspond to best fit value for $b^2$ based on joint analysis of
SNIa+Hubble+BAO+CMB. Here we consider scale factor at present time
as $a_0=1$. The value of $\Omega_D$ at the present time is $0.70$.
In all plots the dark solid line corresponds to the best fit
values determined by joint analysis. For the first model
$\Omega_D=0.7192^{+0.0062}_{-0.0062}$,
$b^2=0.146^{+0.030}_{-0.026}$ and $\zeta=0.104^{+0.044}_{-0.047}$.
For the second model $\Omega_D=0.7209^{+0.0065}_{-0.0065}$,
$b^2=0.0395^{+0.0080}_{-0.0080}$ and $\zeta\le0.0173$ and for the
third  model  $\Omega_D=0.7287^{+0.0062}_{-0.0062}$,
$b^2=0.0109^{+0.0023}_{-0.0023}$ and $\zeta\le0.00764$. }}
\label{wq}
\end{figure}

\section{Non-linear interacting GGDE model in flat universe}\label{nli}
As already mentioned, the main purpose of this paper is to examine
the impact of non-linear interaction terms on the evolution of the
universe.  Beside to the notes mentioned in introduction, it has
been shown in \cite{Golchin:2016yci} that GGDE with specific
non-linear interaction terms is a well behaved model and it shows
a radiation dominated phase at the early times, passes through a
matter dominated phase and finally ends in a stable dark energy
dominated epoch. Moreover considering the non-linear terms, cure
the linearly interacting GGDE which suffers from the absence of
radiation dominated epoch in the early times evolution of the
model \cite{Golchin:2016yci}. In choosing the form of non-linear
interaction terms, we follow the prescription presented in
\cite{Arevalo:2011hh}. The closed form of the interaction which
includes a large variety of choices is
\begin{equation}\label{formnli}
   Q=3Hb^2 \rho_m^p \rho_D^{s}\rho^{r},
\end{equation}
where $\rho=\rho_m+\rho_D$ and $b^2$ is coupling constant factor
also $p$, $s$ and $r$ are integer numbers and one can deduce from
the dimensional analysis that they should satisfy $p+s+r=1$. In
the following we choose three non-linear interactions and
investigate their properties and features.  In the selection of
these interactions we have noted that the model should be
cosmologically accepted which means that the dynamical equations
must be smooth. For instance selecting the interaction terms with
$p=0, s=0, r=1$ or $p=0, s=2, r=-1$, the dynamical equations of
the model is singular and so the model is unphysical. It must
point out that our assumption for functional form of interaction,
$Q$, removes any degeneracy with modified Chaplygin gas model.

\subsection{Model I}
In this model we  consider following non-linear interaction form
\be Q=3b^2H\frac{\rho_D^3}{\rho^2}\,, \ee for this interaction,
noticing (\ref{wdg}) one finds that $Q'=\frac{2b^2
\rho_D^2}{\rho^2}$ and considering the fact that
$\rho=\rho_m+\rho_D$, it can be rewritten as $Q'=2b^2\Omega_D^2$.
By inserting this relation to (\ref{wdg}) and (\ref{decel}) we
find \bea \label{wqnli}
w_D&=&\frac{\zeta-\Omega_D-2b^2\Omega_D^3}{\Omega_D(2-\Omega_D-\zeta)}\,,\\
q&=&\frac12+\frac32\,\frac{\zeta-\Omega_D-2b^2\Omega_D^3}{\,\,2\!-\zeta\!-\Omega_D}\,.
\eea

\begin{figure}
\center
\includegraphics[height=55mm,width=55mm,angle=0]{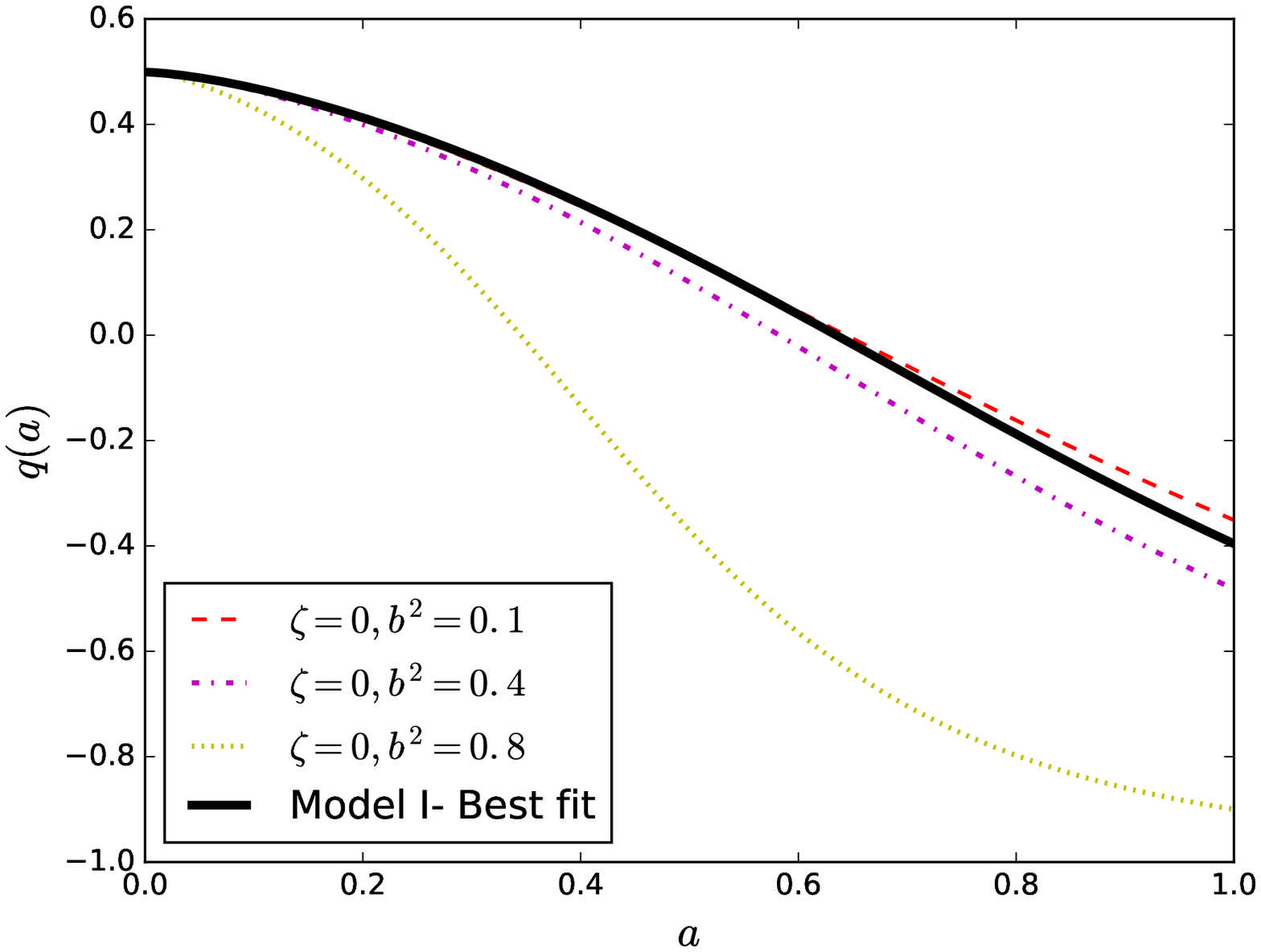}
\includegraphics[height=55mm,width=55mm,angle=0]{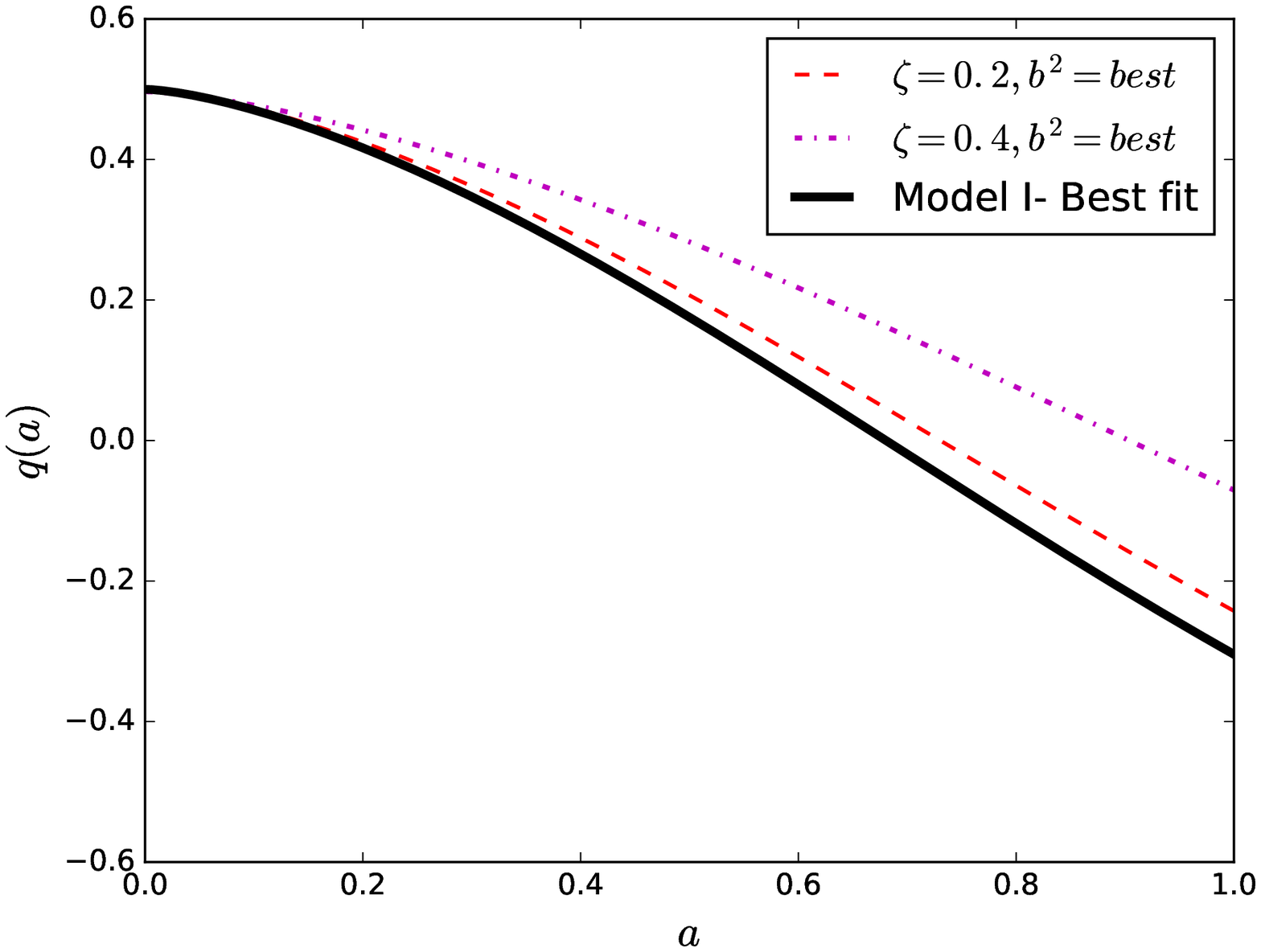}
\includegraphics[height=55mm,width=55mm,angle=0]{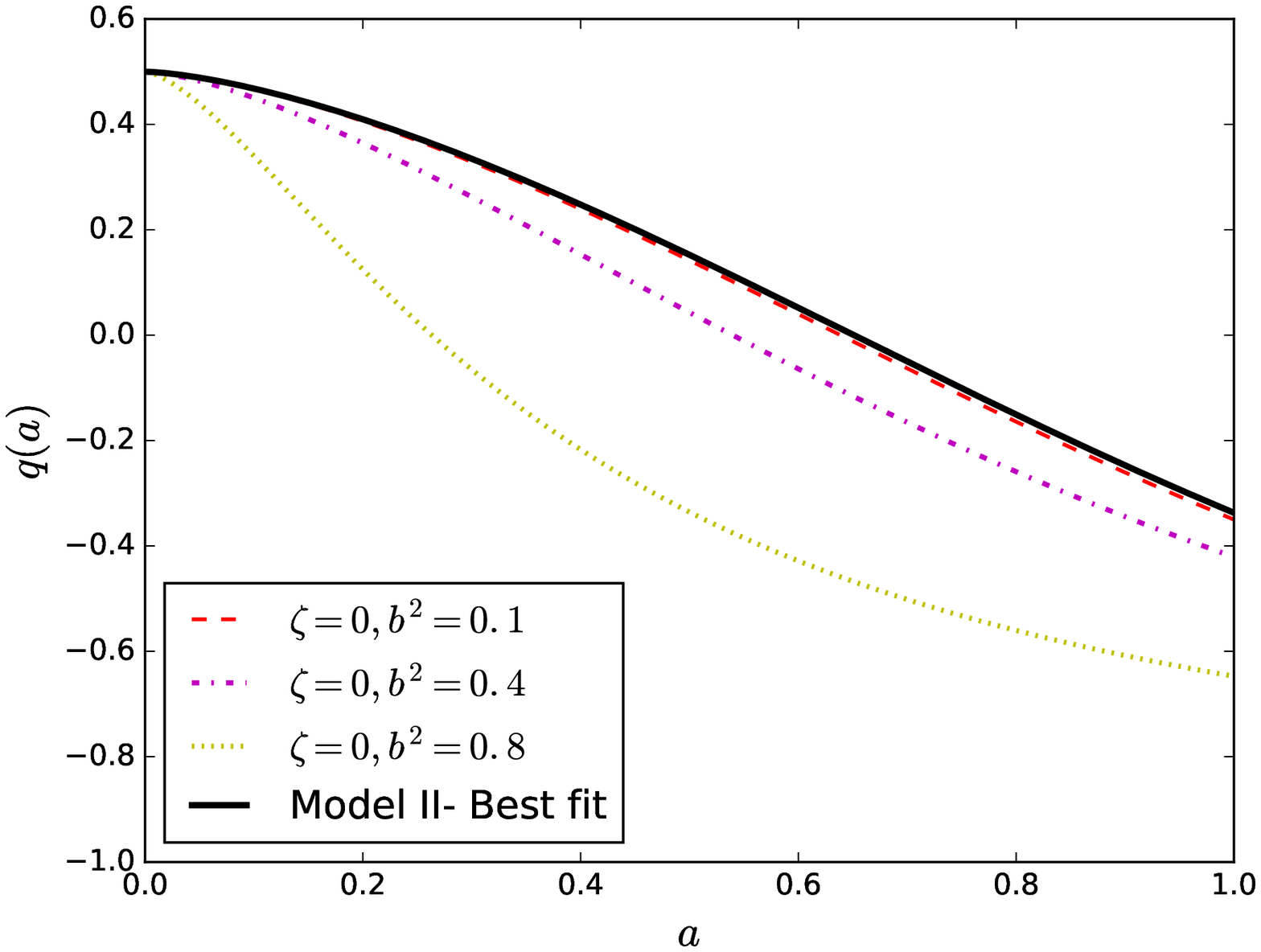}
\includegraphics[height=55mm,width=55mm,angle=0]{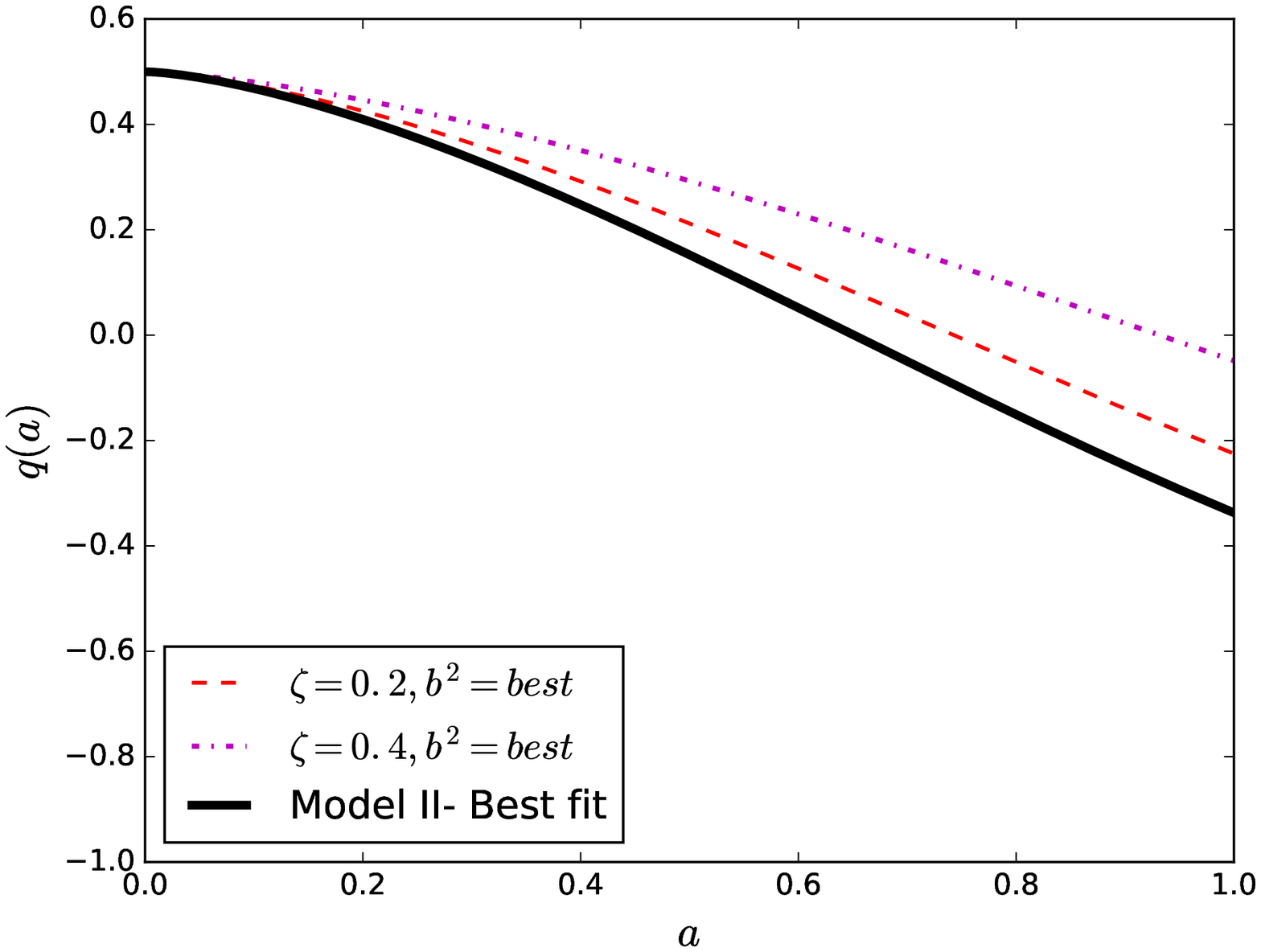}
\includegraphics[height=55mm,width=55mm,angle=0]{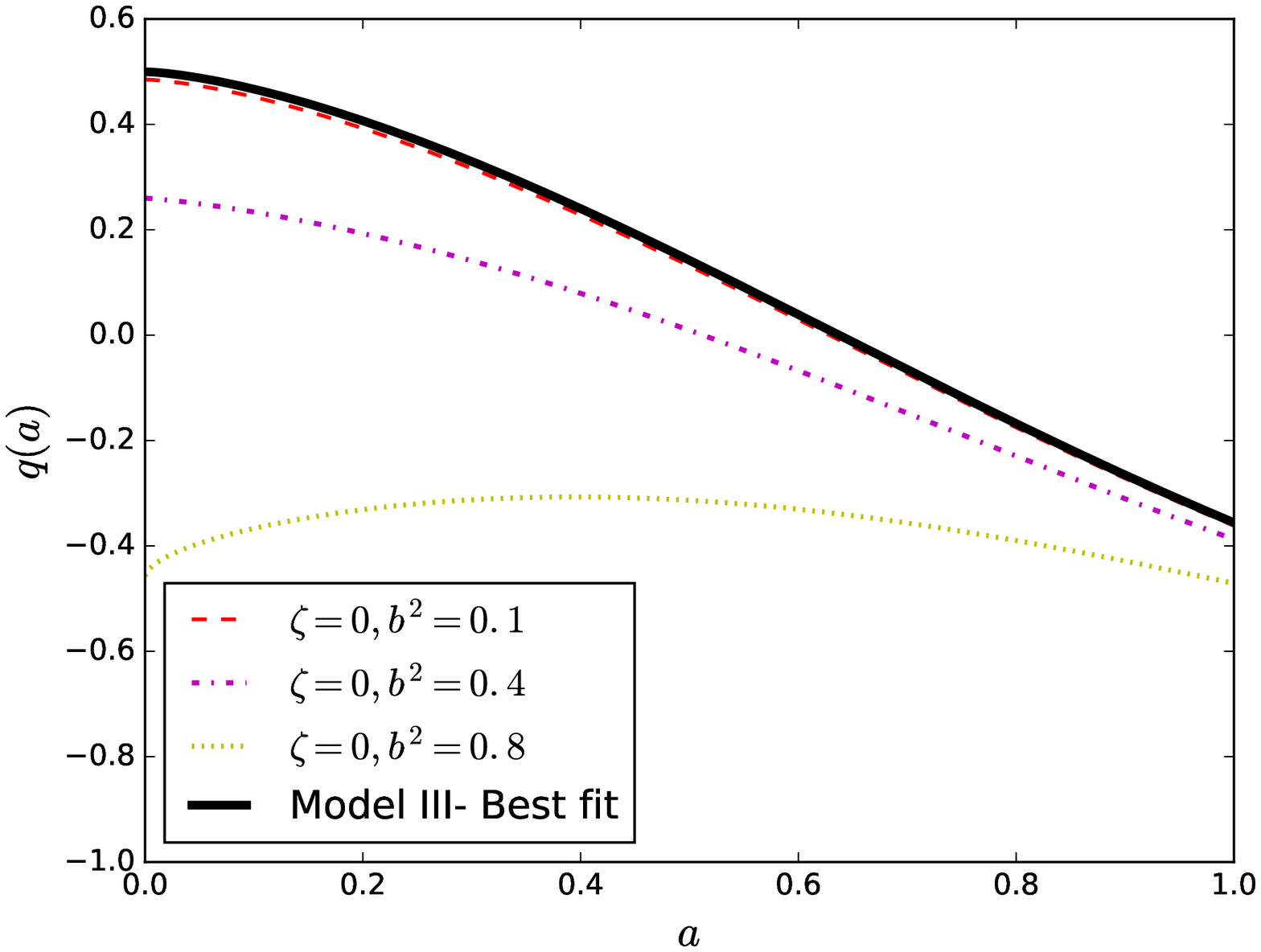}
\includegraphics[height=55mm,width=55mm,angle=0]{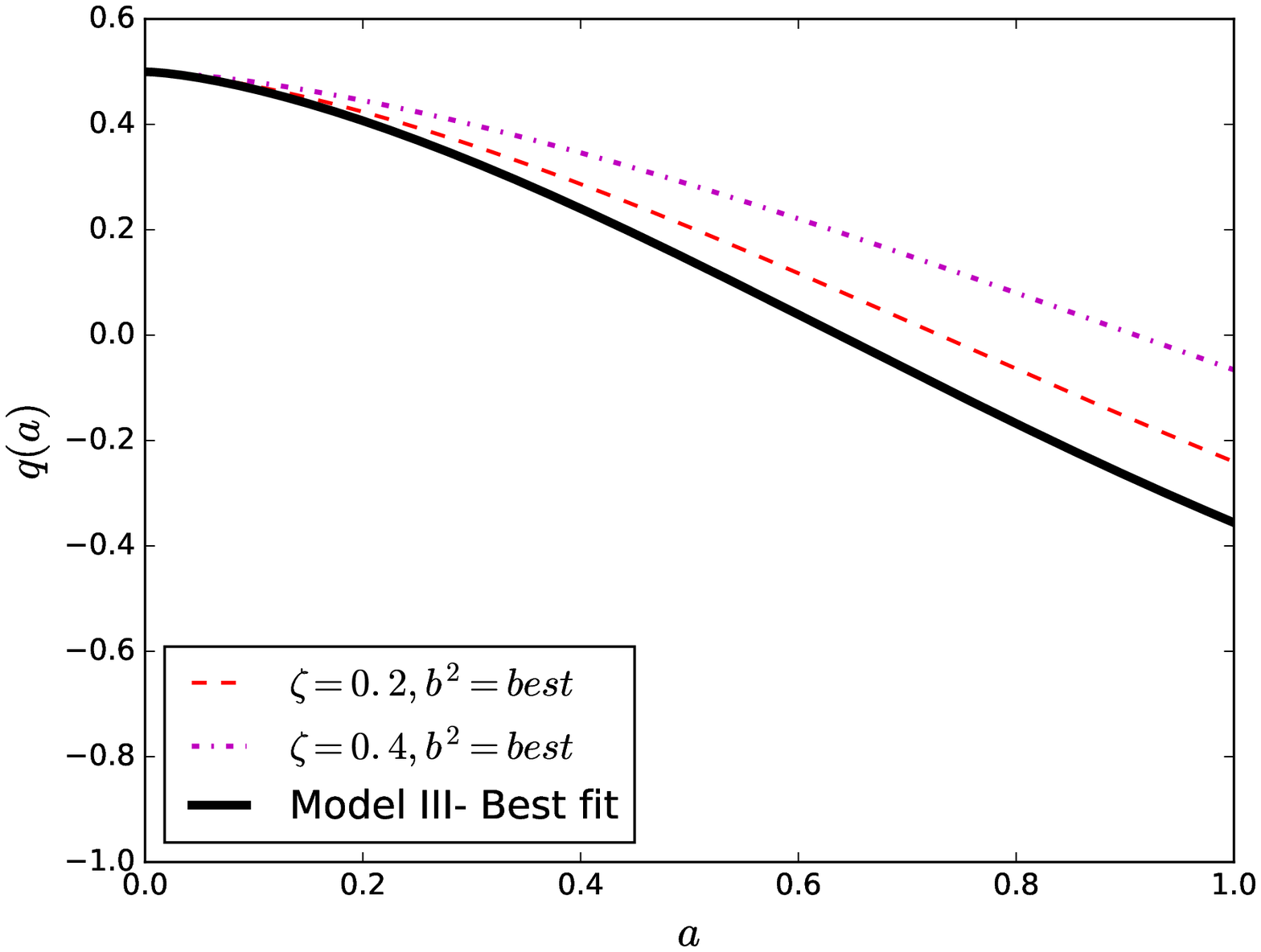}
\vspace{2mm} \caption{{\footnotesize The evolution of $q$ as a
function of scale factor for three models explained in the text.
The rest information is the same as that of mentioned in Fig.
\ref{wq}.}} \label{q}
\end{figure}

\begin{figure}
\center
\includegraphics[height=55mm,width=55mm,angle=0]{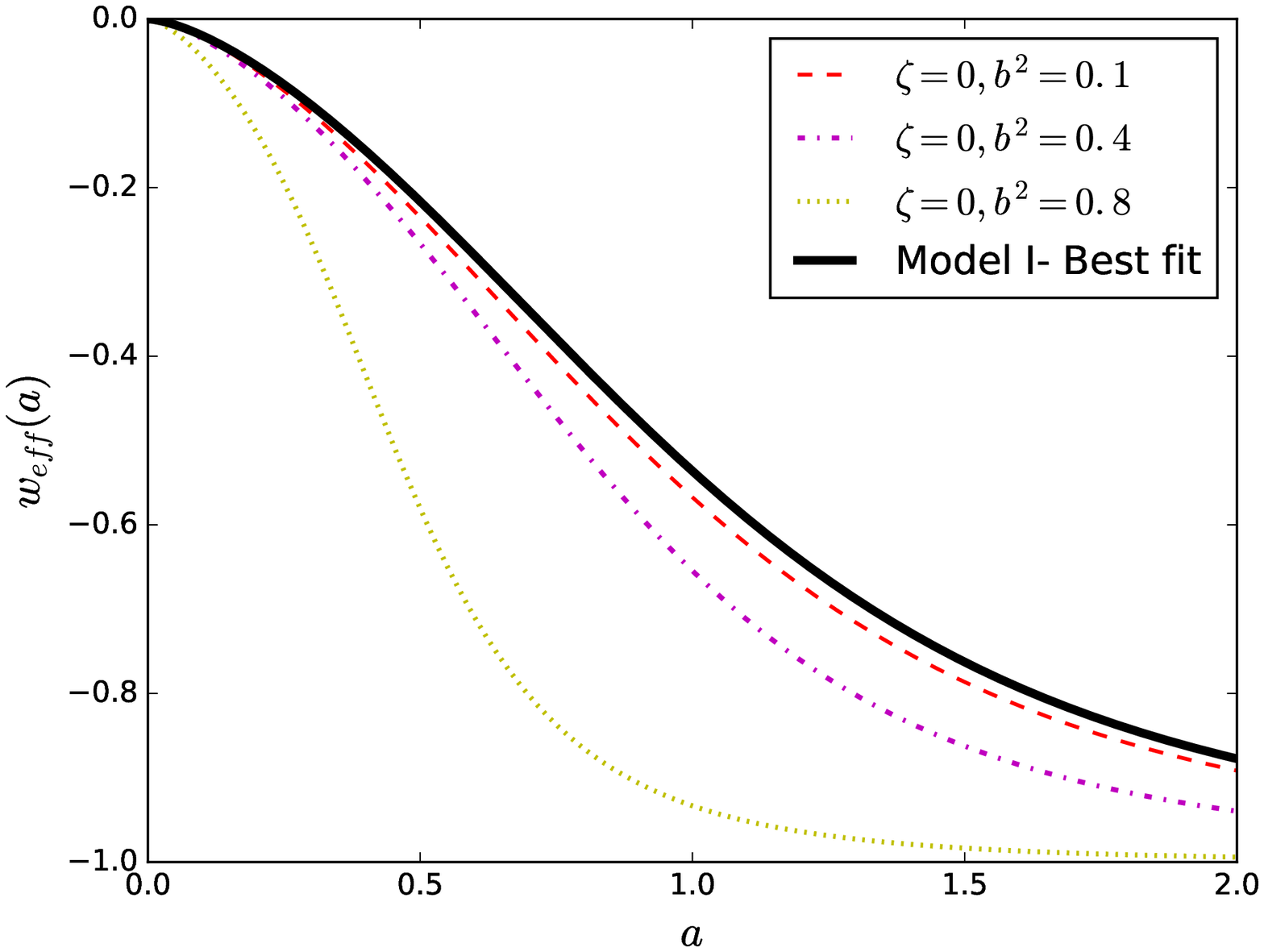}
\includegraphics[height=55mm,width=55mm,angle=0]{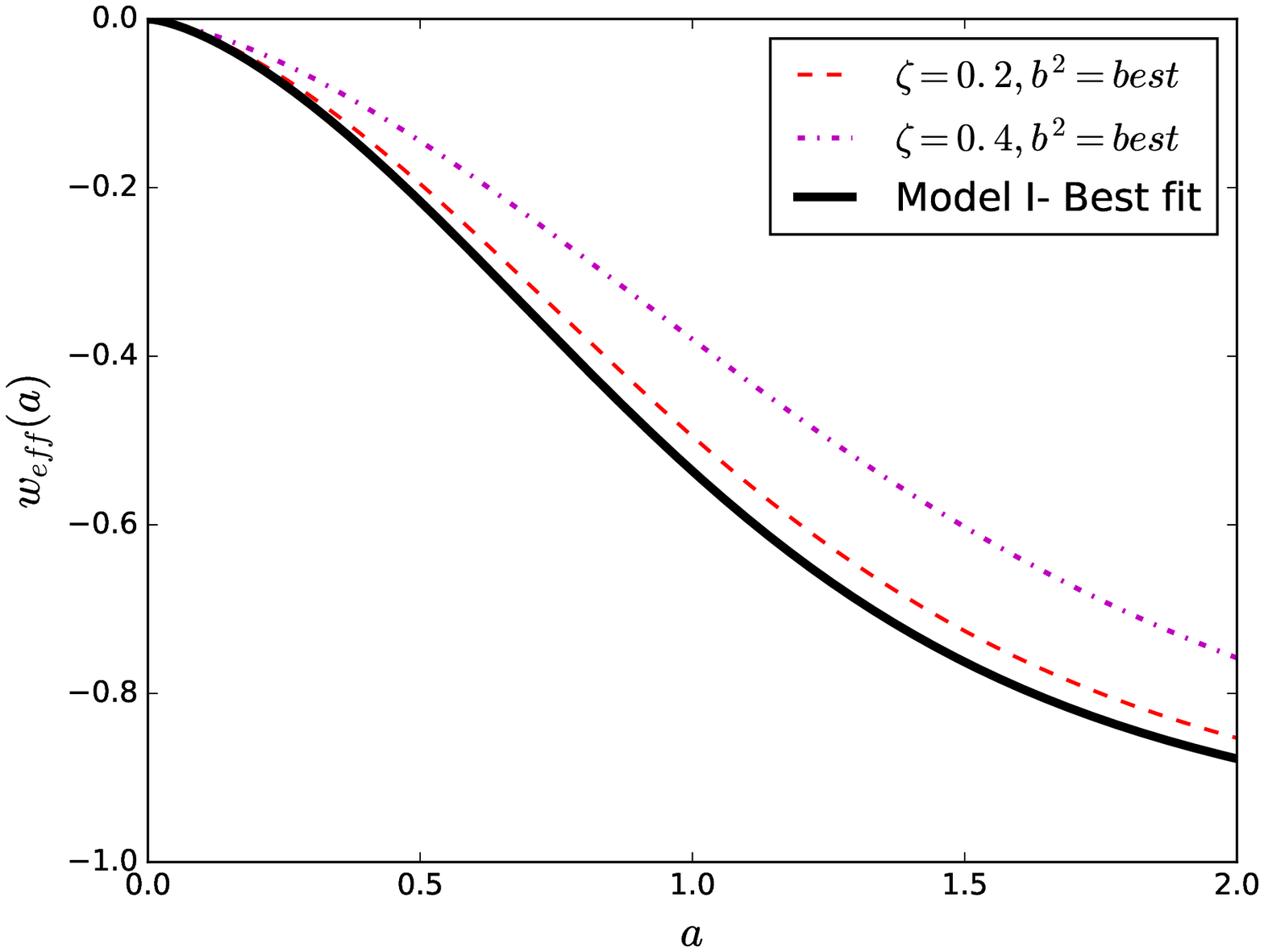}
\includegraphics[height=55mm,width=55mm,angle=0]{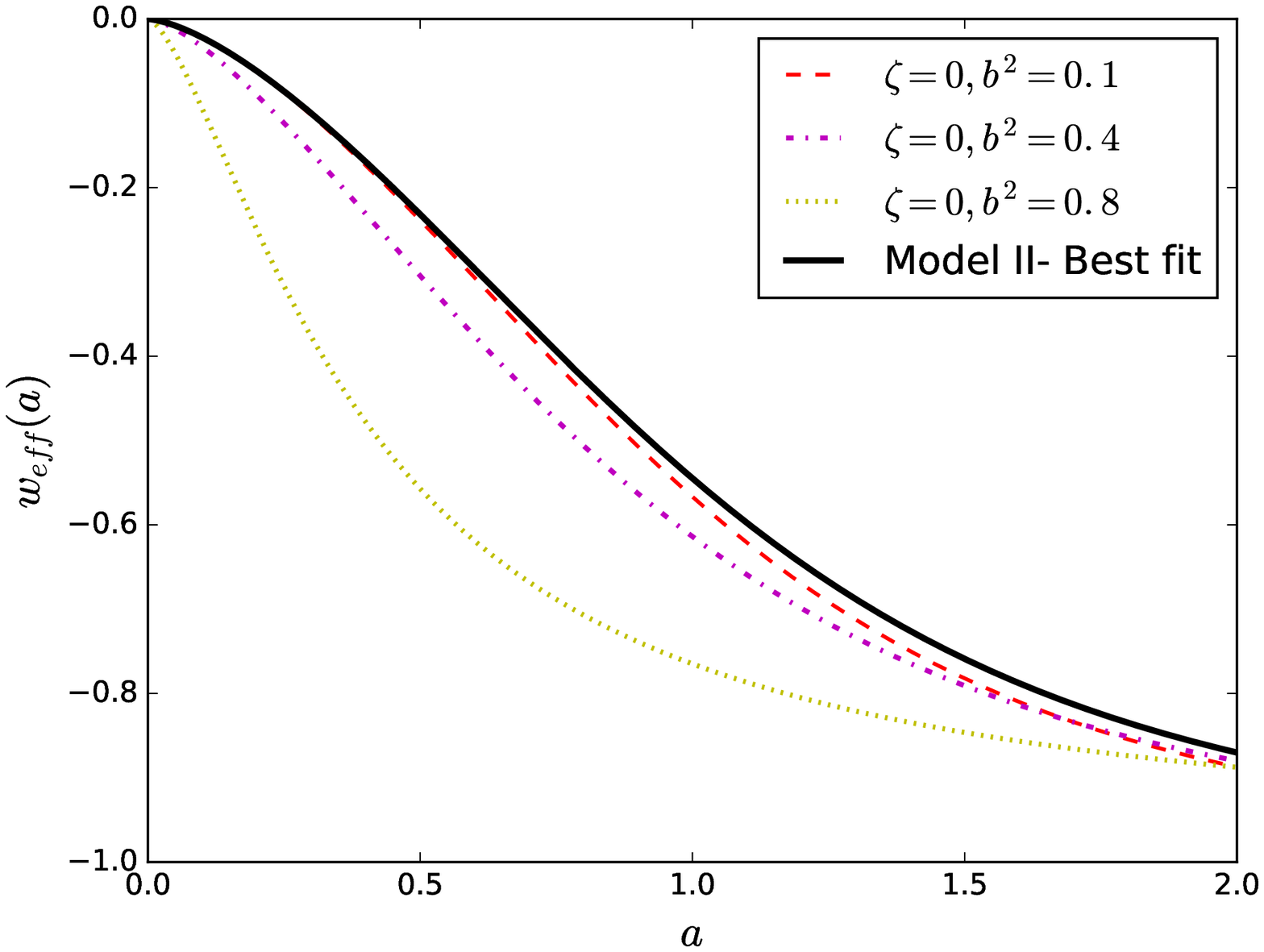}
\includegraphics[height=55mm,width=55mm,angle=0]{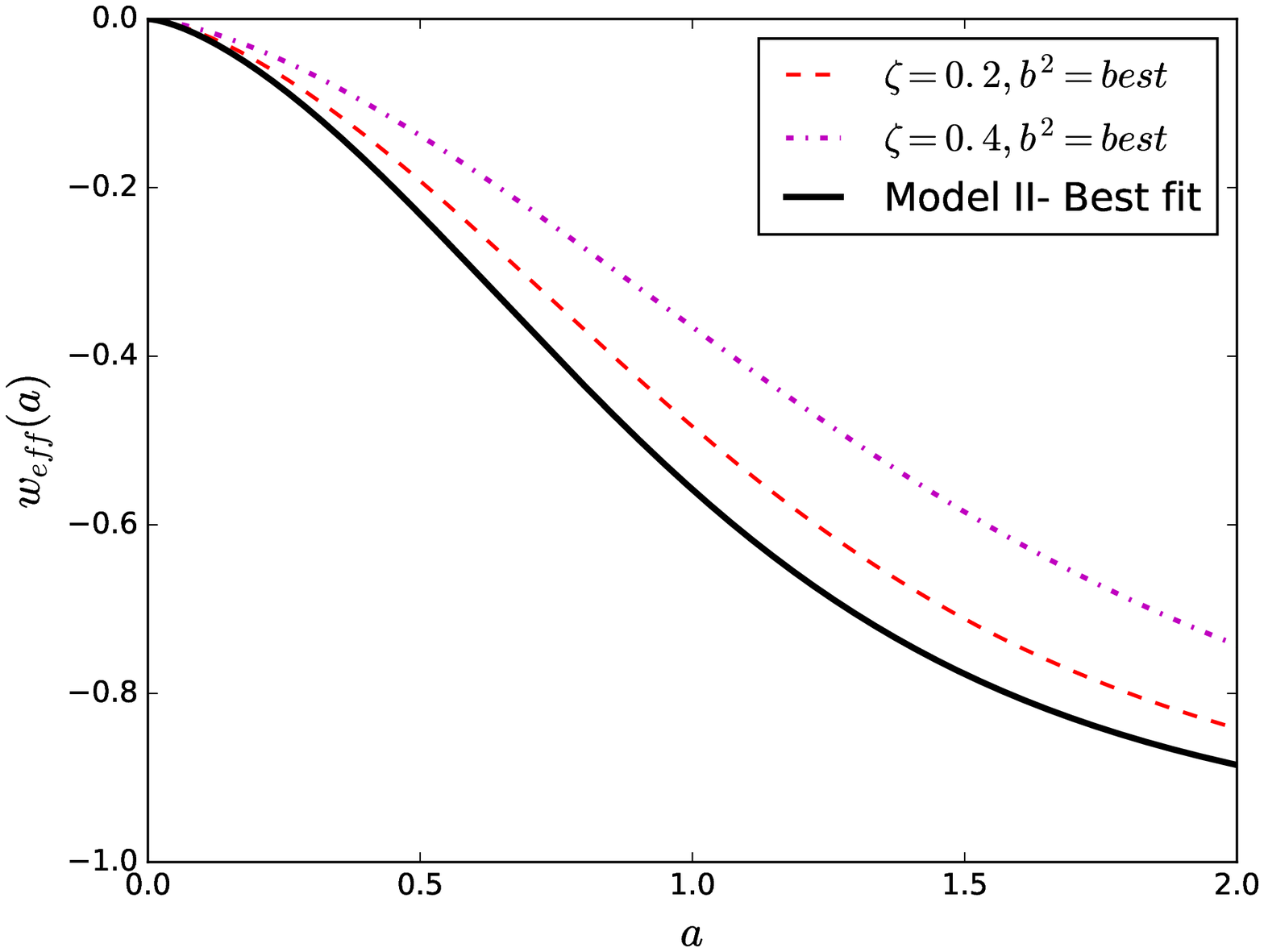}
\includegraphics[height=55mm,width=55mm,angle=0]{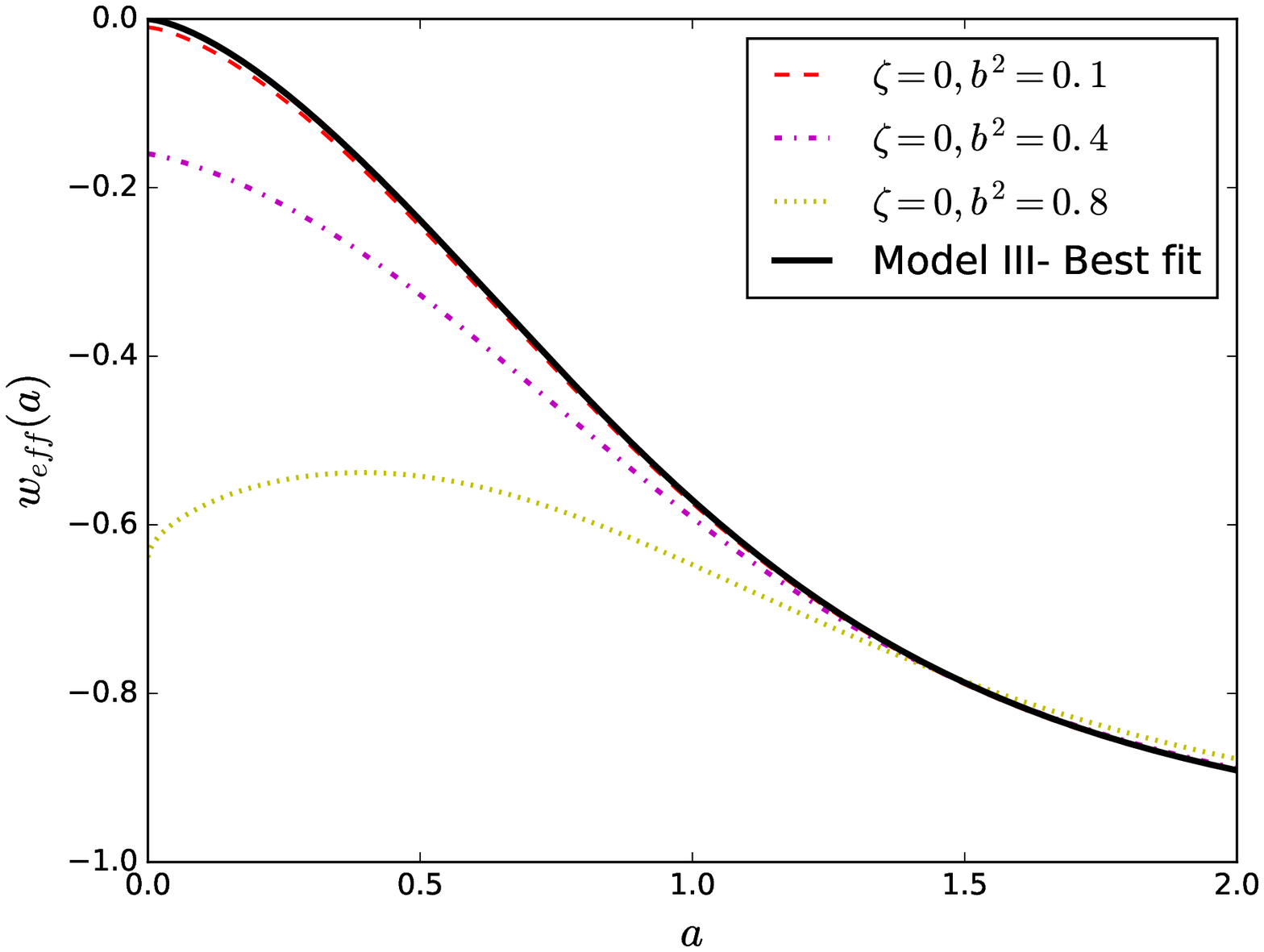}
\includegraphics[height=55mm,width=55mm,angle=0]{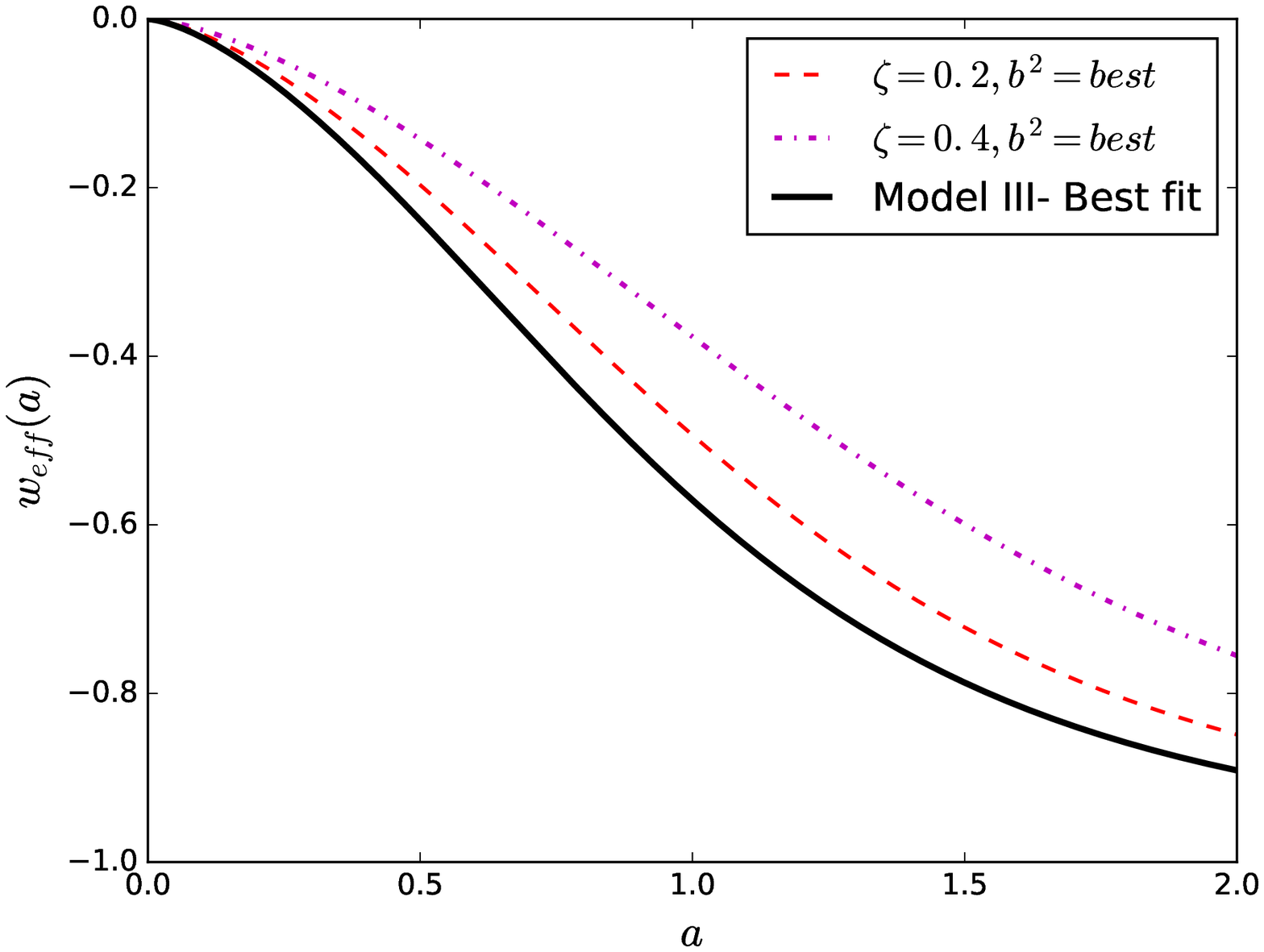}
\vspace{2mm} \caption{{\footnotesize The evolution of $w_{eff}$ as
a function of scale factor for three models explained in the text.
The rest information is the same as that of mentioned in Fig.
\ref{wq}.}} \label{weff}
\end{figure}
It turns out that  the above relations reduce to the corresponding
expressions of non-interacting model (\ref{niwq}) when $b=0$. In
Fig.(\ref{wq}) we have depicted the evolution of $w_D$ as a
function of scale factor  for various values of the free
parameters for this model. It is also clear from Fig. (\ref{wq})
that at late time, for some values of the free parameters, we get
$w_D < -1$, which show a phantom like behavior of this interacting
GGDE model. Fig. (\ref{q}) indicates the deceleration parameter as
a function of scale factor for various choices of $\zeta$ and
$b^2$ parameters. Increasing the value of $b^2$ causes to
decreasing $q$ and getting the negative value at earlier time
while according to right panel of Fig.(\ref{q}), higher values of
$\zeta$ resulting the higher values of decelerating parameter.  In
order to discuss about the fate of the universe filled with DM and
GGDE components we should consider the effective equation of state
parameter:
\begin{equation}\label{wef}
    w_{eff}=\frac{P}{\rho}.
\end{equation}
Taking $P_m=0$, it can be found that $w_{eff}=\Omega_Dw_D$.
Fig.(\ref{weff}) represents $w_{eff}$  as a function of scale
factor. Our results demonstrate that  $w_{eff}\to -1$ at future
implying that the universe will end with a big rip. It can be seen
from Fig.(\ref{weff}) that the universe enters a acceleration
phase earlier for larger values of $b$. In this case the evolution
of DE density parameter can be computed using (\ref{doda}), the
result is

 \be \label{dodlna1} \frac{d\Omega_D}{d\ln
a}=\frac{3(\zeta-\Omega_D)(-1+\Omega_D+b^2\Omega_D^3)}{2-\Omega_D-\zeta}\,.
\ee

\begin{figure}
\center
\includegraphics[height=45mm,width=60mm,angle=0]{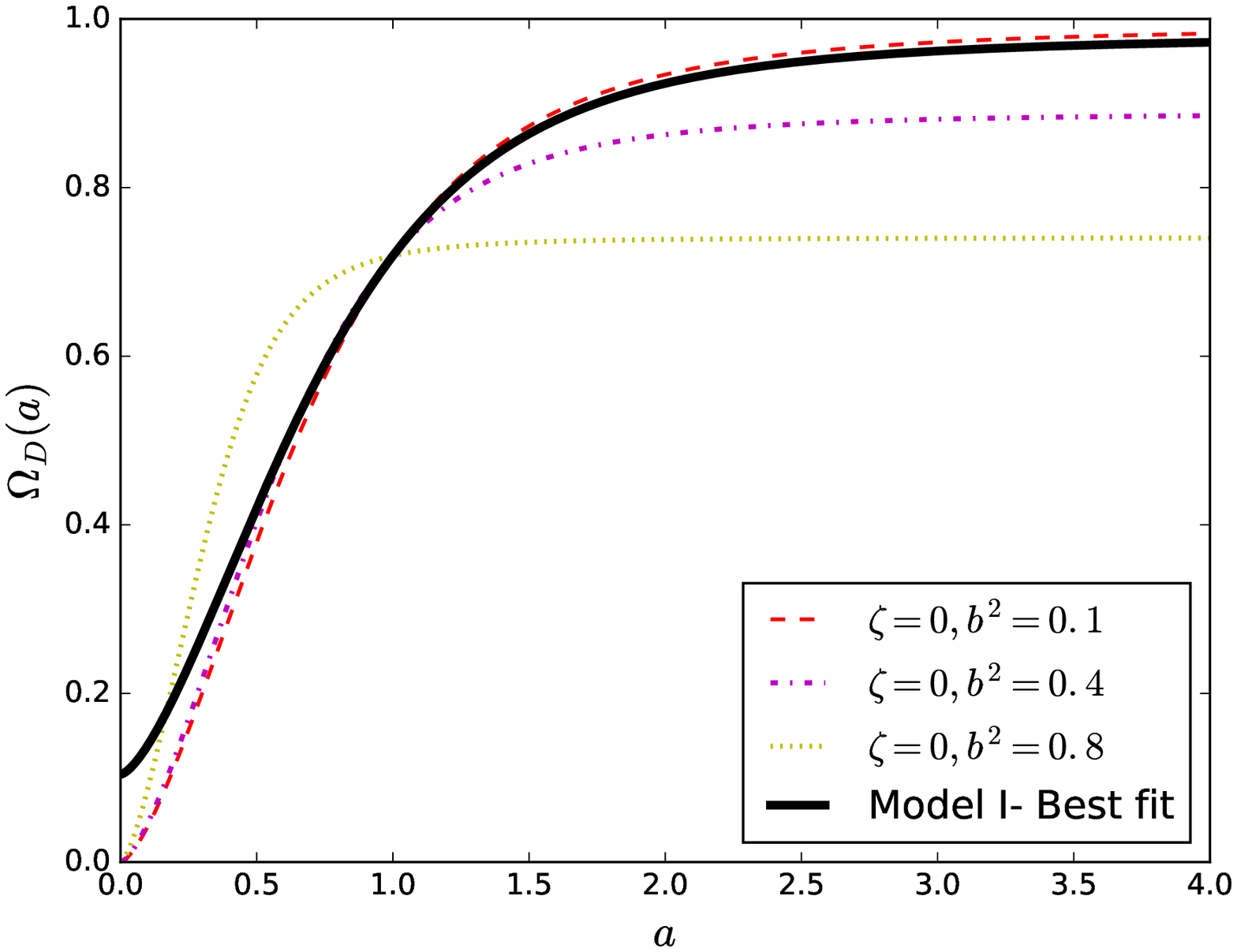}
\includegraphics[height=45mm,width=60mm,angle=0]{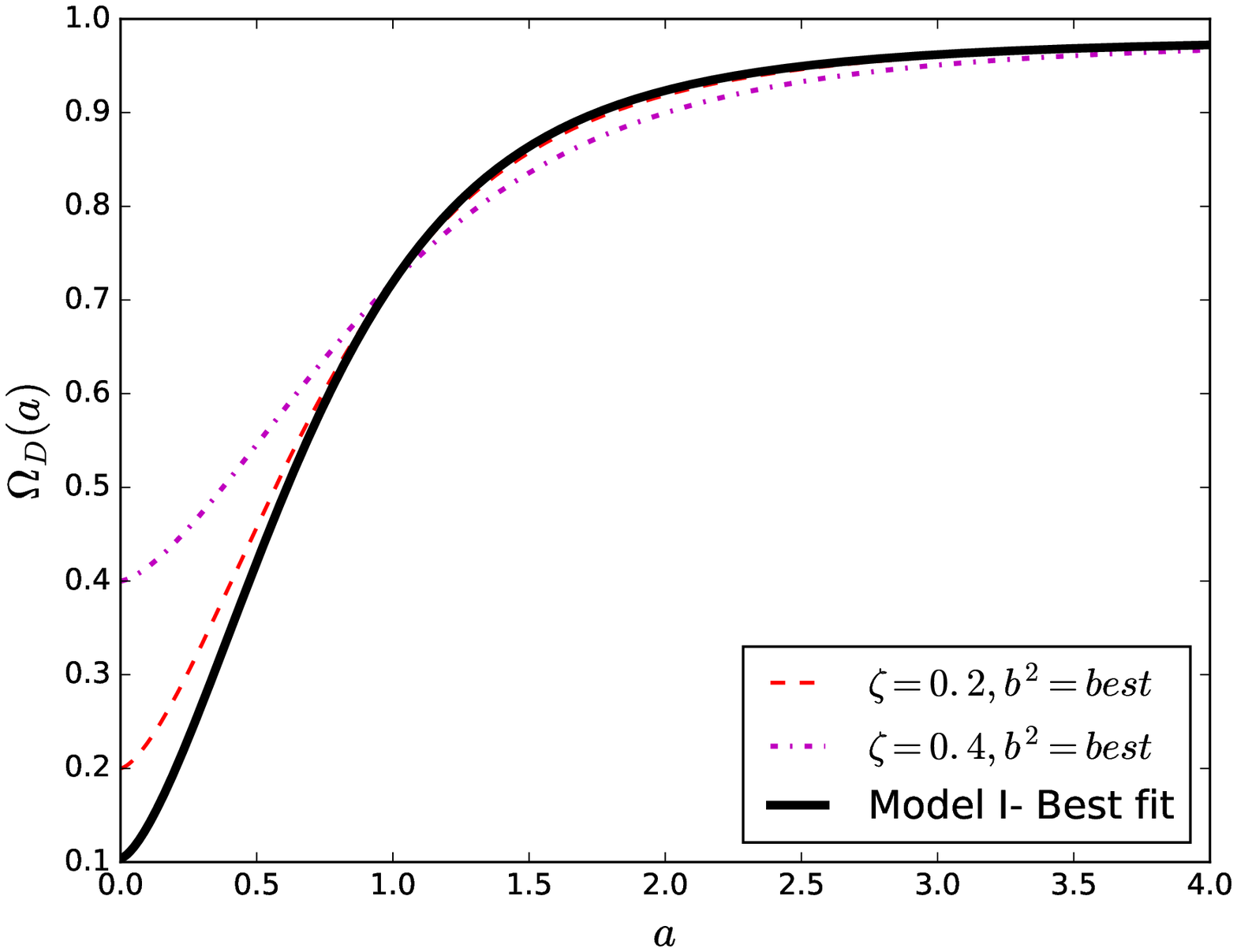}
\includegraphics[height=45mm,width=60mm,angle=0]{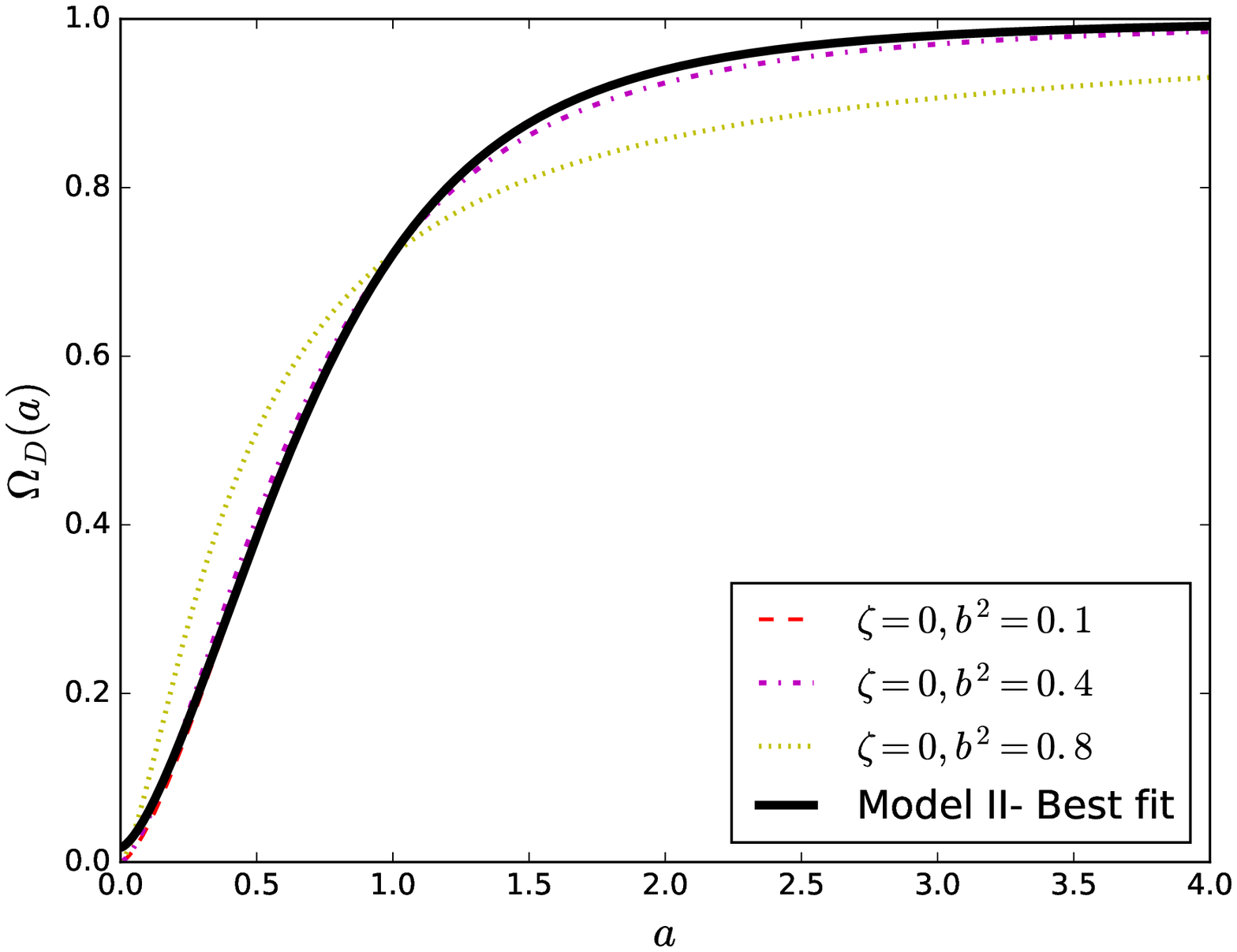}
\includegraphics[height=45mm,width=60mm,angle=0]{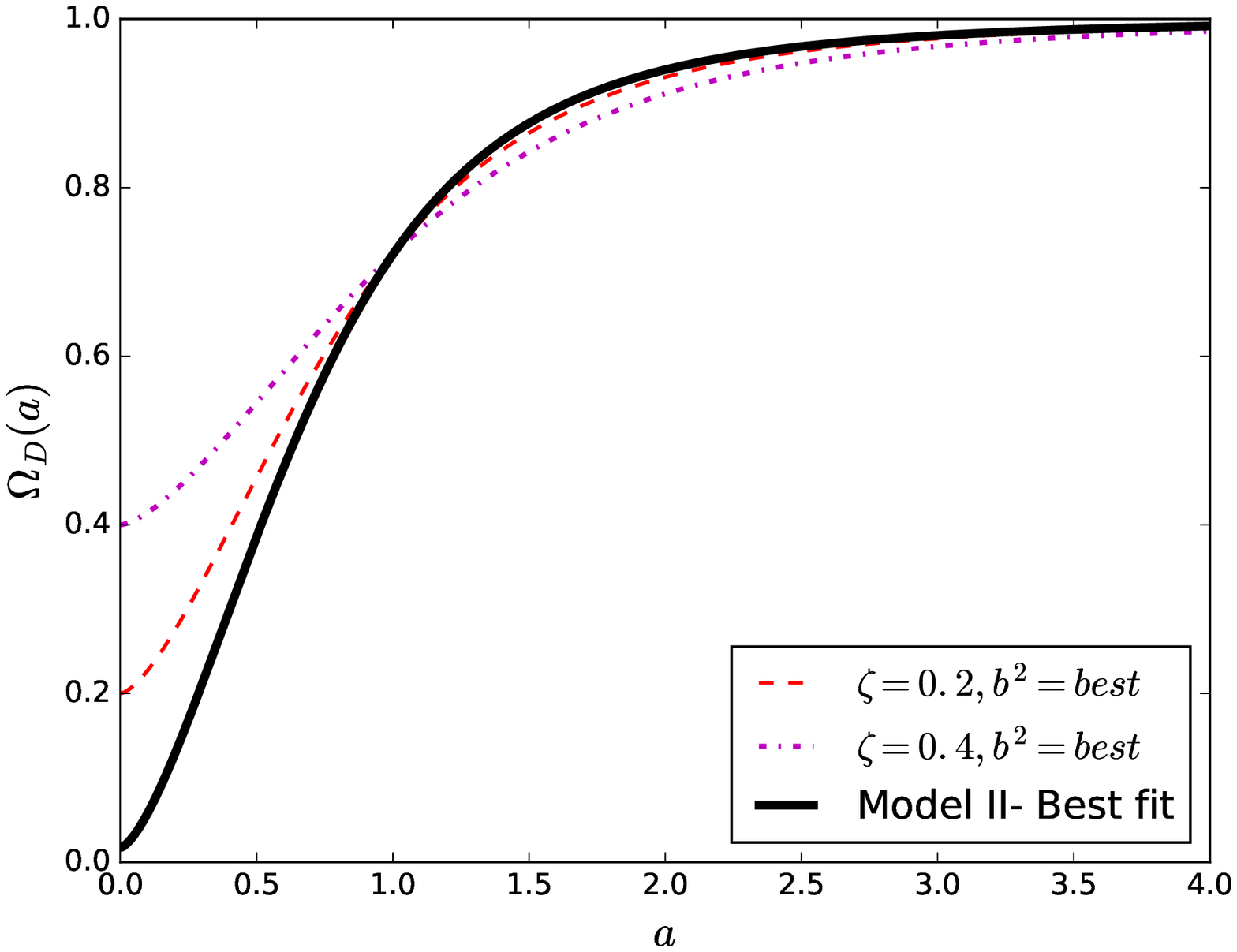}
\includegraphics[height=45mm,width=60mm,angle=0]{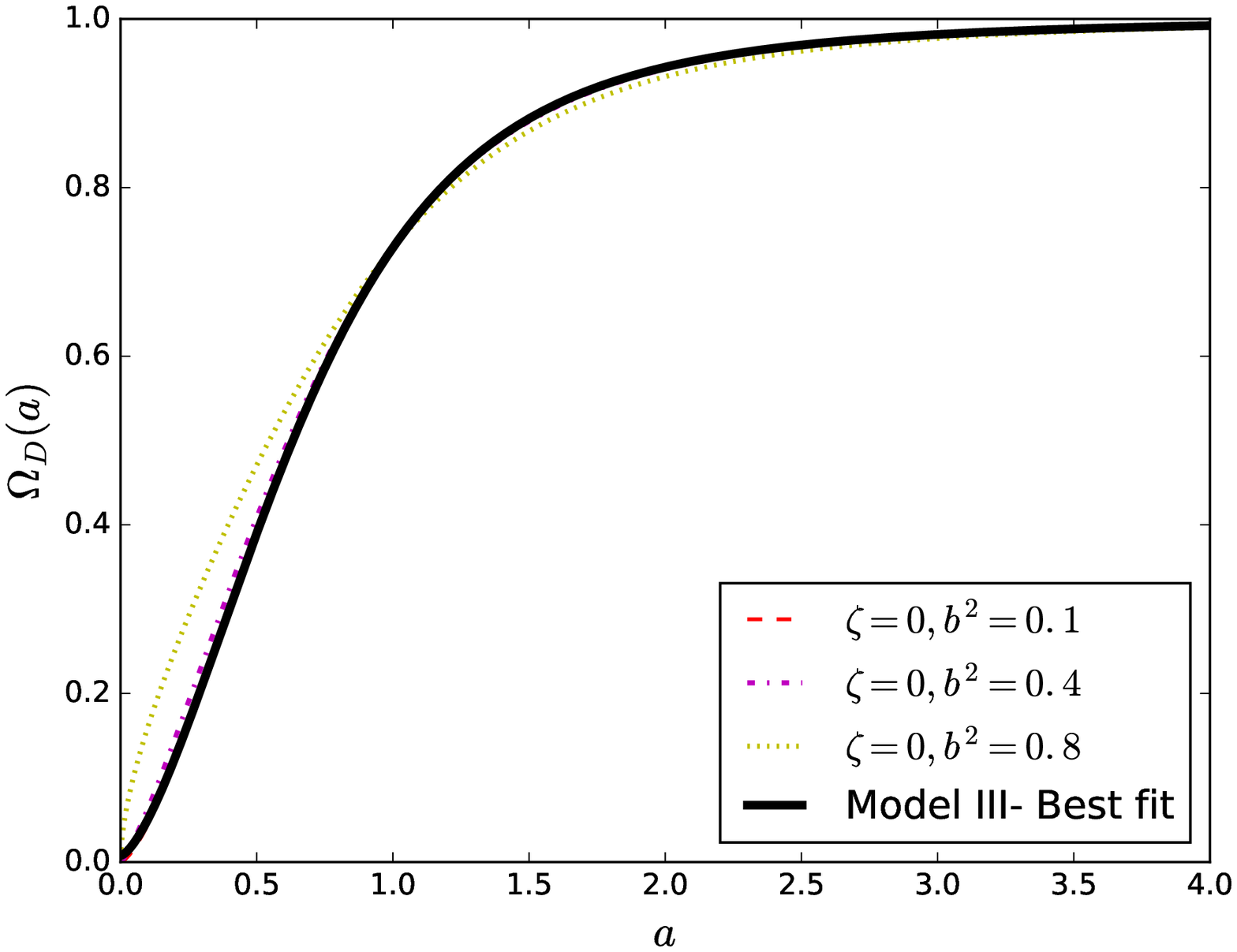}
\includegraphics[height=45mm,width=60mm,angle=0]{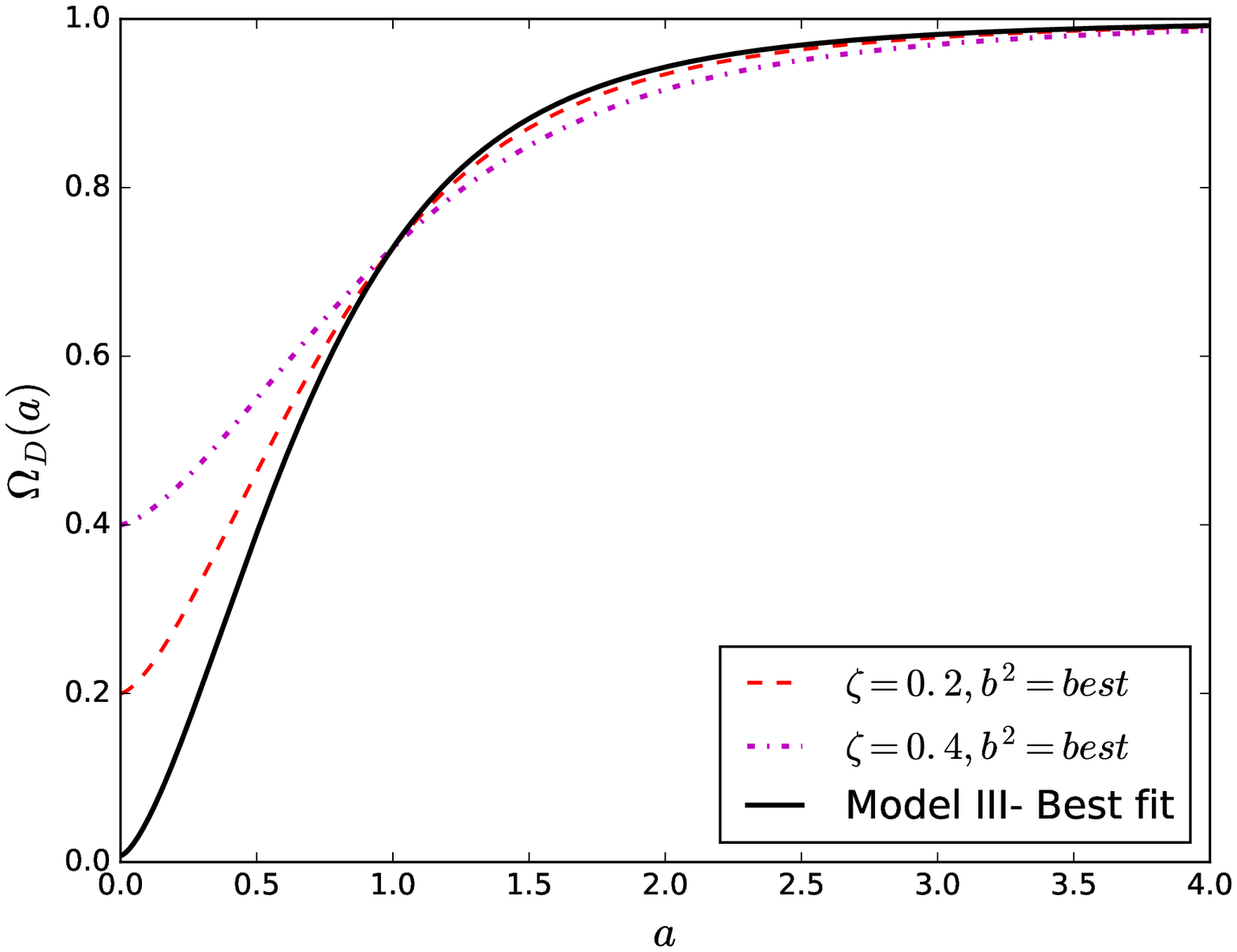}
\vspace{2mm} \caption{{\footnotesize The evolution of $\Omega_D$
as a function of scale factor for three models explained in the
text. The rest information is the same as that of mentioned in Fig. \ref{wq}. }} \label{oev}
\end{figure}

The evolution of DE density parameter versus $a$ is depicted in
Fig.(\ref{oev}). In the  left part of this figure, we turned to
the impact of the coupling constant $b$ on the density parameter
evolution. This figure shows that for larger values of $b$ the
evolution of $\Omega_D$ will be flatter. In the right panel of
Fig.(\ref{oev}), evolution of $\Omega_D$ is plotted for different
choices of $\zeta$ and best value for $b^2$ (see next section).
This figure reveals that the main advantages of the squared term
in the dark energy density happens in early stages of the
universe. This result was expected because the second order term
is very small at late epochs (due to smallness of $H$ at late time
). So it seems that the second term just can affect the late time
acceleration problem through its primary effects on the cosmic
dynamics in earlier stages.

Currently our universe is in dark energy dominated phase, so any
successful model must be able to result in a late time stable
universe in the dark energy dominated phase. In order to consider
the stability issue of a dark energy model we should discuss a
covariant perturbation theory in an expanding background. However
in this step we do not follow such a process. In fact, here we are
interested to do a simple analysis to see if the universe shows
signs of stability in the DE dominated phase.

\begin{figure}
\center
\includegraphics[height=55mm,width=60mm,angle=0]{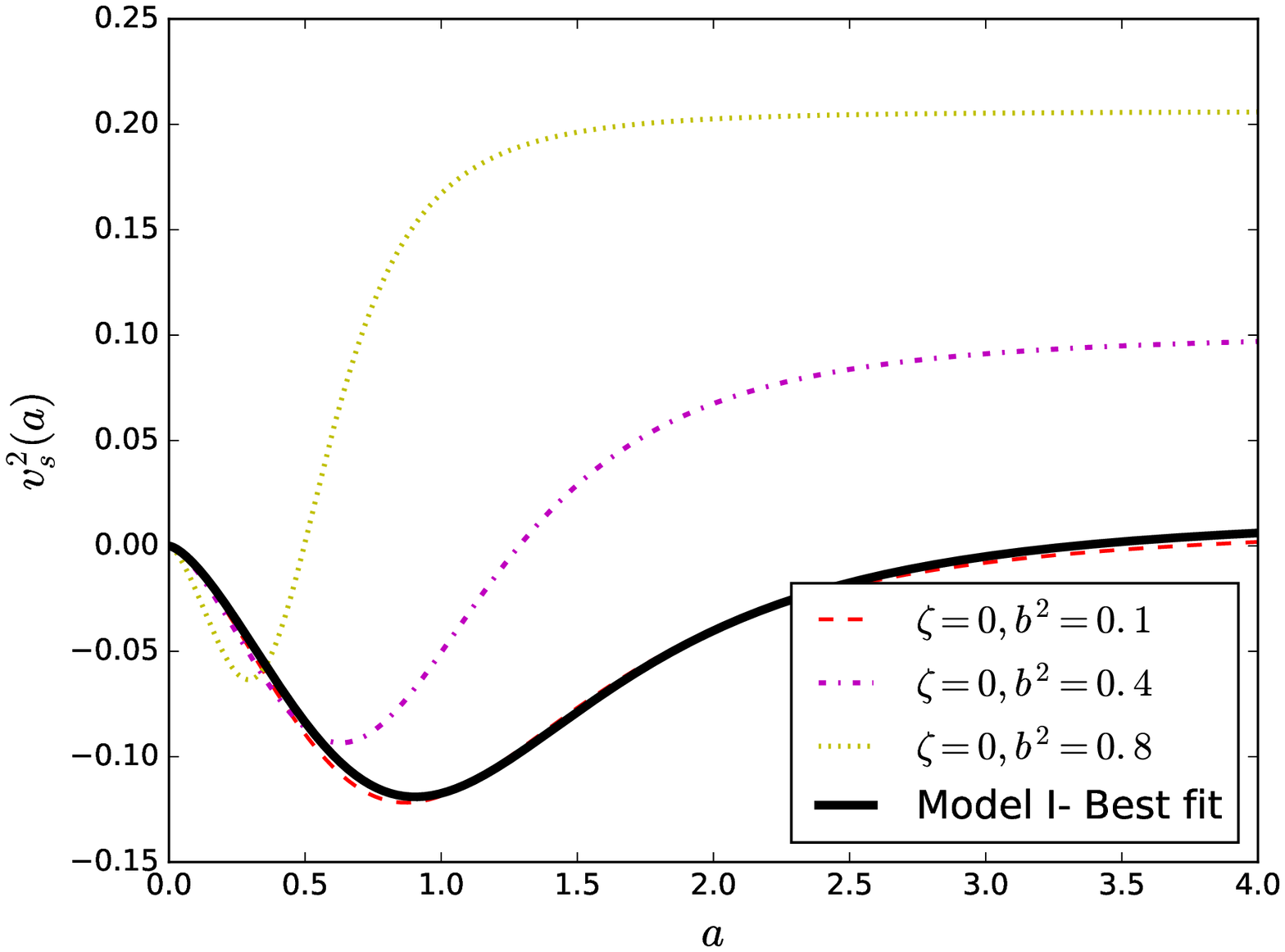}
\includegraphics[height=55mm,width=60mm,angle=0]{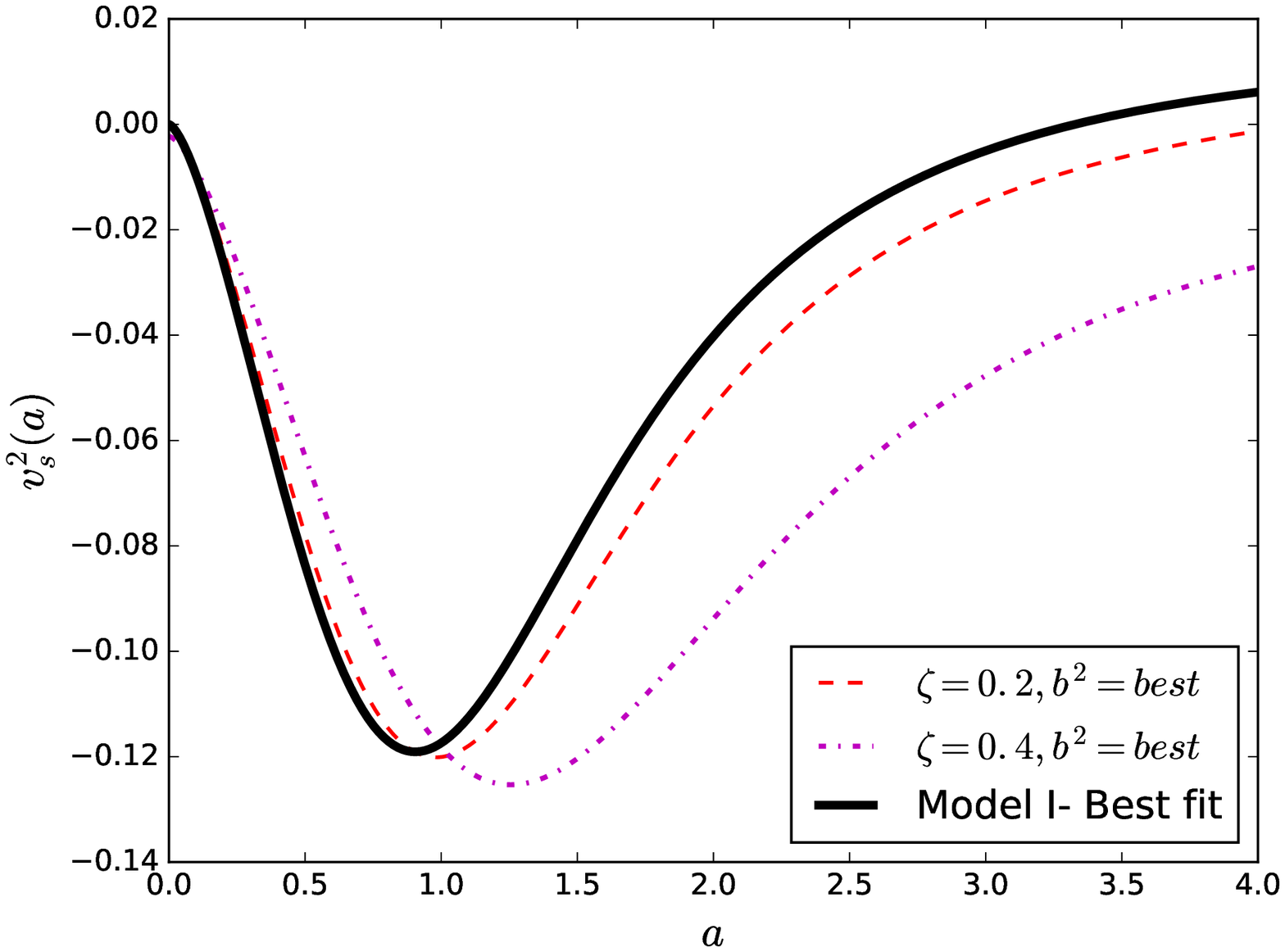}
\includegraphics[height=55mm,width=60mm,angle=0]{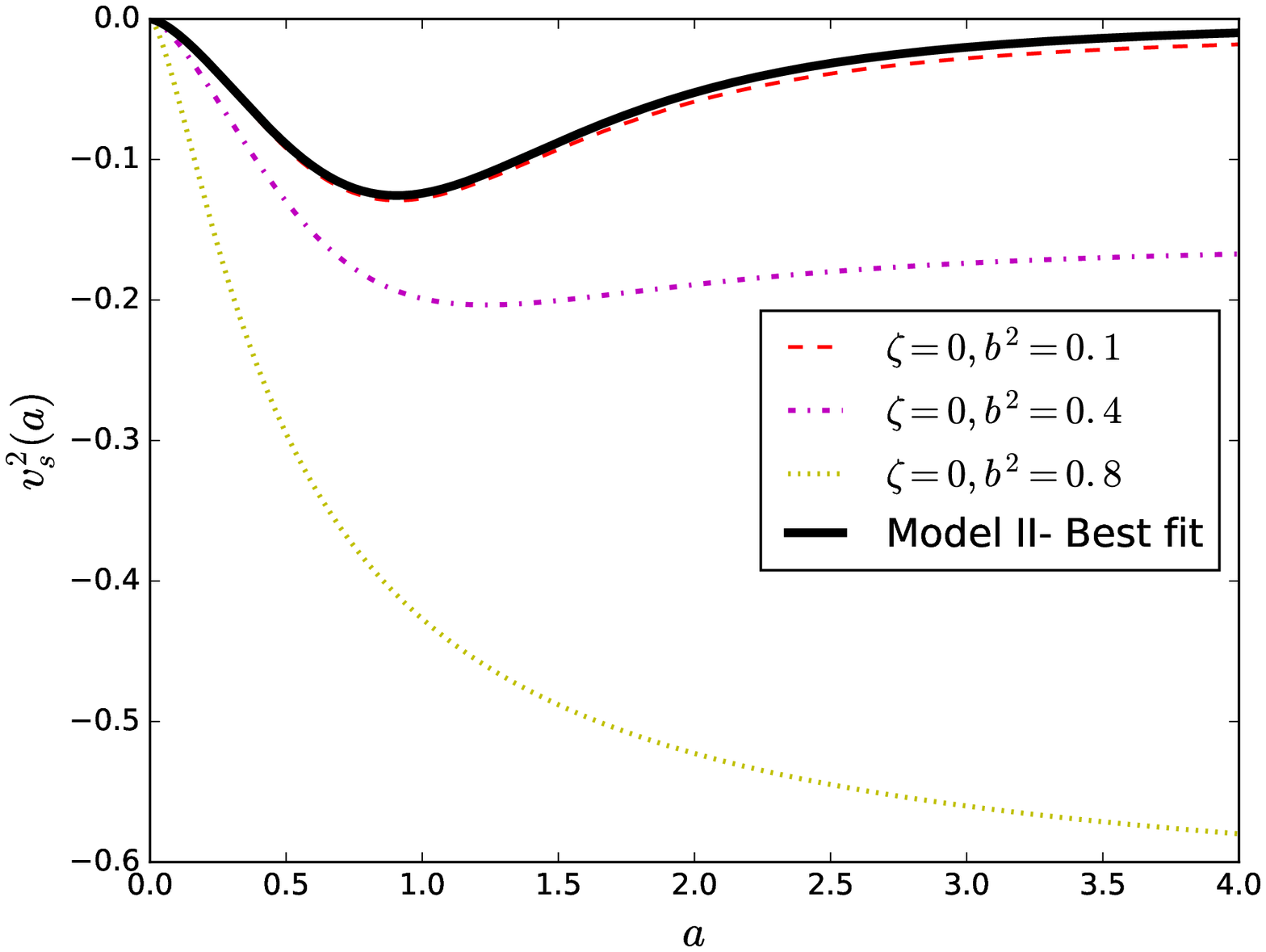}
\includegraphics[height=55mm,width=60mm,angle=0]{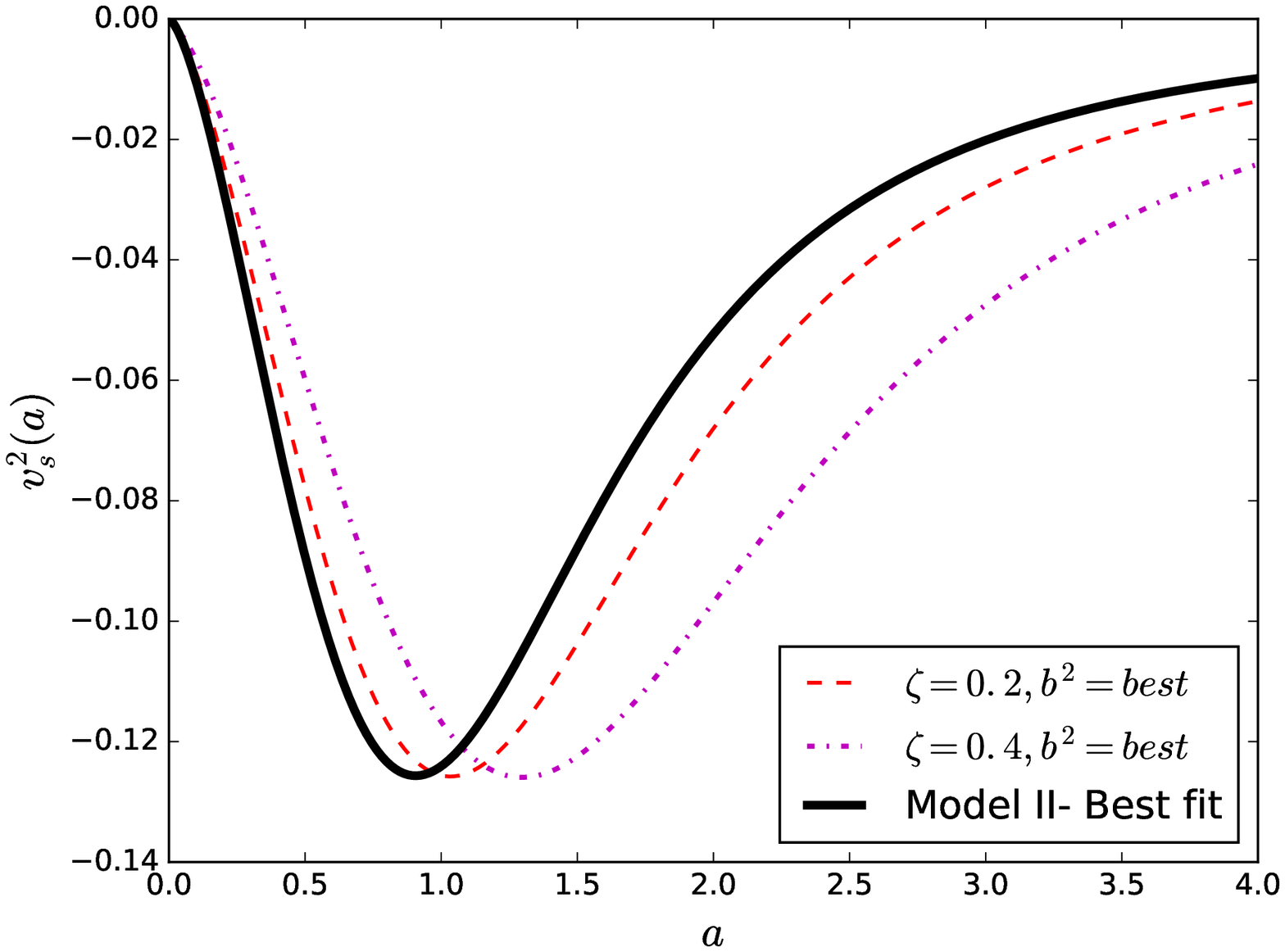}
\includegraphics[height=55mm,width=60mm,angle=0]{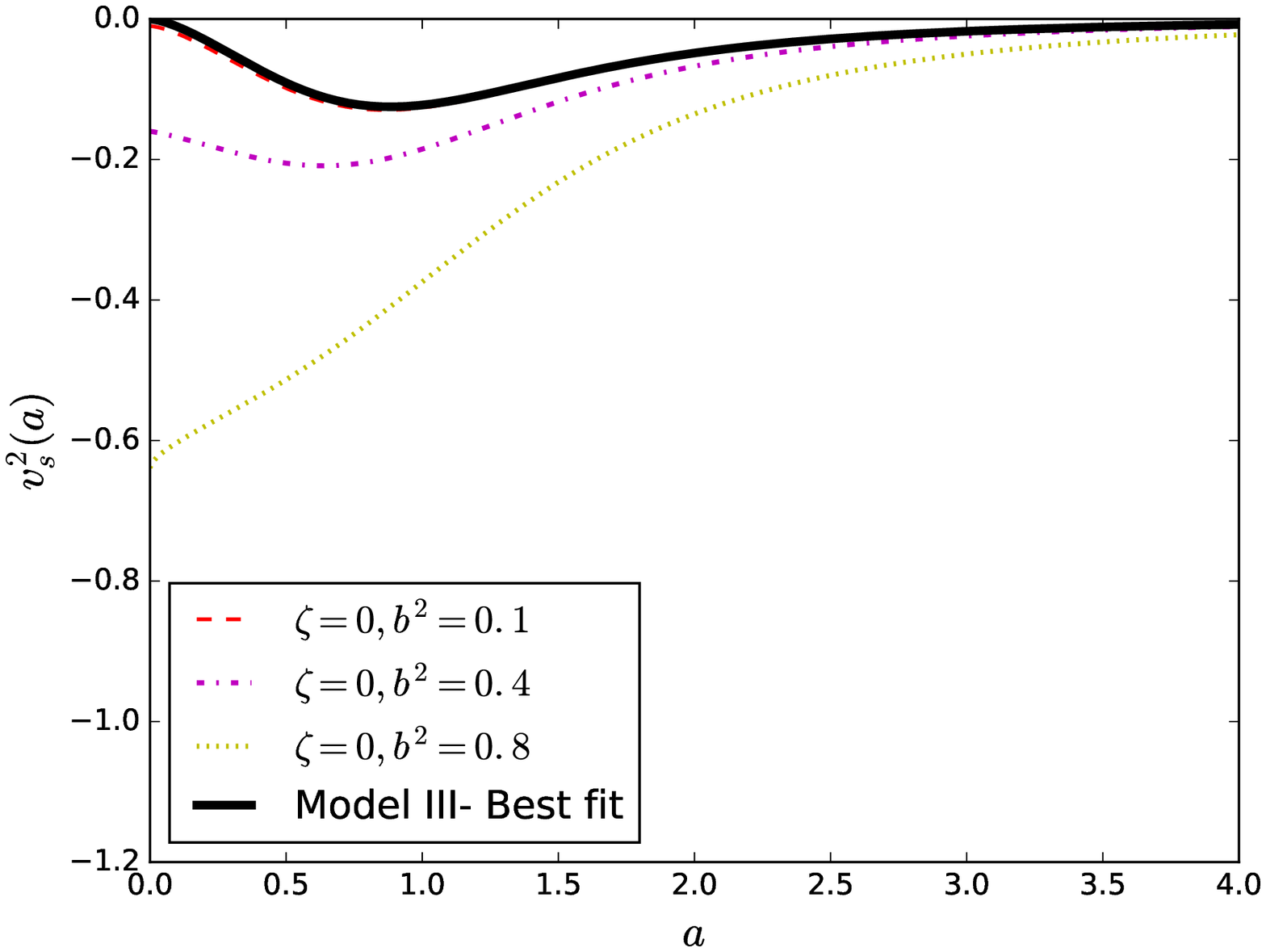}
\includegraphics[height=55mm,width=60mm,angle=0]{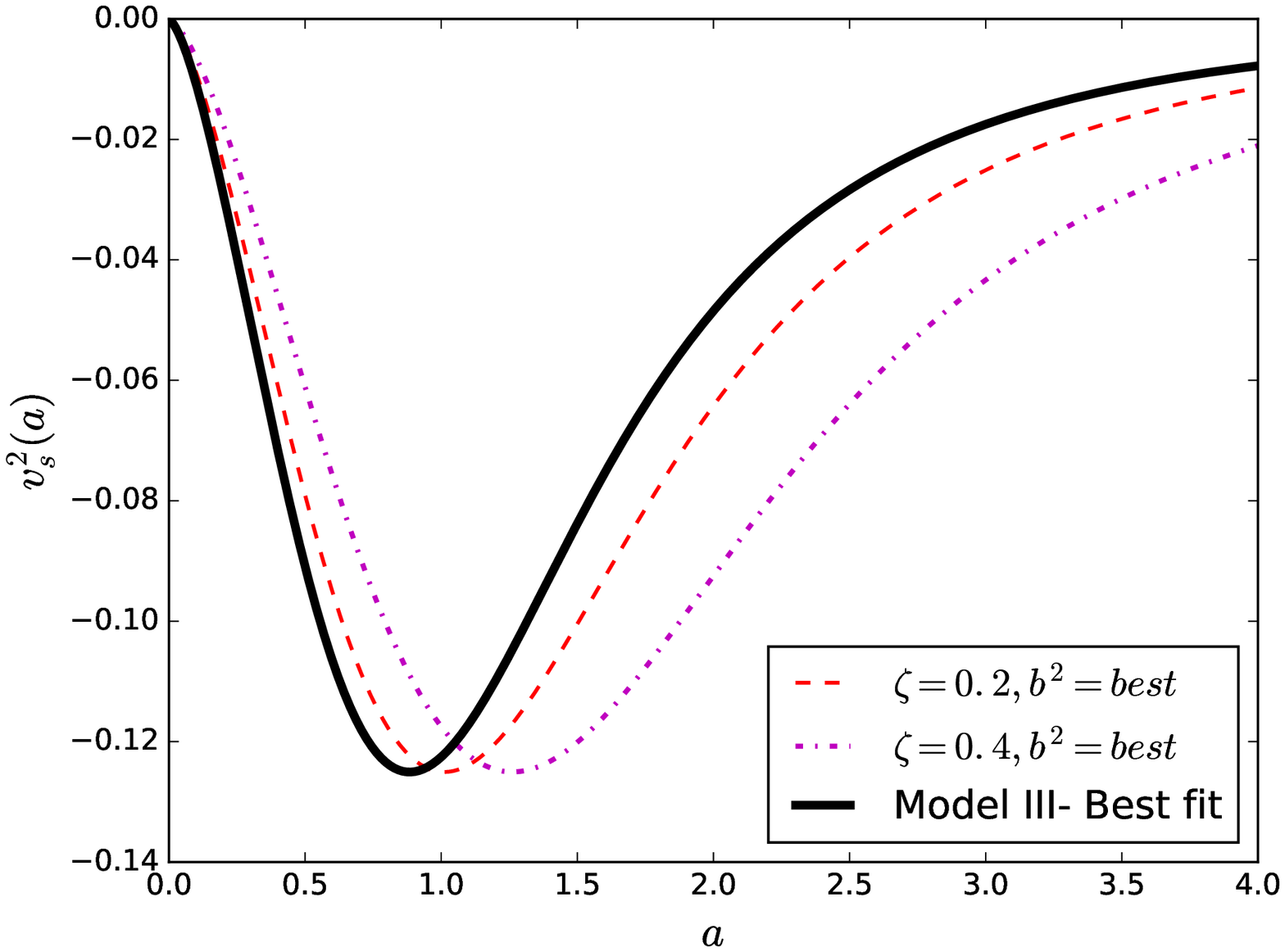}
\vspace{2mm} \caption{{\footnotesize The evolution of $v_s^2$ as a
function of scale factor for three models explained in the text.
The rest information is the same as that of mentioned in Fig.
\ref{wq}.}} \label{vs}
\end{figure}

 To this end we consider the response of the cosmic background
filled with GDE and DM to adiabatic perturbations in linear
regime. The governing equation on the evolution of the universe in
this analysis includes continuity, Euler and Poisson equations.
Inserting a small perturbation in energy density of the background
and obtaining the resulting equations on evolution of the
perturbations up to the first order one finds \be
\ddot{\delta}+2H\dot{\delta}+v_s^2k^2\delta-4\pi G \delta\rho_0=0,
\ee where $\delta$ is energy density perturbation of the fluid.
For short wavelength the solution of above equation are of the
form $\delta\propto e^{\pm i\omega t}$, where $\omega\propto v_s$.
Now it is clear that when $v_s^2<0$, the perturbations can grow
and makes the background unstable while for $v_s^2>0$ the
perturbations propagate oscillatory and the background will be
stable against linear adiabatic perturbations. The way we use
squared sound speed here is that we search for epochs with
positive $v_s^2$ because we need a late time stable DE dominated
universe.
 Here we look for impacts of the non-linear interaction
terms to see if it leaves any chance for a late time positive
$v_s^2>0$. Using \be v_s^2=\frac{dp}{d\rho}=\frac{\dot p}{\dot
\rho}\,,\label{v2nli1} \ee and taking into account the cosmic
dynamic equations with non-linear interactions, after a little
algebra one can find the squared sound speed as
\begin{equation}\label{v2m0n3}
    v_s^2=\frac{(\Omega_D-1)(\Omega_D-\zeta)}{(\Omega_D-2+\zeta)^2}
+\frac{\Omega_D^2(\Omega_D
\zeta+2\Omega_D-6\zeta+3\zeta^2)}{(\Omega_D-2+\zeta)^2}\,b^2.
\end{equation}
In Fig.(\ref{vs}), $v_s^2$ is drawn versus $a$. From the left part
it is obvious that for this class of interaction, $v_s^2$ with
suitable choice of coupling constant $b$, is capable to get
positive values at the late times and this leaves a chance for
stable dark energy dominated universe which favored by
observations. Fig.(\ref{vs}) also indicates that for larger values
of $b$, $v_s^2$ will transits earlier to positive domain. It is
worth mentioning that GDE in non-interacting and linearly
interacting cases suffers from the negativity of $v_s^2$ at the
late times \cite{ebrinsggde} and this is a direct consequence of
this form of non-linear interaction term. The right part of
Fig.(\ref{vs}) reveals that for larger values of $\zeta$, $v_s^2$
will enter to the positive domain later.

The evolution of scale factor as a function of time can be
computed using Eq. (\ref{e1}) as follows: \bea \int_{a_0}^a
(\Omega_D -\zeta)\frac{da'}{a'}=\int_{t_0}^t \frac{8\pi
G\alpha}{3}dt'=B(t-t_0) \eea where $B\equiv \frac{8\pi G\alpha}{3}
$. Fig. (\ref{a_t}) illustrates the behavior of scale factor as a
function of time.

\subsection{Model II}
In this model we consider the following form for non-linear
interaction term \be Q=3b^2H\frac{\rho_D\rho_m}{\rho}\,. \ee In
this case, one finds that $Q'=\frac{2b^2
\rho_m}{\rho}=2b^2\Omega_m$ consequently,  we can rewrite the
equation of state and deceleration parameters as \bea \label{wqi2}
w_D&=&\frac{\zeta-\Omega_D-2b^2\Omega_D(1-\Omega_D)}{\Omega_D(2-\Omega_D-\zeta)}\,,\\
q&=&\frac12+\frac32\,\frac{\zeta-\Omega_D-2b^2\Omega_D(1-\Omega_D)}{2\!-\zeta\!-\Omega_D}.
\eea It is easy to see that these relations in the late and early
time limits approach to the non-interacting case, however they
differ from non-interacting case in middle stages of evolution. It
is also worth to mention that in this case the phantom divide can
not be crossed.
To make more obvious the mentioned behavior, we have plotted the
relevant parameters.

In Fig.(\ref{wq}) we found that the evolution of $w_D$ versus
scale factor. Fig.(\ref{q}) indicates the deceleration parameter
as function of scale factor for various choices of $\zeta$ and
$b^2$ parameters. Increasing the value of $b^2$ causes to
decreasing $q$ and getting the negative value at earlier times
while according to the right panel of Fig.(\ref{q}), larger values
of $\zeta$ leads to the larger deceleration parameters. Note that
to ensure $q<0$ at the present time ($a=1$), we should set
$\zeta<0.4$. This theoretical constraint is consistent to the
other observational constraints (the next section).

Fig.(\ref{weff}) illustrates $w_{eff}$  as a function of scale
factor. Our results demonstrate that  $w_{eff}\to -1$ at future
implying that the universe will end with a big rip as the previous
case. It can be seen from Fig.(\ref{weff}) that $w_{eff}$ gets
more negative values by increasing the coupling $b^2$\,. 
One can also finds the evolution of DE density parameter for this
case as
 \be \label{oi2}
\frac{d\Omega_D}{d \ln a}=
\frac{3(\zeta-\Omega_D)(1-\Omega_D)(b^2\Omega_D-1)}{2-\Omega_D-\zeta}\,.
\ee
 The left part of Fig.(\ref{oev}) corresponds to
the impact of the coupling constant $b^2$ on the density parameter
evolution. This figure shows that larger values of $b^2$, extend
the dominant era of $\Omega_D$. In right panel of this figure, the
evolution of $\Omega_D$ is plotted for different choices of
$\zeta$.

In this case the squared sound speed can be obtained as:
 \be
v_s^2=\frac{(\Omega_D-1)(\Omega_D-\zeta)}{(\Omega_D-2+\zeta)^2}-
\frac{\Omega_D^3+(\zeta-2)\Omega_D^2+(2\zeta^2-5\zeta+2)\Omega_D-\zeta^2+2\zeta}{(\Omega_D-2+\zeta)^2}\,b^2.
\ee Plotting $v_s^2$ versus  $a$ in Fig.(\ref{vs}) reveals that
the squared sound speed is always negative for different choices
of constants implying that in present case, a late time stable
GGDE dominated phase is a matter of doubt. It is also easy to see that
by increasing $b$, the squared sound speed takes more negative
values in contrast with the previous case.

\subsection{Model III}
The functional form of non-linear interaction for this model is
considered as \be
 Q=3b^2H\frac{\rho_m^2}{\rho}\,.
 \ee
Doing the same steps for this interaction, one finds that
$Q'=2b^2\frac{\rho_m^2}{\rho
\rho_D}=2b^2\frac{\Omega_m^2}{\Omega_D}$, and the EoS and
deceleration parameters as
 \bea \label{wqi3}
w_D&=&\frac{\zeta-\Omega_D-2b^2(1-\Omega_D)^2}{\Omega_D(2-\Omega_D-\zeta)}\,\\
q&=&\frac12+\frac32\,\frac{\zeta-\Omega_D-2b^2(1-\Omega_D)^2}{2\!-\zeta\!-\Omega_D}\,.
\eea  In this case the dark energy density parameter evolves as
\be \label{oi3}
\frac{d\Omega_D}{d\ln{a}}=\frac{3(\zeta-\Omega_D)(\Omega_D-1)(b^2\Omega_D+1-b^2)}{2-\Omega_D-\zeta}\,.
\ee The behavior of equation of state, $w_D$, deceleration
parameter, $q$, effective equation of state, $w_{eff}$ and energy
density of DE have been plotted in Figs. (\ref{wq}), (\ref{q}),
(\ref{weff}) and (\ref{oev}), respectively.  Finally the squared
sound speed for this case reads
 \be
v_s^2=\frac{(\Omega_D-1)(\Omega_D-\zeta)}{(\Omega_D-2+\zeta)^2}
+\frac{(\Omega_D-1)\left[\Omega_D^2+(\zeta-3)\Omega_D+(2\zeta-5)\zeta+4
\right]}{(\Omega_D-2+\zeta)^2}\,b^2. \ee Fig.(\ref{vs}) shows that
in this case $v_s^2$ can not get positive values but approaches
the border line at the future. The model with smaller value of $b$
will approaches the border line earlier.
\begin{figure}
\center
\includegraphics[height=60mm,width=78mm,angle=0]{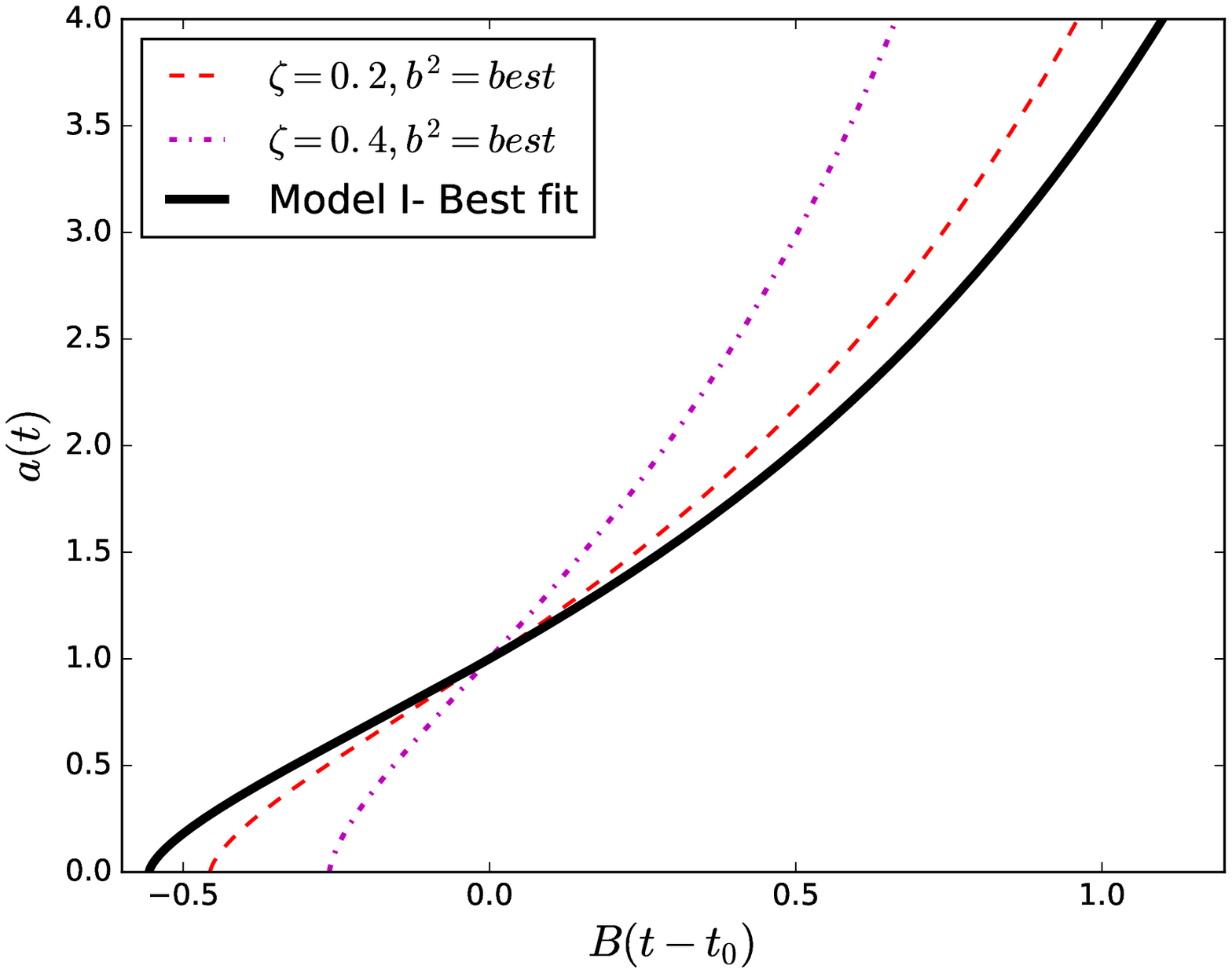} \quad
\includegraphics[height=60mm,width=78mm,angle=0]{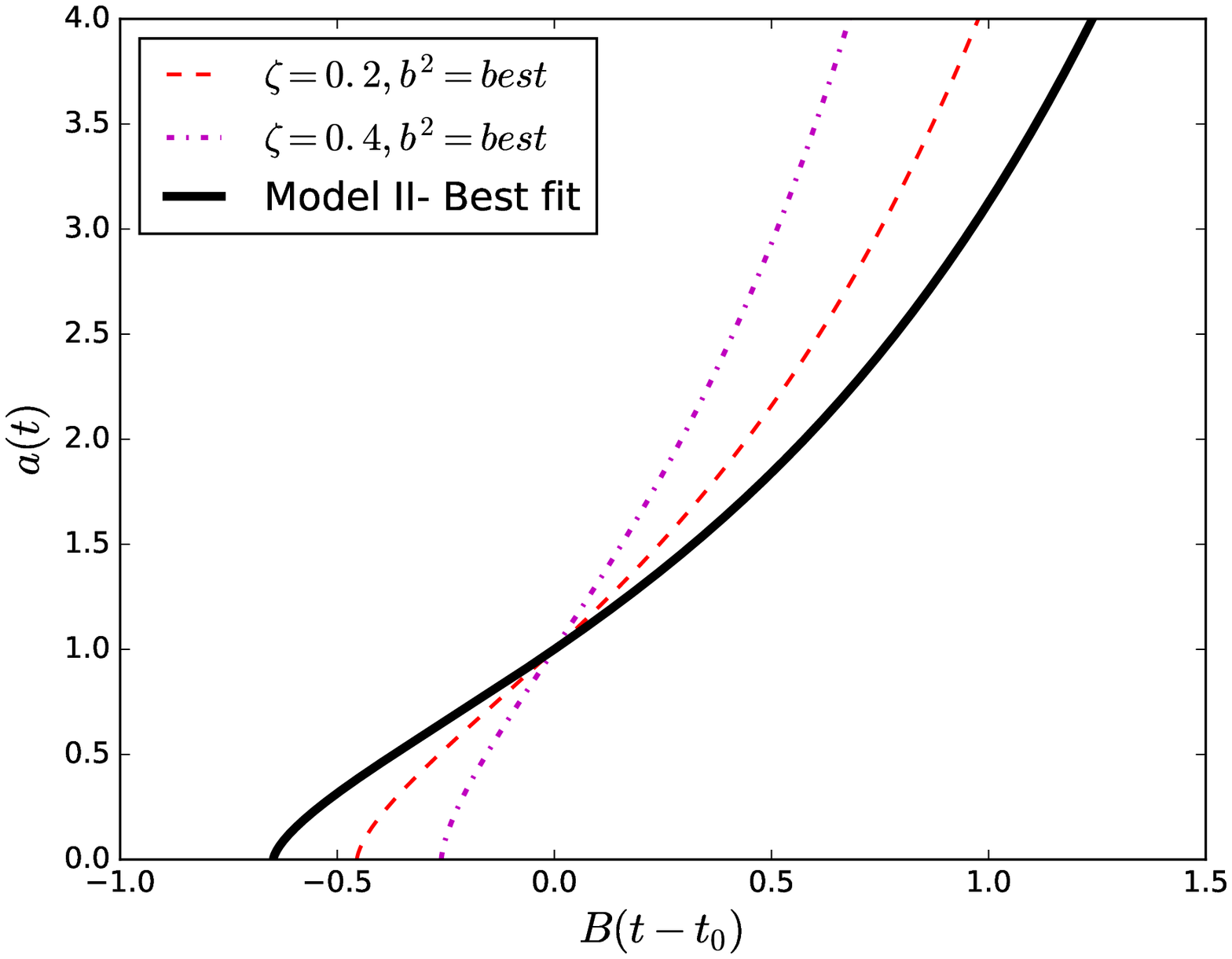}\quad
\includegraphics[height=60mm,width=78mm,angle=0]{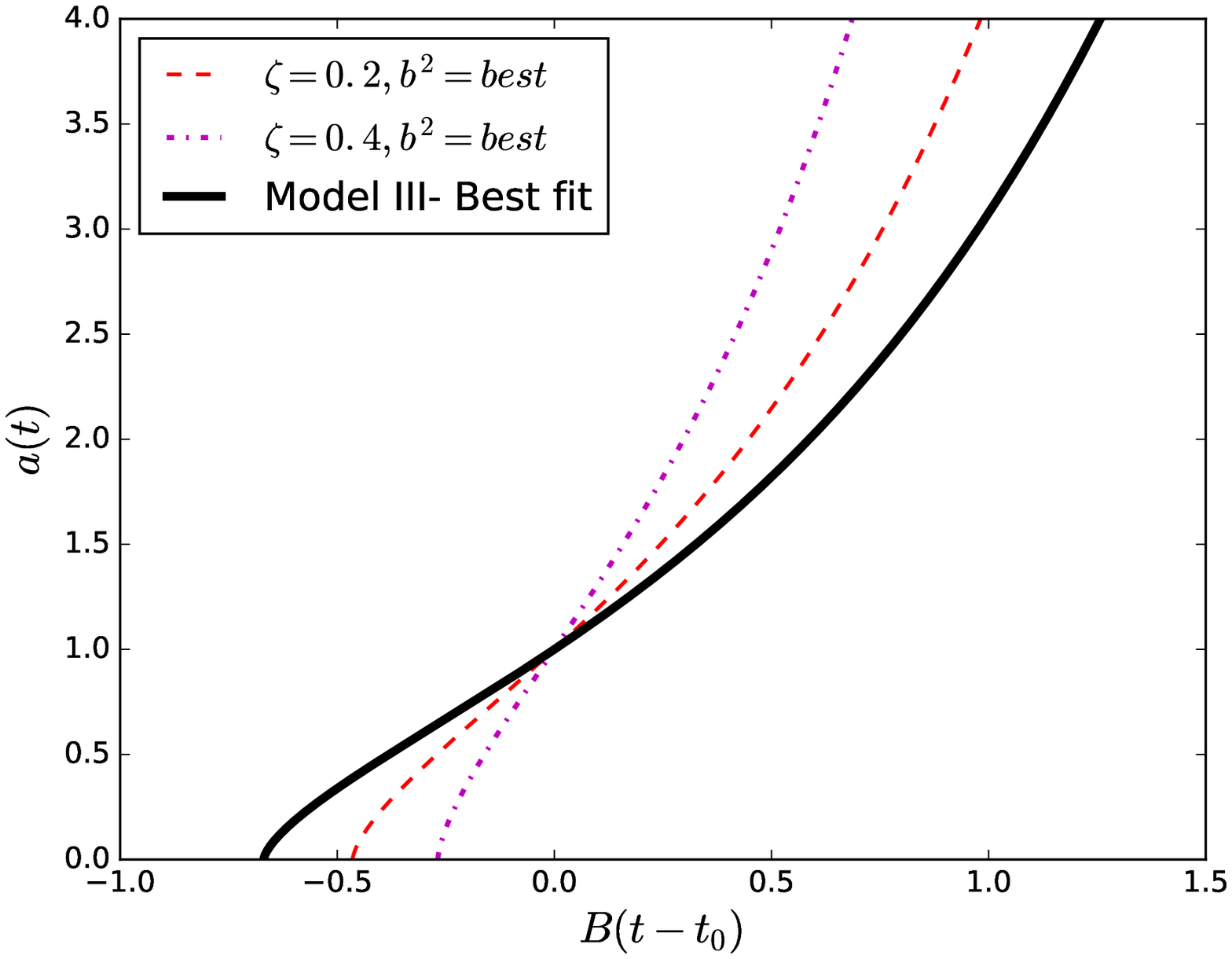}
\vspace{2mm} \caption{\footnotesize  The evolution of scale factor
as a function of time. Here $B\equiv \frac{8\pi G\alpha}{3} $, the
value of $\Omega_D$ at the present time and $b^2$ have
been fixed according to their best values determined by joint
analysis of SNIa+Hubble+BAO+CMB. In all plots the dark solid line
corresponds to the best fit values determined by joint analysis.}
\label{a_t}
\end{figure}
Finally, the behavior of scale factor as a function of time is
represented in Fig.(\ref{a_t}). This plot shows that the model in
principle, can explain the late accelerating expansion of the
universe.
In the next section we are
going to carry out the comparison between results given by our
models and that of given directly from observation.

\section{Observational consistency check}\label{observational}
In this section, we rely on the most recent observational data sets to
check the consistency of generalized ghost dark energy model
accompanied by non-linear interactions between DM and DE. This
approach enables us to put observational constraints on the free
parameters of our model and asses how the presence of interaction is suitable as an alternative paradigm.
The following observables for examination of the nature of dark energy can be used:  \\
I) Expansion history of the Universe\\
II) Changing the evolution of gravitational potentials, causing to ISW effect \\
III) Cross-correlation of CMB and large scale structures\\
IV) Growth of structures\\
V) Weak lensing\\
In this paper we limit ourselves on almost evolution expansion
indicators such as distance modulus of Supernovae Type Ia and the Hubble parameter,
Baryon acoustic oscillations (BAO) and
CMB including the position of first peak, shift parameter and
redshift of recombination. Throughout this paper we have supposed
that our Universe is flat, so $\Omega_{tot} =
\Omega_m+\Omega_b+\Omega_r+\Omega_D=1$, where $\Omega_m ,
\Omega_b$ denote the density parameters of dark matter and
baryonic matter respectively. Energy density of radiation and
baryonic matter are fixed by other relevant observations. For
more convenience, we indicate free parameters by
$\Theta:\{\Omega_D,\zeta,b^2\}$. The expected priors on these
parameters are listed in Table \ref{prior}. Radiation density is
fixed to $\Omega_r=2.469\times 10^{-5}h^{-2}(1.6903)$
\cite{Hinshaw:2012aka}. The baryonic density is set to
$\Omega_{b}h^2=0.02222\pm0.00023$ \cite{Ade:2015xua}


\begin{table}[ph]
\tbl{Priors on free parameter space, used in the
observational constraints analysis in this paper. }
{\begin{tabular}{llccc}
\hline
\hline Parameter & Prior & Shape of PDF \\
\hline $\Omega_{tot}$ &1.000& Fixed&\\
 $\zeta$ &$[0.050-0.020]$& Top-Hat&\\
 $H_0$ &$[40.0-100.0]$& Top-Hat&\\
 $\Omega_D$ &$[0.0-1.0]$& Top-Hat&\\
 $b^2$ &$[0.0-1.0]$& Top-Hat&\\
\hline \hline
\end{tabular}
\label{prior}}
\end{table}

\subsection{Supernovae Ia}

In order to compute cosmological distance, we must look for
standard candles. Such brilliant cosmological objects enable us to
determine cosmological distance for a given redshift corresponding
to mentioned object. Type Ia supernovae is a category of
cataclysmic variable starts produces due to explosion of a white
dwarf star. Its thermonuclear explosion is well known,
consequently, it can be used as standard candle. In 1998, Riess et
al., could discover the acceleration of expanding universe using
16 distant and 34 nearby SNIa from the Hubble space telescope
(HST) project \cite{Riess:1998cb}.
 After that, Perlmutter et al., using SNIa from Calan-Tololo and 42 high-redshift SN,
  confirmed acceleration expansion \cite{Perlmutter:1998np}.
  The surveys of SNIa have drawn more attention. Higher-Z Team \cite{ries04,ries07},
   the Supernova Legacy Survey (SNLS) \cite{ast06,bau08,reg09,guy10},
   the ESSENCE \cite{mik07,wood07}, the Nearby Supernova Factory (NSF) \cite{cop06,scal09},
    the Carnegie Supernova Project (CSP) \cite{fol10,fol101},
    the Lick Observatory Supernova Search (LOSS) \cite{lea10,li10}, the Sloan Digital Sky Survey (SDSS)
    SN Survey \cite{holtz08,kess09}, Union2.1 SNIa dataset \cite{Suzuki12,Cao:2014jza}.
     More recently a catalog made from SNLS and SDSS SNIa compilation which is called
      Joint Light-curve Analysis (JLA) was produced \cite{Betoule14}.
      In this paper we  consider constraints on model parameters coming from SNIa
      observations using JLA dataset including 740 Ia supernovae
      in the redshift range of $z\in [0.01,1.30]$ \cite{Betoule14}. 

Observation of SNIa does not provide
standard ruler but rather gives distance modulus. This quantity is
defined by:
\begin{eqnarray}
\mu (z;\{\Theta\}) \equiv m - M = 5\log \left(
{\frac{{{d_L}(z;\{\Theta\})}}{{Mpc}}} \right) + 25
\end{eqnarray}
where $m$ and $M$ are apparent and absolute magnitude,
respectively. For the spatially flat universe the luminosity
distance defined in above equation reads as:
\begin{eqnarray}
{d_L}(z) = \frac{c}{{{H_0}}}(1 + z)\mathop \int_0^z
\frac{dz'}{\mathcal{H}(z';\{\Theta\})}
\end{eqnarray}
where ${\mathcal{H}}=H/H_0$. To compare the observational data and
that determined by our model, we apply the likelihood analysis.
To this end, we use $\chi^2$ which is defined as
\begin{eqnarray}
\chi^2_{SNIa}\equiv
\Delta\mu^{\dag}\cdot\mathcal{C}^{-1}_{SNIa}\cdot\Delta\mu
\end{eqnarray}
where $ \Delta\mu\equiv \mu_{obs}(z)-\mu(z;\{\Theta\})$ and
correlation information of SNIa data sets is encoded in covariance
matrix, $\mathcal{C}_{SNIa}$. The observed distance moduli and the relevant covariance matrix can be found on the web site \cite{JLAdata1,JLAdata2}.  
 Here $\mu_{obs}(z)$ is observed
distance modulus for a SNIa located at redshift $z$. Marginalizing
over $H_0$ as a nuisance parameter yields \cite{li11, Ade:2015rim}
\begin{eqnarray}
\chi^2_{SNIa}=\mathcal{M}^{\dag}\cdot\mathcal{C}_{SNIa}^{-1}\cdot\mathcal{M}+\mathcal{A}_{SNIa}
+\mathcal{B}_{SNIa}
\end{eqnarray}
where $\mathcal{M}(z)\equiv
\mu_{obs}(z)-25-5\log_{10}[H_0d_L(z;\{\Theta\})/c]$, and
\begin{eqnarray}
\mathcal{A}_{SNIa}&\equiv&-\frac{\left[\sum_{i,j}\mathcal{M}(z_i)\mathcal{C}^{-1}_{SNIa}(z_i,z_j)-\ln
10/5\right]^2}{\sum_{i,j}\mathcal{C}^{-1}_{SNIa}(z_i,z_j)}\\
\mathcal{B}_{SNIa}&\equiv&-2\ln\left( \frac{\ln
10}{5}\sqrt{\frac{2\pi}{\sum_{i,j}\mathcal{C}^{-1}_{SNIa}(z_i,z_j)}}\right)
\end{eqnarray}

\subsection{OHD dataset}
In addition to SNIa standard candles, we use Observational Hubble parameter
Data (OHD) \cite{farooq13}. In this case the chi-square is:
\begin{equation}
\chi_H^2=\sum_i\frac{[H_{obs}(z_i)-H(z_i,\{\Theta\})]^2}{\sigma^2(z_i)}
\end{equation}
where $H_{obs}$ is observed Hubble parameter and $H$ is given by
Eq. (\ref{Fried}).

\subsection{Baryon acoustic oscillations}
The footprint of oscillations in the baryon-photon plasma on the
matter power spectrum is recognized by Baryon Acoustic Oscillation
(BAO).  The acoustic scale is so large therefore BAO are largely
unaffected by nonlinear evolution. Usually, the BAO data is
applied to measure both the angular diameter distance
$D_A(z;\{\Theta\})$, and the expansion rate of the Universe
$H(z;\{\Theta\})$ \cite{Ade:2015rim}:
\begin{equation}
D_V(z;\{\Theta\}) = \left[ (1+z)^2 D_A^2 (z;\{\Theta\})
\frac{cz}{H(z;\{\Theta\})} \right]^{1/3}
\end{equation}
where $D_V (z;\{\Theta\})$ is volume-distance. The distance ratio
is defined by
\begin{eqnarray}
d(z;\{\Theta\})\equiv
\frac{r_{s}(z;\{\Theta\})}{D_V(z;\{\Theta\})}
\end{eqnarray}
here $r_{s}(z;\{\Theta\})$ is the comoving sound horizon. In order
to take into account various aspects of BAO observations, we use
six BAO indicators including Sloan Digital Sky Survey (SDSS) data
release 7 (DR7) \cite{pad12},  SDSS-III Baryon Oscillation
Spectroscopic Survey (BOSS) \cite{anderson12}, WiggleZ survey
\cite{blake12} and 6dFGS survey \cite{beu11}. These measurements
include redshifts in the range $z\in [0.1,0.7]$. Table
\ref{baodata}
 reports the observed value of mentioned parameters.

\begin{table}[ht]
\tbl{ Observed data for BAO
\cite{Hinshaw:2012aka}. }
{\begin{tabular}{llcccc}
\hline \hline {\rm Redshift} & {\rm Data Set} &
{$r_s/D_V(z;\{\Theta\})$} & {\rm Ref.}\\ \hline
0.10   &  6dFGS     & $0.336\pm0.015$       &   \cite{beu11} \\
0.35 &  SDSS-DR7-rec    & $0.113\pm0.002$&  \cite{pad12} \\
0.57 &  SDSS-DR9-rec    & $0.073\pm0.001$ & \cite{anderson12} \\
0.44 &  WiggleZ & $0.0916\pm0.0071$     & \cite{Blake:2011en} \\
0.60 &  WiggleZ & $0.0726\pm0.0034$     &  \cite{Blake:2011en} \\
0.73 &  WiggleZ & $0.0592\pm0.0032$     &  \cite{Blake:2011en} \\
\hline
\end{tabular}\label{baoval}\label{baodata}}
\end{table}
The fisher information matrix is given by  \cite{Hinshaw:2012aka}
\begin{eqnarray}\label{covbao}
{\mathcal C}^{-1}_{BAO} = \left(\begin{array}{rrrrrr}
4444.4 & 0 & 0 & 0 & 0 & 0 \\
0 & 34.602 & 0 & 0 & 0 & 0 \\
0 & 0 & 20.661157 & 0 & 0 & 0 \\
0 & 0 & 0 & 24532.1  & -25137.7 & 12099.1 \\
0 & 0 & 0 & -25137.7 & 134598.4 & -64783.9 \\
0 & 0 & 0 & 12099.1 & -64783.9 & 128837.6 \\
\end{array}
\right).
\end{eqnarray}
Accordingly,  $\chi^2_{BAO}$ is written by
\begin{eqnarray}
\chi^2_{BAO}\equiv \Delta d^{\dag}\cdot {\mathcal
C}^{-1}_{BAO}\cdot \Delta d\,.
\end{eqnarray}
In above equation $\Delta d(z;\{\Theta\})\equiv
d_{obs}(z)-d(z;\{\Theta\})$, where ${\mathcal C}^{-1}_{BAO}$ is
given in (\ref{covbao}) and $d_{obs}(z)$ is reported in table
\ref{baoval}. In the next subsection for sake of clarity, we will
give proper explanation about the main constraints with CMB
observations mainly affecting the background evolution.

\subsection{CMB observations}
The most popular quantities devoted to CMB power spectrum are the
position of sound horizon in power spectrum ($l_a$), CMB shift
parameter ($R$) and the redshift of recombination ($z_*$). The
sound horizon angular scale at recombination in the flat Universe
read as
\begin{equation}
l_a=\pi\frac{\int_0^{z_{\ast}}\frac{dz}{H(z;\{\Theta\})}}{r_s(z_{\ast};\{\Theta\})}.
\end{equation}
The CMB shift parameter represented by $R$ is defined by
\begin{equation}
R(z_\ast;\{\Theta\})=\sqrt{\Omega_{\rm
m}}\int_0^{z_\ast}\frac{dz}{\mathcal{H}(z;\{\Theta\})}.
\end{equation}
here, the redshift at the recombination expresses as
$z_{\ast}=1048[1+0.00124(\Omega_bh^2)^{-0.738}(1+g_{1}(\Omega_{\mathrm{m}}h^2)^{g_2})]$.
Where the relevant parameters $g_1$ and $g_2$ can be found in
Ref.~\cite{Hu96}.
\begin{eqnarray}
g_1 &= &\frac{0.0783\, (\Omega_b h^2)^{-0.238}}
{1+39.5\, (\Omega_b h^2)^{0.763}}\,,\\
g_2 &= &\frac{0.560}{1+21.1\, (\Omega_b h^2)^{1.81}}\,.
\end{eqnarray}
To consider the contribution of mentioned observational
quantities, we use following likelihood function for CMB
observation \cite{Hinshaw:2012aka}:
\begin{eqnarray}
\chi^2_{CMB}=\Delta {\mathcal D}^{\dag}\cdot \mathcal
{C}_{CMB}^{-1}\cdot \Delta {\mathcal D}\,,
\end{eqnarray}
where $\Delta {\mathcal D}\equiv {\mathcal D}_{obs}-{\mathcal
D}(\{\Theta\})$, in which ${\mathcal D}_{\rm obs}$ for the 9-year
WMAP (WMAP9) observation with flat prior are
\cite{Hinshaw:2012aka}
\begin{eqnarray}
\hspace{-.5cm}{\mathcal D}_{obs} &\equiv&
\left(\begin{array}{c}
{l_a} \\
{ R}\\
{z_{\ast}}\end{array}
  \right)=
  \left(\begin{array}{c}
302.40\\
1.7246\\
1090.88\end{array}
  \right),
 \end{eqnarray}
and  $\mathcal {C}_{CMB}$ is as follows
\begin{eqnarray}
{\mathcal C_{{CMB}}}^{-1}=\left(
\begin{array}{ccc}
3.182 & 18.253  & -1.429  \\
18.253& 11887.879& -193.808\\
-1.429& -193.808& 4.556
\end{array}
\right).
\end{eqnarray}
According to Planck TT+LowP data the most important parameters to
use are \cite{Ade:2015rim}
\begin{eqnarray}
\hspace{-.5cm}{\mathcal D}_{obs} &\equiv&
\left(\begin{array}{c}
{l_a} \\
{ R}\\
{\Omega_bh^2}\end{array}
  \right)=
  \left(\begin{array}{c}
301.76 ,\\
1.7488\\
0.02228\end{array}
  \right),
 \end{eqnarray}
and  $\mathcal {C}_{CMB}$ is as follows
\begin{eqnarray}
{\mathcal C_{{CMB}}}=\left(
\begin{array}{ccc}
1 & 0.54  & -0.63  \\
0.54& 1& -0.43\\
-0.63& -0.43& 1
\end{array}
\right).
\end{eqnarray}
In this paper we use the former observational quantities for CMB.
Finally, we perform a MCMC analysis using joint likelihood of both
standard rulers and standard candles. The combined chi-square
function is:
\begin{equation}
\chi^2_{SHBC}={\chi^{2}_{Sn}} + {\chi^{2}_{H}}+{\chi^{2}_{BAO}}
+ {\chi^{2}_{CMB}}\,.
\end{equation}
The abbreviation (SHBC) stands in above equation is replaced by
\underline{S}NIa+\underline{H}ubble+\underline{B}AO+\underline{C}MB.
In the next subsection we will summarize the results achieved for
best fit values of the free parameters and compare them.

\subsection{Results}

In the following, we present the observational constraints on the
free parameters of ghost dark energy model with three interaction
terms introduced in section \ref{nli}.

\begin{figure}[t]
\center
\includegraphics[height=75mm,width=62mm,angle=0]{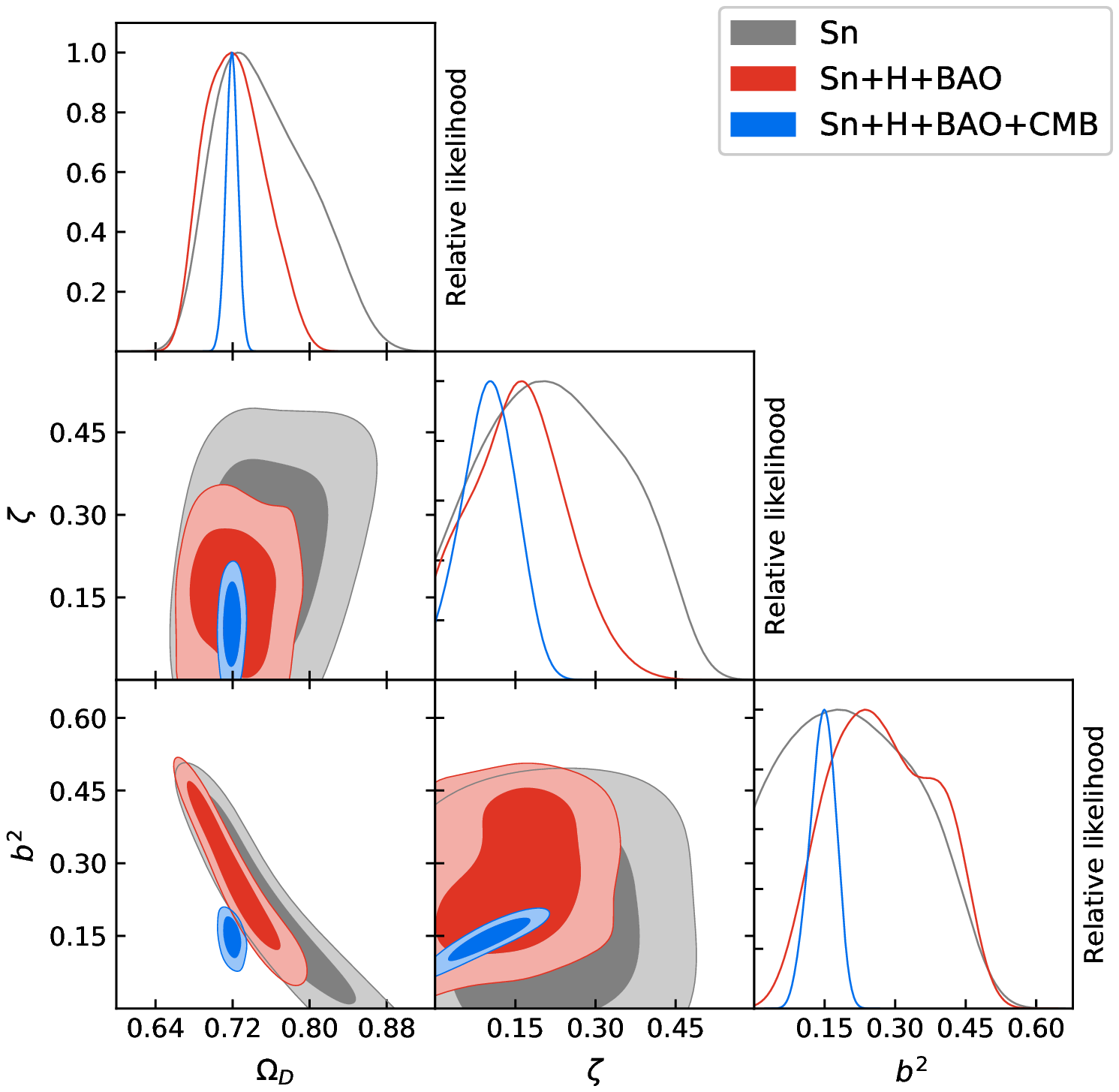}
\includegraphics[height=75mm,width=62mm,angle=0]{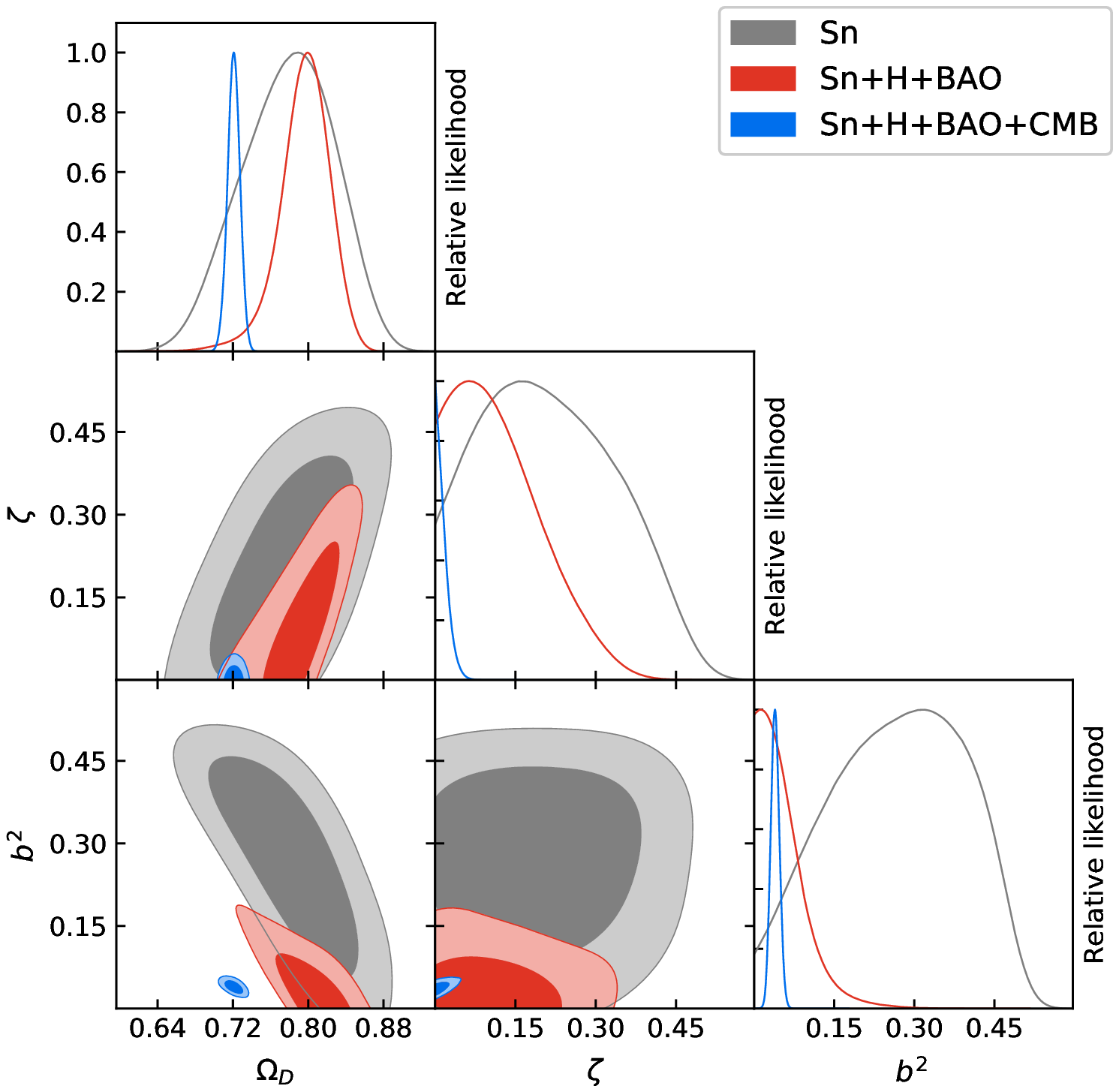}
\includegraphics[height=75mm,width=62mm,angle=0]{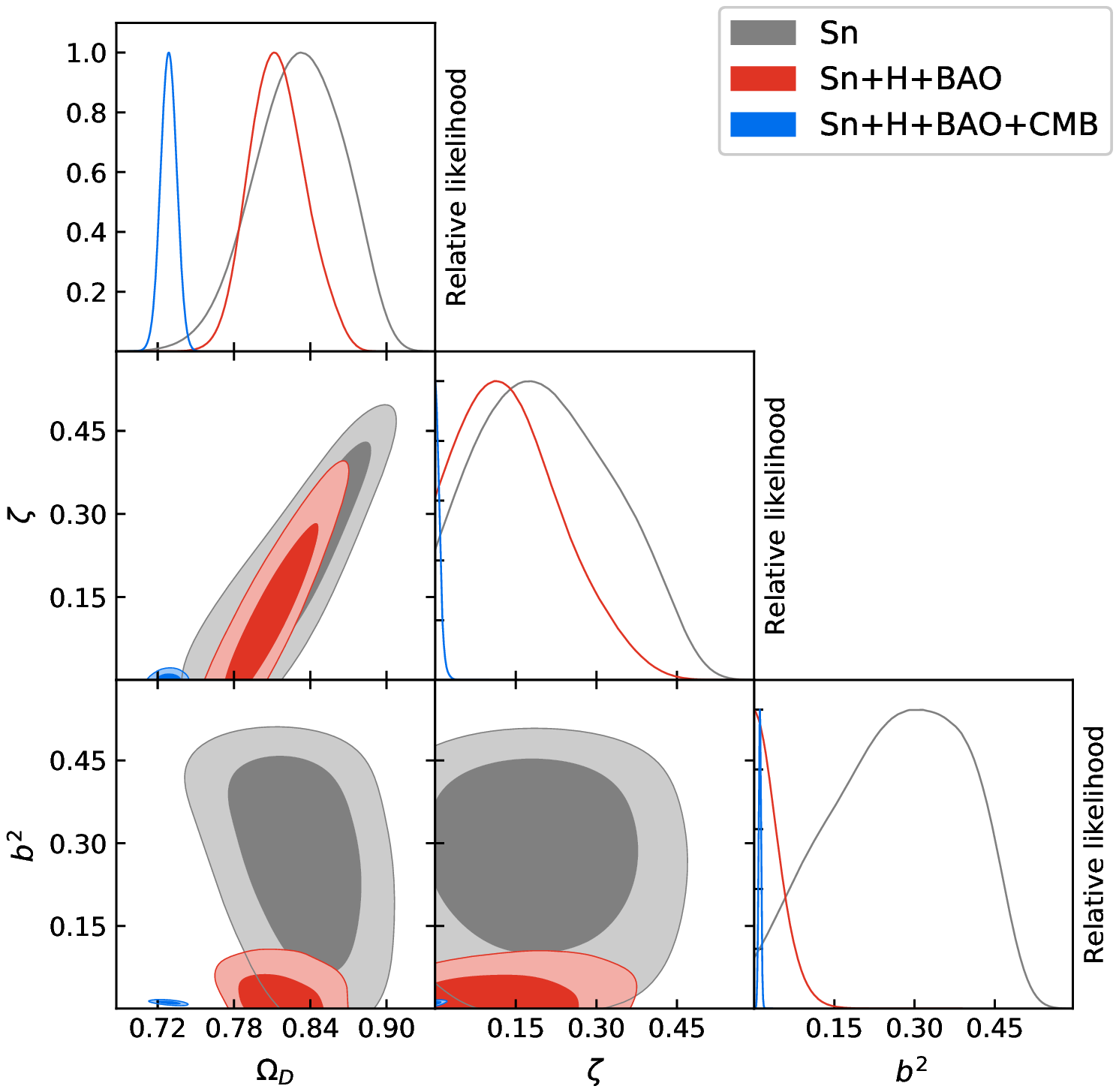}
\vspace{2mm}\caption{\footnotesize Upper left panel corresponds to
marginalized and contour plots of joint analysis base on
SNIa, SNIA+Hubble+BAO and SNIA+Hubble+BAO+CMB observations. The
upper right panel shows the same for second model while the lower
panel is devoted to model III of non-linear interacting dark
energy.} \label{jointall}
\end{figure}

{\bf Model I:} $Q=3b^2H\frac{\rho_D^3}{\rho^2}$ \\
SNIa observational constraint shows that
$\Omega_D=0.753^{+0.038}_{-0.060}$, $\zeta=0.23^{+0.12}_{-0.15}$ and
$b^2=0.23^{+0.10}_{-0.18}$ at $1\sigma$ confidence interval. The best fit values
at $68\%$ and $95\%$ intervals for SNIa using JLA catalog are reported in Table
\ref{tbl:model11}. 
Combining the other observational data sets, namely
Hubble parameters, BAO and CMB with SNIa observation results in
$\Omega_D=0.7192^{+ 0.0062}_{-0.0062}$, $\zeta=0.104^{+0.047}_{-0.047}$ and $b^2=0.146^{+0.030}_{-0.026}$
at $1\sigma$ confidence interval. Table \ref{tbl:model11} reports
the value at $68\%$ and $95\%$ optimal variance errors. Upper left
panels of Fig. \ref{jointall} illustrates the
marginalized likelihood functions and contours for free parameter
of first dark energy model using SNIa, SNIa+Hubble+BAO and
SNIa+Hubble+BAO+CMB respectively.

\begin{table}[h]
\tbl{ Best fit values for non-linear interacting dark energy model I using SNIa,
 SNIa+Hubble+BAO and joint combination of SNIa+Hubble+BAO+CMB (SHBC) at $68 \%$ and $95\%$ confidence intervals.} 
{\begin{tabular}{|c|c c c|}
\hline
\hline Parameters & SNIa & SNIa+Hubble+BAO & SHBC  \\
of model I&&&\\
\hline&&&\\
 $\Omega_{D}$& $0.753^{+0.038+0.094}_{-0.060-0.081}$&  $0.722^{+0.026+0.060}_{-0.036-0.054}$ &$0.7192^{+ 0.0062+0.012}_{-0.0062-0.012}$  \\  &&&\\  \hline &&&\\

$\zeta  $ & $0.23^{+0.12+0.20}_{-0.15-0.23}$ &
$0.161\pm 0.084<0.304$
 &$0.104^{+0.047+0.085}_{-0.047-0.096}$\\ &&&\\  \hline   &&&\\

$b^2$ & $0.23^{+0.10}_{-0.18}<0.443$& $0.28^{+0.11+0.19}_{-0.11-0.19}$&$0.146^{+0.030+0.050}_{-0.026-0.057}$\\  &&&\\
\hline \hline
\end{tabular}
\label{tbl:model11}}
\end{table}

\begin{table}[h]
     \tbl{ Best fit values for non-linear interacting dark energy
    model II using SNIa,SNIa+Hubble+BAO and joint combination of SNIa+Hubble+BAO+CMB (SHBC)
    at $68 \%$ and $95\%$ confidence intervals.}
  {  \begin{tabular}{|c|c c c|}
        \hline
        \hline Parameter & SNIa & SNIa+Hubble+BAO & SHBC  \\
        of model II &&&\\
        \hline&&&\\
        $\Omega_D $ &$0.779^{+0.054+0.088}_{-0.046-0.096}$
        & $0.796^{+0.029+0.052}_{-0.021-0.055}$ &$0.7209^{+0.0065+0.0130}_{-0.0065-0.0130}$
          \\  &&&\\  \hline &&&\\

        $\zeta  $ & $0.22^{+0.10}_{-0.16}<0.430$ & $<0.154<0.285$
         &$< 0.0173 < 0.0373$\\ &&&\\  \hline   &&&\\

        $b^2$ & $0.27^{+0.15+0.21}_{-0.12-0.23}$& $< 0.0669 < 0.1420$
        &$0.0395^{+0.0080+0.0160}_{-0.0080-0.0150}$\\  &&&\\
        \hline
        \hline
    \end{tabular}
    \label{tbl:model21}}
\end{table}
\begin{table}[h]
 \tbl{ Best fit values for non-linear interacting dark energy model III
    using SNIa, SNIa+Hubble+BAO and joint combination of SNIa+Hubble+BAO+CMB (SHBC) at $68 \%$ and $95\%$
    confidence intervals. }
  {\begin{tabular}{|c|c c c|}
        \hline
        \hline Parameter & SNIa & SNIa+Hubble+BAO & SHBC  \\
        of model III &&&\\
        \hline&&&\\
        $\Omega_D $ &  $0.831^{+0.038+0.064}_{-0.032-0.066}$& $0.814^{+0.020+0.046}_{-0.024-0.040}$
         &$0.7287^{+0.0062+0.0120}_{-0.0062-0.0120}$  \\  &&&\\  \hline &&&\\

        $\zeta  $ & $0.216^{+0.099}_{-0.160}<0.424$ & $0.150^{+0.054}_{-0.130}<0.325$
         &$< 0.00764 <0.0173$\\ &&&\\  \hline   &&&\\

        $b^2$ & $0.28^{+0.15+0.20}_{-0.11-0.24}$& $<0.0433< 0.0897$
        &$0.0109^{+0.0023+0.0044}_{-0.0023-0.0045}$\\  &&&\\
        \hline
        \hline
    \end{tabular}
    \label{tbl:model31}}
\end{table}

{\bf Model II:} $Q=3b^2H\frac{\rho_D\rho_m}{\rho}$ \\
For this model, SNIa observations constrain the value of
parameters as: $\Omega_D=0.779^{+0.054}_{-0.046}$,
$\zeta=0.22^{+0.10}_{-0.16}$ and $b^2=0.27^{+0.15}_{-0.12}$ at
$1\sigma$ confidence interval.  The best fit values at $68\%$ and
$95\%$ intervals for SNIa and their joint analysis are
reported in table \ref{tbl:model21}. 
Combining Hubble parameters, BAO and CMB observational data sets with
SNIa, results in $\Omega_D=0.7209^{+0.0065}_{-0.0065}$,
$\zeta< 0.0173$ and $b^2=0.0395^{+0.0080}_{-0.0080}$ at $1\sigma$
confidence interval. Table \ref{tbl:model21} reports the value at
$68\%$ and $95\%$ levels. Upper right panels of Fig.
 \ref{jointall} illustrate the marginalized
likelihood functions and contours for free parameters of the
second dark energy model.

{\bf Model III:} $Q=3b^2H\frac{\rho_m^2}{\rho}$ \\
For the third interacting dark energy model, SNIa
observations confine the value of parameters as:
$\Omega_D=0.831^{+0.038}_{-0.032}$, $\zeta=0.216^{+0.099}_{-0.160}$
and $b^2=0.28^{+0.15}_{-0.11}$ at $1\sigma$ confidence interval.
The best fit values at $68\%$ and $95\%$ intervals for SNIa
and their joint analysis are reported in table \ref{tbl:model31}.
Combining the
other observational data sets, namely Hubble parameters,  BAO and  CMB with SNIa
observations leads to  $\Omega_D=0.7287^{+0.0062}_{-0.0062}$,
$\zeta < 0.00764$ and $b^2=0.0109^{+0.0023}_{-0.0023}$ at $1\sigma$
confidence interval. Table \ref{tbl:model31} reports these values
at $68\%$ and $95\%$ levels. Bottom panels of Fig. \ref{jointall} illustrates the marginalized likelihood
functions and contours for free parameters of this model.

In order to give a quantitative and robust measure  to compare
three interacting dark energy models with $\Lambda$CDM model as a reference model, we must go beyond computing
$\chi^2$ and take into account the number of degrees of freedom
(dof). Referring to the fact that higher degrees of freedom
usually cause
to more consistency of the model and observations, we use
AIC~\cite{H.Akaike:1974} and BIC~\cite{G.schwarz:1978} criteria.
AIC is defined by:
\begin{equation}
{\rm AIC}=-2\ln{\mathcal{L}_{\rm{max}}}+2g,
\end{equation}
and BIC~\cite{G.schwarz:1978} is given by:
\begin{equation}
{\rm BIC}=-2\ln{\mathcal{L}_{\rm{max}}}+g\ln{N},
\end{equation}
In the above equations, $g$ is the number of free parameters and
$N$ is the number of observational data sets used for
constraining. The relative form of mentioned criteria are in
principle utilized, namely $\Delta {\rm
AIC}=\Delta\chi^{2}_{\rm{min}}+2\Delta g$ and $\Delta{\rm
BIC}=\Delta\chi^{2}_{\rm{min}}+\Delta g\ln{N}$. The lower values
of $\Delta$AIC and $\Delta$BIC, the more favored model by
observational data sets. Regarding the capability of AIC and BIC
measures, the former does not consider the number of observational
data set, while the latter takes into accounts observations
\cite{Xu:2016grp}. Since three models introduced in this paper
have same degrees of freedom, therefore  $\Delta$AIC and
$\Delta$BIC don't make sense. Subsequently we compute $\chi^2_{min}$ for $\Lambda$CDM model considering two free parameters and suppose to be a reference model. In such case $\Delta$AIC less than 10 corresponds to consistent model with respect to the reference model. Our results demonstrate that our three interacting dark energy models are supported by observations, but they are worse than $\Lambda$CDM model but still are good models. If we consider only SNIa dataset, the corresponding $\chi^2_{min}$ for models are $682.247$ for model I, $682.117$ for model II, $682.057$ for model III and $682.895$ for $\Lambda$CDM. Therefore the value of minimum $\chi^2$ for all interacting dark energy models are smaller than  $\Lambda$CDM due to their one more free parameter. If we take into account SNIa+Hubble+BAO the values of minimum chi-square are $705.537$ for model I, $706.942$ for model II, $707.122$ for model III and $706.424$ for $\Lambda$CDM.  The relevant results for our three
cases accompanying $\Lambda$CDM are summarized in
Tabs.\ref{aicsn}, \ref{aicsnhbao} and \ref{aicall} .  
For SNIa observational constraint and its combination with Hubble+BAO, we find that all non-linear interacting dark energy models are relatively good models. While for SNIa+Hubble+BAO+CMB, all non-linear interacting models are worse than $\Lambda$CDM but they are relatively good models in the sense of declaring observations.

\begin{table}
\tbl{The minimum value of $\chi^2$, AIC and BIC criteria for
our models and $\Lambda$CDM when we use SNIa datasets.}
{\begin{tabular}{|c| c cc c|}
\hline \hline
$\rm{Model}$ & $\chi^2_{min}$ & $\Delta\rm{AIC}$ & $\Delta\rm {BIC}$ & $\chi^2_{min}$/dof \\
\hline
$\Lambda$CDM  &682.895 & 0& 0& 0.925\\
\hline
$\rm{I}$  & 682.247& 1.352  & 5.959& 0.926\\
\hline
$\rm{II}$  &682.117 &1.222 &5.829& 0.925 \\
\hline
$\rm{III}$  & 682.057& 1.162&5.769 &0.925 \\
\hline
\end{tabular}
\label{aicsn}}
\end{table}

\begin{table}
\tbl{The minimum value of $\chi^2$, AIC and BIC criteria for
our models and $\Lambda$CDM when we use SNIa+Hubble+BAO datasets.}
{\begin{tabular}{|c| c cc c|}
\hline \hline
$\rm{Model}$ & $\chi^2_{min}$ & $\Delta\rm{AIC}$ & $\Delta\rm {BIC}$ & $\chi^2_{min}$/dof \\
\hline
$\Lambda$CDM  & 706.424& 0& 0& 0.908\\
\hline
$\rm{I}$  & 705.537&  1.113 & 5.772& 0.908\\
\hline
$\rm{II}$  &706.942 &2.518 &7.177& 0.910 \\
\hline
$\rm{III}$  & 707.122& 2.698& 7.357&0.910 \\
\hline
\end{tabular}
\label{aicsnhbao}}
\end{table}

\begin{table}
\tbl{The minimum value of $\chi^2$, AIC and BIC criteria for
our models and $\Lambda$CDM when we use SNIa+Hubble+BAO+CMB.}
{\begin{tabular}{|c| c cc c|}
\hline \hline
$\rm{Model}$ & $\chi^2_{min}$ & $\Delta\rm{AIC}$ & $\Delta\rm {BIC}$ & $\chi^2_{min}$/dof \\
\hline
$\Lambda$CDM  & 708.565& 0& 0& 0.907\\
\hline
$\rm{I}$  & 710.891&  4.326 & 8.989& 0.911\\
\hline
$\rm{II}$  &711.765 &5.200 &9.863& 0.912 \\
\hline
$\rm{III}$  & 712.726& 6.161& 10.824&0.914 \\
\hline
\end{tabular}
\label{aicall}}
\end{table}


\begin{figure}[t]
\center
\includegraphics[height=50mm,width=60mm,angle=0]{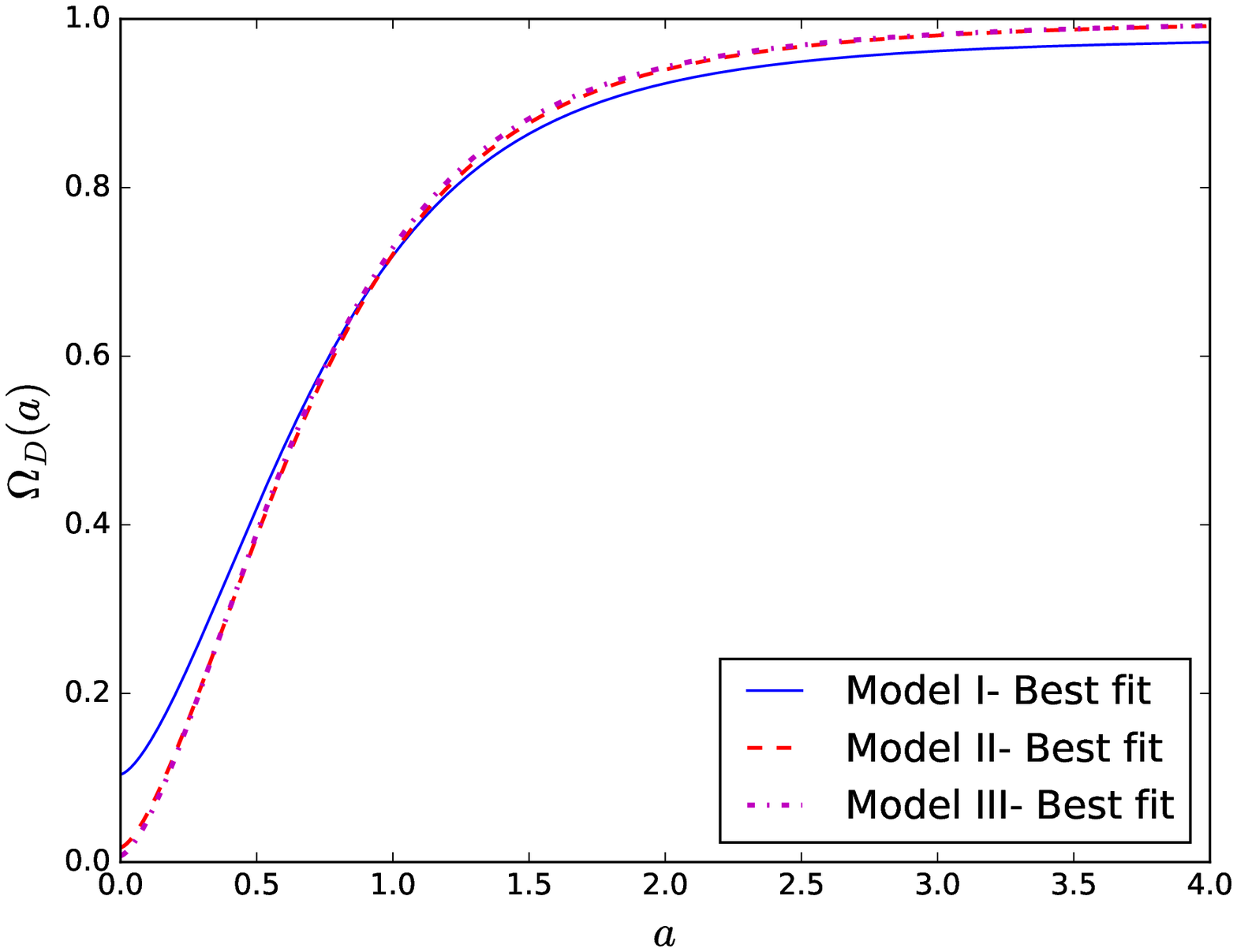}
\includegraphics[height=50mm,width=60mm,angle=0]{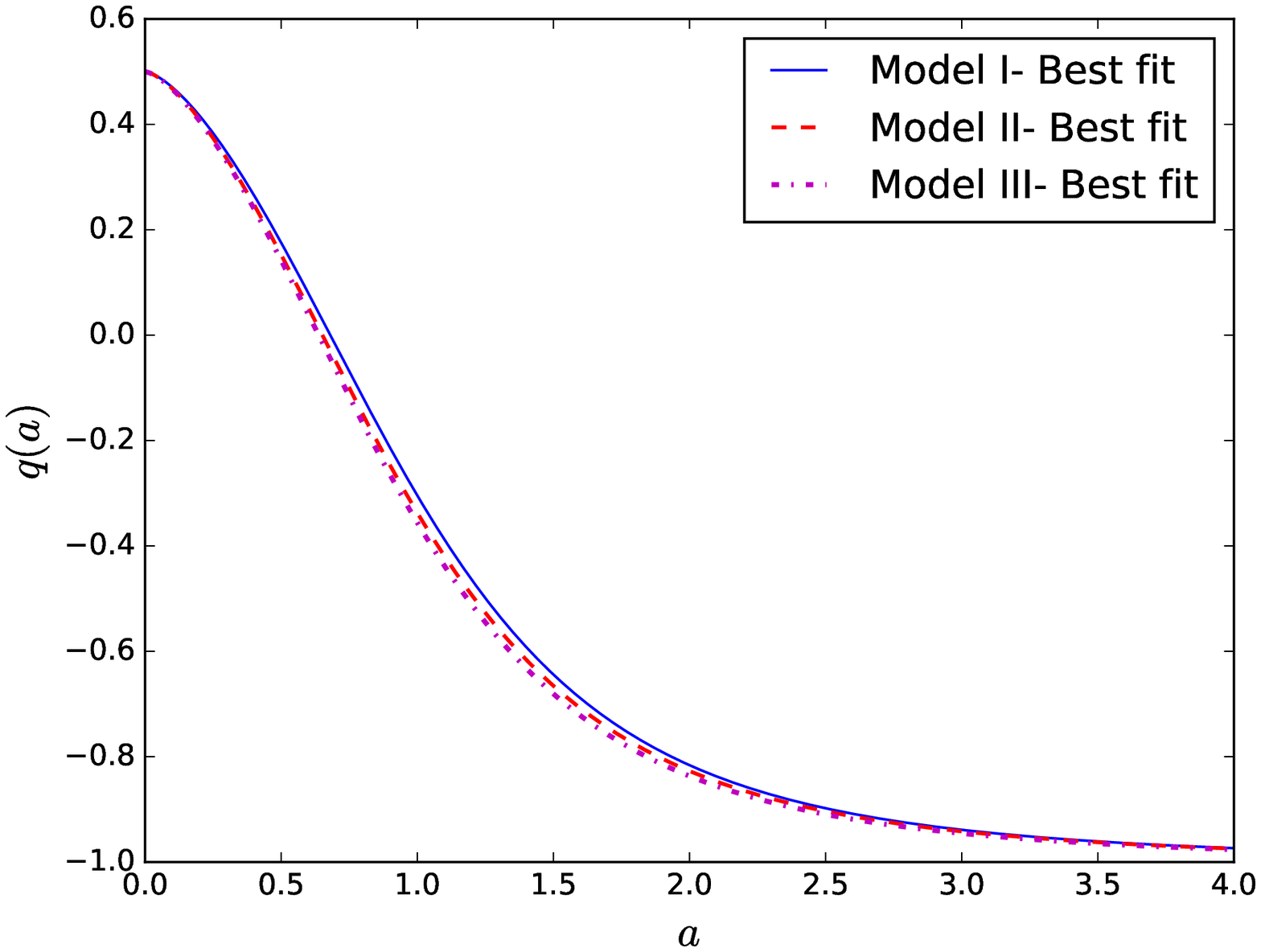} \\
\includegraphics[height=50mm,width=60mm,angle=0]{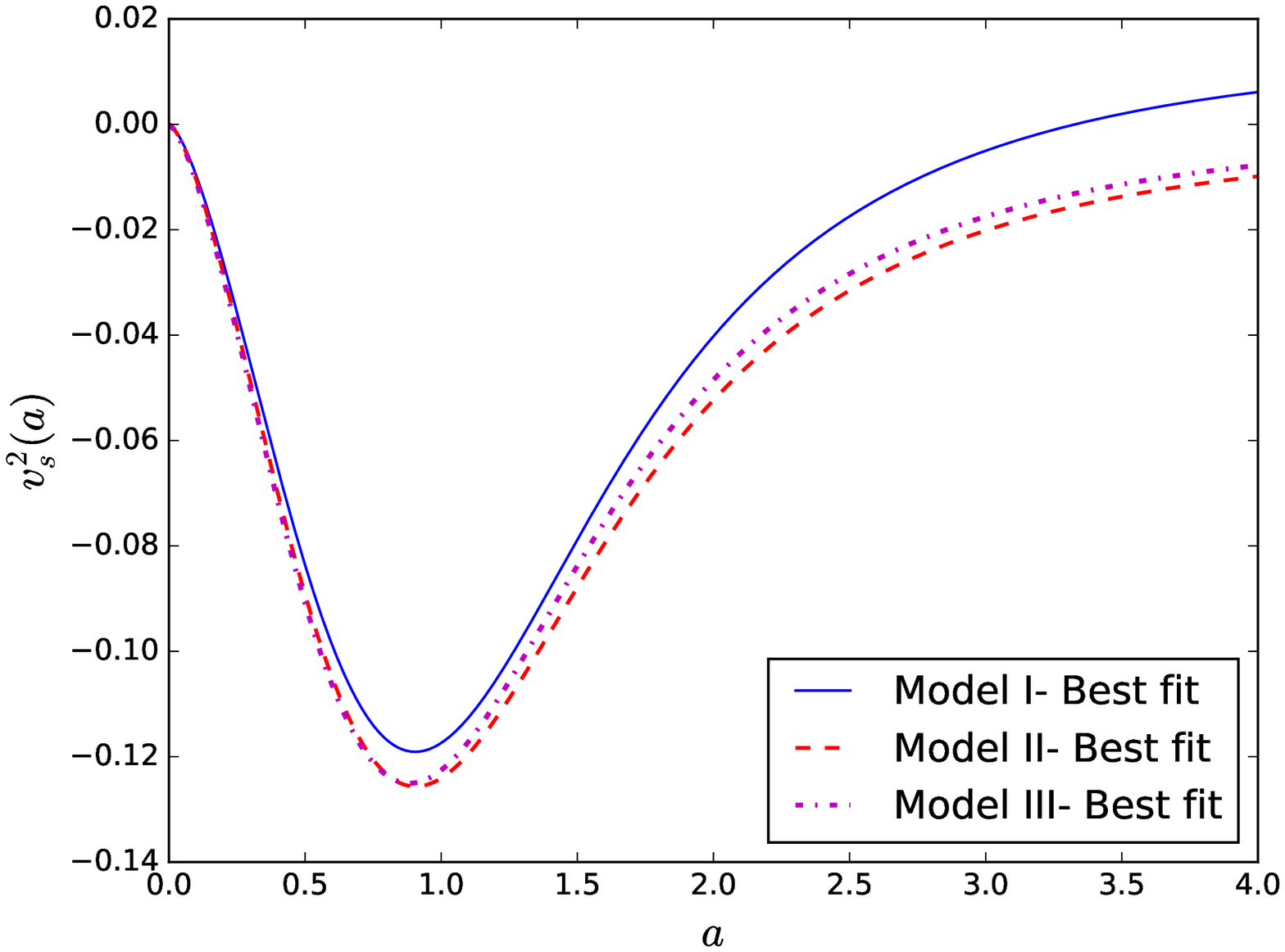}
\includegraphics[height=50mm,width=60mm,angle=0]{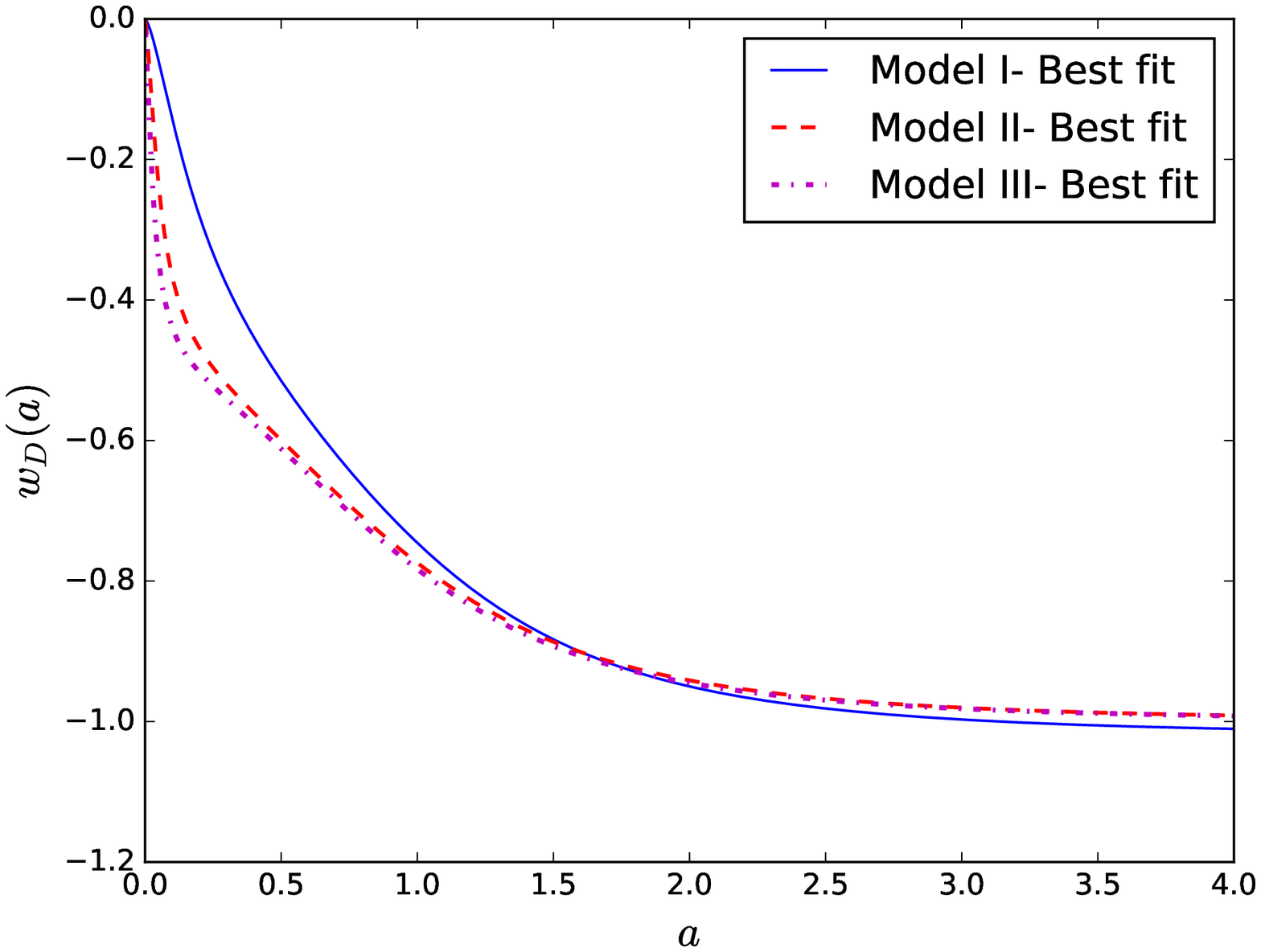}
\includegraphics[height=50mm,width=60mm,angle=0]{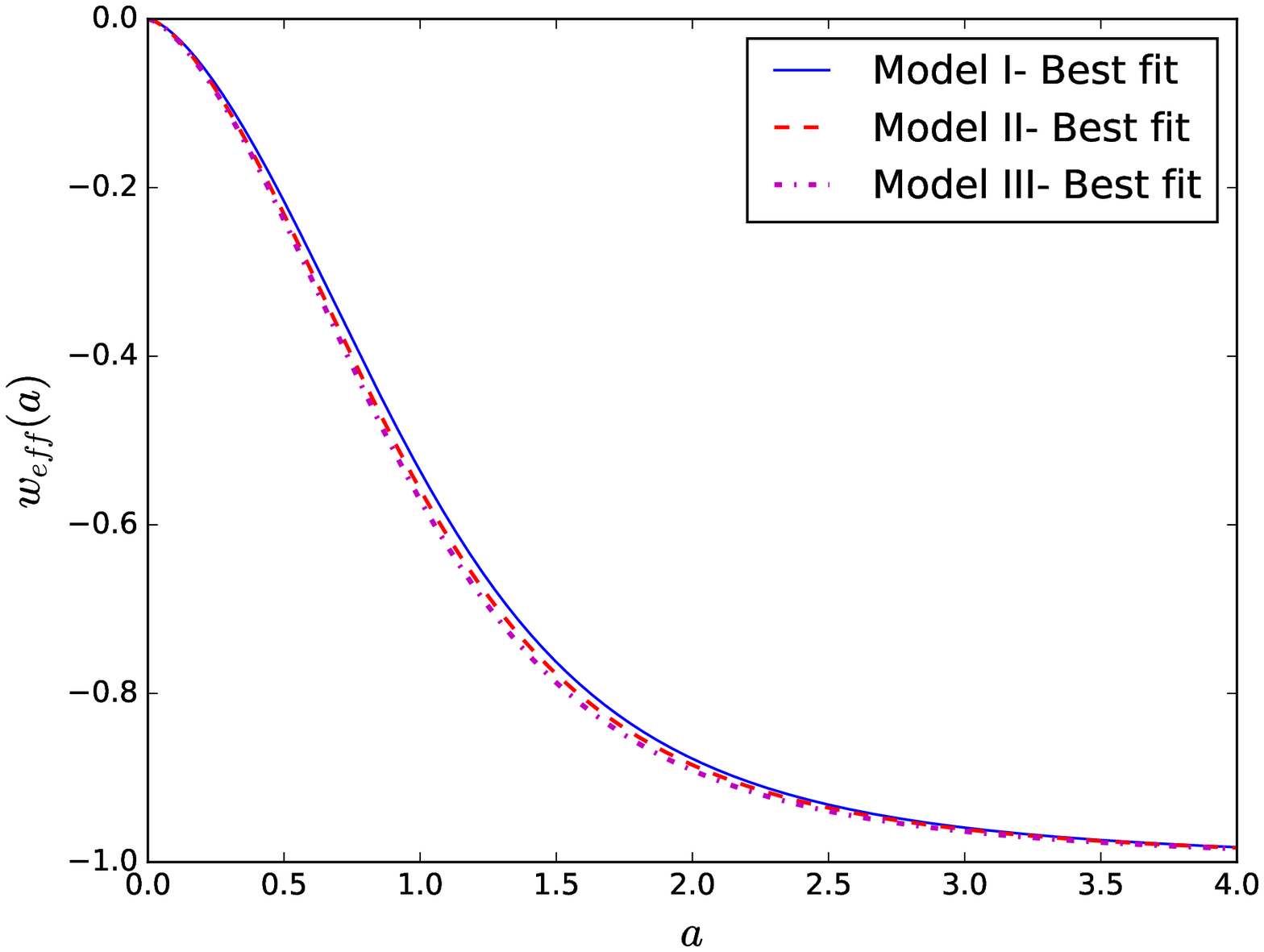}
\vspace{2mm} \caption{{\footnotesize The functional behavior of
$\Omega_D$, $q$, $v_s^2$, $w_d$ and $w_{eff}$ for model I, II and
III for the best fit value of free parameters. For the first model
$\Omega_D=0.7192^{+0.0062}_{-0.0062}$,
$b^2=0.46^{+0.030}_{-0.026}$ and $\zeta=0.104^{+0.047}_{-0.047}$.
For the second model $\Omega_D=0.7209^{+0.0065}_{-0.0065}$,
$b^2=0.0395^{+0.0080}_{-0.0080}$ and $\zeta\le0.0173$ and for the
third  model  $\Omega_D=0.7287^{+0.0062}_{-0.0062}$,
$b^2=0.0109^{+0.0023}_{-0.0023}$ and $\zeta\le0.00764$. Here we
choose the scale factor at present time as $a_0=1$.}} 
\label{all}
\end{figure}

Finally, in order to compare three models of dark energy
introduced in this paper, we indicate the behavior of  $\Omega_D$,
$q$, $v_s^2$, $w_d$ and $w_{eff}$ for the best fit values of free
parameters as a function of scale factor in Fig.\ref{all}. As
indicated in this plot, from the view point of $q$ and $w_{eff}$
the three models are rather similar and they show the expected
behavior of accelerating expansion of the universe. There is also
some features that distinct these models, for instance $\Omega_D$
for model I takes non-zero value at very early universe while this
quantity for model II and III goes to zero. The value of $v_s^2$
for first model takes positive value for $a>1$ but for second and
third models this quantity always remains in the negative domain.
The $w_{D}$ for third model goes to large negative values while
for model I and II this quantity remains in interval
$w_D\in[-1,0]$.

\section{Summary and Conclusion}\label{sum}
Ghost dark energy (GDE) is one of interesting models that solves
the recent acceleration problem. A great point about GDE is that
this model is based on the Veneziano ghost field which has already
presented in the literature and there is no need to introduce any
additional degree of freedom in physics. This model also solves
the coincidence problem \cite{CaiGhost}. However, GDE shows signs
of instability (due to negativity of adiabatic squared sound speed
$v_s^2$) in non-interacting and linear interacting domain. It is
possible to switch from GDE to the
generalized  ghost dark energy (GGDE) which $\rho_D=\alpha H+\beta H^2$. We also extended the interaction
between DE and DM to non-linear regime seeking for better
agreement with observations and improving the negativity of
squared sound speed of the model.

To this end, we proposed three forms of non-linear interaction
terms and investigated different features of these models.
Generally, one can consider the functional form $Q\equiv
f(H,\rho_m,\rho_D,\rho_{tot})$ for interaction between dark
sectors of the universe with a coupling coefficient denoted by
$b^2$. We considered GGDE with non-linear interaction terms in the
form $Q=3b^2H\rho_D^3/\rho^2$, $Q=3b^2H\rho_D\rho_m/\rho$ and
$Q=3b^2H\rho_m^2/\rho$ as model I, II and III, respectively. If we
suppose the universe is flat, the free parameters of the models
are $\{\Omega_D,\zeta,b^2\}$ .

{We evaluated the behavior of $w_D(a)$, $q(a)$, $w_{eff}(a)$,
$\Omega_D(a)$ and $v_s^2(a)$. As an interesting result we found
that $v_s^2$ takes positive values at $a>1$ for the best fit
values of parameters of the model I (Fig.\ref{all}), which is a
sign of stability of this model. However the squared sound speed
of models II, III remains negative. Considering the variations of
$q$, $w_{eff}$ (Fig.\ref{all}), one can see that all models show
the expected behavior of deceleration in the past and acceleration
in the present time and future, in fact there is a transition from
decelerating to accelerating phase at $z=0.69$, $z=0.67$ and
$z=0.66$ for models I, II and III respectively.
Our results also show the phantom behavior of DE in the model I at
late times. In the case of third model, $w_D$ takes large negative
values at very early times but it does not affect the model due to
the negligible value of $\Omega_D$ at the early times. It is also
notable that $\Omega_D$ of the first model takes non-zero values
in the early universe. This is another difference of model I and
models II and III.}

For observational consistency check, we used Joint Light-curve
analysis (JLA) for SNIa and used Hubble parameter for different redshifts. 
Six observational indicators of BAO and
3 well-defined CMB parameters have been used for that purpose.
For the first model, the joint analysis indicates that
$\Omega_D=0.7192\pm 0.0062$,
$b^2=0.146^{+0.030}_{-0.026}$ and $\zeta=0.104\pm0.047$
at 1 optimal variance error. For the second model the best fit
values at $1\sigma$ confidence are
$\Omega_D=0.7209\pm0.0065$,
$b^2=0.0395\pm0.0080$ and $\zeta\le0.0173$. According to
the combination of all observational data sets considered in this
paper, the best fit values for third non-linear interacting dark
energy model are $\Omega_D=0.7287^{+0.0062}_{-0.0062}$,
$b^2=0.0109^{+0.0023}_{-0.0023}$ and $\zeta\le0.00764$ at $1\sigma$
confidence interval.  
Based on $\Delta$AIC and $\Delta$BIC criteria reported in tables
\ref{aicsn}, \ref{aicsnhbao} and \ref{aicall} and considering
$\Lambda$CDM model as reference cosmological model, we found that
all non-linear interacting dark energy models are compatible with
current observations. Extending our approach to constraint on the
models based on large scale structures and those observations
beyond zero order perturbations could be interesting especially
for discrimination between different models of interacting dark
energy. This part of research is in progress and we will be
addressing them later.

\section*{Acknowledgment}
The work of E. Ebrahimi has been supported financially by Research
Institute for Astronomy and Astrophysics of Maragha (RIAAM) under
project No. 1/4165-55.


\begin{thebibliography}{0}    

\bibitem{Riess:1998cb}
  A.~G.~Riess {\it et al.} [Supernova Search Team Collaboration],
  Astron.\ J.\  {\bf 116}, 1009 (1998)
  doi:10.1086/300499
  [astro-ph/9805201].



\bibitem{Perlmutter:1998np}
  S.~Perlmutter {\it et al.} [Supernova Cosmology Project Collaboration],
  Astrophys.\ J.\  {\bf 517}, 565 (1999)
  doi:10.1086/307221
  [astro-ph/9812133].

\bibitem{bernardis}   P. de Bernardis, et al.,  Nature  404 (2000)
955.
\bibitem{perl2} S. Perlmutter, et al.,  Astrophys. J.  598 (2003) 102;
\bibitem{hanany} S. Hanany et al., Astrophys. J. Lett. 545, L5
(2000);
\bibitem{netter} C. B. Netterfield et al., Astrophys. J. 571, 604 (2002);
\bibitem{spergel} D.N. Spergel et al., Astrophys. J. Suppl. 148, 175 (2003).

\bibitem{Ade:2015rim}
  P.~A.~R.~Ade {\it et al.} [Planck Collaboration],
  arXiv:1502.01590 [astro-ph.CO].

\bibitem{coll} M. Colless et al., Mon. Not. R. Astron. Soc. 328, 1039
(2001);
\bibitem{teg} M. Tegmark et al., Phys. Rev. D 69, 103501 (2004);
\bibitem{cole} S. Cole et al.,
Mon. Not. R. Astron. Soc. 362, 505 (2005);
\bibitem{springel} V. Springel, C.S. Frenk, and S.M.D. White, Nature (London) 440, 1137
(2006).

  \bibitem{Percival:2009xn}
  W.~J.~Percival {\it et al.} [SDSS Collaboration],
  Mon.\ Not.\ Roy.\ Astron.\ Soc.\  {\bf 401}, 2148 (2010)
  doi:10.1111/j.1365-2966.2009.15812.x
  [arXiv:0907.1660 [astro-ph.CO]].


  \bibitem{Blake:2011rj}
  C.~Blake {\it et al.},
  Mon.\ Not.\ Roy.\ Astron.\ Soc.\  {\bf 415}, 2876 (2011)
  doi:10.1111/j.1365-2966.2011.18903.x
  [arXiv:1104.2948 [astro-ph.CO]].

  \bibitem{Reid:2012sw}
  B.~A.~Reid {\it et al.},
  Mon.\ Not.\ Roy.\ Astron.\ Soc.\  {\bf 426}, 2719 (2012)
  doi:10.1111/j.1365-2966.2012.21779.x
  [arXiv:1203.6641 [astro-ph.CO]].


  \bibitem{Sahni:1999gb}
  V.~Sahni and A.~A.~Starobinsky,
  Int.\ J.\ Mod.\ Phys.\ D {\bf 9}, 373 (2000)
  doi:10.1142/S0218271800000542
  [astro-ph/9904398].

  \bibitem{Carroll:2000fy}
  S.~M.~Carroll,
  Living Rev.\ Rel.\  {\bf 4}, 1 (2001)
  doi:10.12942/lrr-2001-1
  [astro-ph/0004075].


  \bibitem{Copeland:2006wr}
  E.~J.~Copeland, M.~Sami and S.~Tsujikawa,
  Int.\ J.\ Mod.\ Phys.\ D {\bf 15}, 1753 (2006)
  doi:10.1142/S021827180600942X
  [hep-th/0603057].

\bibitem{luon}
O. Luongo, H. Quevedo, Int. J. Mod. Phys. D, 23, 1450012, pp. 8, (2014).

  \bibitem{DeFelice:2010aj}
  A.~De Felice and S.~Tsujikawa,
  Living Rev.\ Rel.\  {\bf 13}, 3 (2010)
  doi:10.12942/lrr-2010-3
  [arXiv:1002.4928 [gr-qc]].

  \bibitem{Tsujikawa:2010zza}
  S.~Tsujikawa,
  Lect.\ Notes Phys.\  {\bf 800}, 99 (2010)
  doi:10.1007/978-3-642-10598-2-3
  [arXiv:1101.0191 [gr-qc]].

  \bibitem{Clifton:2011jh}
  T.~Clifton, P.~G.~Ferreira, A.~Padilla and C.~Skordis,
  Phys.\ Rept.\  {\bf 513}, 1 (2012)
  doi:10.1016/j.physrep.2012.01.001
  [arXiv:1106.2476 [astro-ph.CO]].

  \bibitem {Amendola12a} Amendola,  L.,  Appleby,  S.,  Bacon,  D.,  et  al.,  Cosmology  and  fundamental
physics with the Euclid satellite. 2012a, ArXiv e-prints,
arXiv:1206.1225



\bibitem{sola} J. Sola and H. Stefancic, Phys. Lett. B 624, 147 (2005).

\bibitem{wett} C. Wetterich, Nucl. Phys. B 302 (1988) 668;

\bibitem{Ratra} B. Ratra, J. Peebles, Phys. Rev. D 37 (1988) 321.

\bibitem{duta} S. Dutta, E. N. Saridakis and R. J. Scherrer, Phys. Rev. D 79,
103005 (2009).

\bibitem{saridakis} E. N. Saridakis and S. V. Sushkov, Phys. Rev. D 81, 083510
(2010).

\bibitem{Rahvar:2006tm}
  S.~Rahvar and M.~S.~Movahed,
  Phys.\ Rev.\ D {\bf 75}, 023512 (2007)
  doi:10.1103/PhysRevD.75.023512
  [astro-ph/0604206].


\bibitem{caldwell} R. R. Caldwell, Phys. Lett. B 545, 23 (2002).

\bibitem{nojiri1} S. Nojiri and S. D. Odintsov, Phys. Lett. B 562, 147 (2003)
\bibitem{singh} P. Singh, M. Sami and N. Dadhich, Phys. Rev. D 68, 023522
(2003).
\bibitem{hu} W. Hu, Phys. Rev. D 71, 047301 (2005).

\bibitem{setare} M. R. Setare and E. N. Saridakis, JCAP 0903, 002
(2009).

\bibitem{saridakis2} E. N. Saridakis, Nucl. Phys. B 819, 116
(2009).

\bibitem{hsu}S. D. H. Hsu, Phys. Lett. B 594, 13 (2004).

\bibitem{li} M. Li, Phys. Lett. B 603, 1 (2004).

\bibitem{li1} Q. G. Huang and M. Li, JCAP 0408, 013 (2004).

\bibitem{nojiri2} E. Elizalde, S. Nojiri, S. D. Odintsov and P. Wang, Phys. Rev.
D 71, 103504 (2005)

\bibitem{saridakis3} E. N. Saridakis, Phys. Lett. B 660, 138 (2008)

\bibitem{saridakis4} E. N. Saridakis, JCAP 0804, 020 (2008).

\bibitem{cai} R.G. Cai, Phys. Lett. B 657, 228 (2007).

\bibitem{weicai} H. Wei and R.G. Cai, Phys. Lett. B 660, 113 (2008).

\bibitem{weicai2} H. Wei and R.G. Cai, Eur. Phys. J. C 59, 99 (2009)

\bibitem{urban} F. R. Urban and A. R. Zhitnitsky, Phys. Lett. B 688 (2010) 9
;Phys. Rev. D 80 (2009) 063001; JCAP 0909 (2009) 018; Nucl. Phys.
B 835 (2010) 135.

\bibitem{ohta} N. Ohta, Phys. Lett. B 695 (2011) 41,
arXiv:1010.1339.

\bibitem{kawar} K. Kawarabayashi and N. Ohta, Nucl. Phys. B 175, 477 (1980).

\bibitem{witten} E. Witten, Nucl. Phys. B 156 (1979) 269;

 \bibitem{Veneziano}G. Veneziano,
Nucl. Phys. B 159 (1979) 213.

\bibitem{rosen} C. Rosenzweig, J. Schechter and C. G. Trahern, Phys. Rev. D 21
(1980) 3388.

\bibitem{nath} P. Nath and R. L. Arnowitt, Phys. Rev. D 23 (1981) 473.

\bibitem{CaiGhost} R.G. Cai, Z.L. Tuo, H.B. Zhang, arXiv:1011.3212.

\bibitem{sheykhi1} A. Sheykhi, A. Bagheri, Europhys. Lett. 95, 39001 (2011).

\bibitem{sheykhi2} E. Ebrahimi, A. Sheykhi, Phys. Lett. B 705, 19 (2011).

\bibitem{sheykhi3} E. Ebrahimi, A. Sheykhi, Int. J. Mod. Phys. D 20, 2369 (2011)
\bibitem{sheykhi4} A. Sheykhi, M. Sadegh Movahed, Gen. Relativ. Gravit.
[DOI:10.1007/s10714-011-1286- 3].

\bibitem{chao1} Chao-Jun Feng, Xin-Zhou Li, Ping Xi, JHEP 1205(2012) 046.

\bibitem{chao2} Chao-Jun Feng, Xin-Zhou Li, Xian-Yong Shen, Phys.Rev. D87 (2013)
023006.
\bibitem{chao3} Chao-Jun Feng, Xin-Zhou Li, Xian-Yong Shen,
Mod. Phys. Lett. A27 (2012) 1250182.

\bibitem{caighost2} R. G. Cai, Z. L. Tuo, Y. B. Wu, Y. Y. Zhao, Phys. Rev. D 86 (2012) 023511.

\bibitem{ebrsheyggde} E.~Ebrahimi, A.~Sheykhi and H.~Alavirad,
 Central Eur.\ J.\ Phys.\  {\bf 11}, no. 7, 949 (2013).



\bibitem{interact1} Bertolami O, Gil Pedro F and Le Delliou M 2007 Phys. Lett. B
654 165.

\bibitem{oli} G. Olivares, F. Atrio, D. Pavon, Phys. Rev. D 71 (2005) 063523.

\bibitem{mangano} G. Mangano, G. Miele, and V. Pettorino, Mod.Phys.Lett. A 18,
831 (2003), arXiv:astroph/ 0212518.

\bibitem{baldi} M. Baldi, Mon.Not.Roy.Astron.Soc. 411,1077 (2011),
arXiv:1005.2188

\bibitem{jian} Jian-Hua He and Bin Wang, JCAP 0806, 010 (2008).

\bibitem{Arevalo:2011hh}
  F.~Arevalo, A.~P.~R.~Bacalhau and W.~Zimdahl,
  Class.\ Quant.\ Grav.\  {\bf 29}, 235001 (2012)


\bibitem{afsh}
N. Afshordi, C. Coriano, L. Delle Rose, E. Gould, K. Skenderis, Phys. Rev. Lett. 118, 041301 (2017)

\bibitem{qiang}
C.-Qiang Geng, C.-Chi Lee, K. Zhang, Phys. Lett. B 757 (2016) 422-427

\bibitem{cuzin}
R. R. Cuzinatto, L. G. Medeiros, E. M. de Morais, Astroparticle Physics 73 (2016) 52

\bibitem{avil}
A. Aviles, C. Gruber, O. Luongo, H. Quevedo, Phys.Rev.D86, 123516 (2012)


\bibitem{rani}
S. Rani, A. Altaibayeva, M. Shahalam, J. K. Singh, R. Myrzakulov, JCAP 03 (2015) 031

\bibitem{mores}
M. Moresco, L. Verde, L. Pozzetti, R. Jimenez, A. Cimatti, JCAP, 07, 053, (2012)


\bibitem{dela}
A. de la Cruz-Dombriz, P. K. S. Dunsby, O. Luongo, L. Reverberi, Jour. Cosmo. Astrp. Phys., 12, 042, pp. 32, (2016).


\bibitem{chen}
Y. Chen, S. Kumar, B. Ratra, Astrophys.J. 835 (2017) 86


\bibitem{semiz}
I. Semiz, A. Kazim Camlibel, JCAP 1512 (2015) no.12, 038


\bibitem{gavela09} M. B. Gavela, D. Hernandez, L. L. Honorez, O. Mena, and S. Rigolin, JCAP
0907 (2009) 034, [0901.1611].
\bibitem{Majerotto09}E. Majerotto, J. Valiviita, and R. Maartens, 0907.4981.









\bibitem{ebrinsggde} E. Ebrahimi, A. Sheykhi, Int J Theor Phys (2013)
52:2966"1¤776.

\bibitem{Golchin:2016yci}
  H.~Golchin, S.~Jamali and E.~Ebrahimi,
  arXiv:1605.05068 [gr-qc].


  \bibitem{Hinshaw:2012aka}
  G.~Hinshaw {\it et al.} [WMAP Collaboration],
  Astrophys.\ J.\ Suppl.\  {\bf 208}, 19 (2013)


\bibitem{Ade:2015xua}
  P.~A.~R.~Ade {\it et al.} [Planck Collaboration],
  arXiv:1502.01589 [astro-ph.CO].

\bibitem{ries04}A. G. Riess et al., ApJ. 607 (2004) 665.
\bibitem{ries07}A. G. Riess et al., ApJ. 659 (2007) 98.

\bibitem{ast06}P. Astier et al., Astron. Astrophys. 447 (2006) 31;
\bibitem{bau08} S. Baumont et al., Astron. Astrophys. 491 (2008) 567.

\bibitem{reg09}N. Regnault et al., arXiv:0908.3808
 \bibitem{guy10}J. Guy et al., arXiv:1010.4743.
\bibitem{mik07}G. Miknaitis et al., ApJ. 666 (2007) 674
\bibitem{wood07} W. M. Wood-Vasey et al., ApJ.
666 (2007) 694.
\bibitem{cop06}Y. Copin et al., New Astronomy Rev. 50 (2006) 436
\bibitem{scal09} R. A. Scalzo et al.,
ApJ. 713 (2009) 1073
\bibitem{fol10}G. Folatelli et al., AJ. 139 (2010) 120
\bibitem{fol101} G. Folatelli et al., AJ. 139 (2010)
519.

\bibitem{lea10}J. Leaman et al., arXiv:1006.4611
\bibitem{li10} W. D. Li et al., arXiv:1006.4612; W. D.
Li et al., arXiv:1006.4613.
\bibitem{holtz08}J. A. Holtzman et al., AJ. 136 (2008) 2306
\bibitem{kess09} R. Kessler et al., ApJS. 185
(2009) 32.




\bibitem{Suzuki12} Suzuki, N., et al. 2012, ApJ, 746, 85

\bibitem{Cao:2014jza}
  S.~Cao and Z.~H.~Zhu,
  Phys.\ Rev.\ D {\bf 90}, no. 8, 083006 (2014)


\bibitem{Betoule14} M. Betoule et al. [SDSS Collaboration], Astron. Astrophys. 568
(2014) A22 [arXiv: 1401.4064 [astro-ph.CO]].
\bibitem{JLAdata1} \texttt{http://supernova.lbl.gov/Union/}
\bibitem{JLAdata2}\texttt{http://supernovae.in2p3.fr/sdss\char`_snls\char`_jla/ReadMe.html}

\bibitem{li11} Li, Z. X., Wu, P. X., \& Yu, H. W. 2011, PLB, 695, 1
\bibitem{farooq13} O. Farooq and B. Ratra, Astrophys. J. 766, L7 (2013)


\bibitem{pad12} Padmanabhan, N., et al., Mon.\ Not.\ Roy.\ Astron.\ Soc.\  {\bf 427}, 2132
(2012)[arXiv:1202.0090].
\bibitem{anderson12} Anderson, L., et al.,Mon.\ Not.\ Roy.\ Astron.\ Soc.\  {\bf 427}, 3435
(2013)[arXiv:1203.6594].
\bibitem{blake12} Blake, C., et al. 2012, MNRAS, 425, 405
\bibitem {beu11} Beutler, F., et al. 2011, MNRAS, 416, 3017





\bibitem{Blake:2011en}
  C.~Blake {\it et al.}, Mon.\ Not.\ Roy.\ Astron.\ Soc.\  {\bf 418}, 1707 (2011)

\bibitem{Hu96} Hu, W., \& Sugiyama, N. 1996, ApJ, 471, 542









\bibitem{H.Akaike:1974} H. Akaike, A new look at the statistical model identica-
tion, IEEE Trans. Automatic Control 19, 716 (1974).

\bibitem{G.schwarz:1978} G. Schwarz, Estimating the dimension of a model. Ann.
Stat. 6, 461 (1978)

\bibitem{Xu:2016grp}
  Y.~Y.~Xu and X.~Zhang,
  arXiv:1607.06262 [astro-ph.CO].


\end{thebibliography}
\end{document}